\documentclass[12pt,a4paper]{article}

\usepackage{soul} 
\usepackage{macros}
\usepackage{tikz-feynman, geometry, epstopdf, epsfig, amsmath, amssymb, slashed, bbold, graphicx, bm}
\usepackage{pdfpages, float, cite, lscape, hyperref}
\tikzfeynmanset{compat=1.0.0}
\usepackage[toc,page]{appendix}
\usepackage[export]{adjustbox}
\usepackage{authblk}
\usepackage{caption}
\usepackage{todonotes} 
\graphicspath{{figures/}}
\DeclareGraphicsExtensions{.pdf}
\newcommand\livpool{Theoretical Physics Division, Department of Mathematical Sciences,
  University of Liverpool, Liverpool L69 3BX, UK}
\newcommand\hope{School of Mathematics, Computer Science and Engineering, 
 Liverpool Hope University, Hope Park, Liverpool L16 9JD, UK}

\begin{document}
\title{Non-perturbative renormalisation with interpolating momentum schemes}
\author[1,2]{N. Garron}
\affil[1]{\hope}
\author[2]{C. Cahill}
\author[2]{M. Gorbahn}
\author[2]{\\ J. A. Gracey} 
\author[2]{P. E. L. Rakow}
\affil[2]{\livpool}

\date{\today }

\maketitle

\begin{center}
\section*{Abstract}
\end{center}
Hadronic matrix elements evaluated on the lattice can be converted
to a continuum scheme such as $\MSbar$ using intermediate non-perturbative
renormalisation schemes.
Discretisation effects on the lattice and convergence of the  continuum
perturbation  theory are both scheme dependent  and we explore 
this dependence in the framework of the Rome-Southampton method
for generalised  kinematics.
In particular, we implement several non-exceptional {\em interpolating}
momentum schemes, where the momentum transfer is {\em not} restricted
to the symmetric point defined in RI/SMOM. 
Using flavour non-singlet quark bilinears, we compute the renormalisation
factors of the quark mass and wave function for $N_f=3$ flavours of
dynamical quarks. 
We investigate the perturbative and non-perturbative scale-dependencies.
Our numerical results are obtained from lattice simulations performed with
Domain-Wall fermions, based on ensembles generated by RBC-UKQCD collaborations;
we use two different lattice spacings $1/a \sim 1.79 $ and $2.38$ GeV.
We also give the numerical values for the relevant anomalous dimensions
and matching coefficients at next-to-next-to-leading order. 

\vspace{1.cm}{\footnotesize{\mbox{}\hfill {LTH 1290}}}
\newpage 

\sect{Introduction}
\label{sec:Intro}

One major goal of lattice QCD is to determine the hadronic matrix elements of 
hadronic operators, with particular emphasis on flavour non-singlet quark 
bilinear operators, such as the local scalar, vector, axial vector or 
pseudoscalar operators. To produce physically meaningful results we need 
renormalisation factors to relate the bare lattice results to quantities in 
standard renormalisation schemes,
such as the modified minimal subtraction scheme, $\MSbar$.
This conversion can be done in lattice perturbation theory but it is well known
that it results in large systematic error because the perturbative series
converges poorly. 
Instead, in the last couple of decades, more efficient methods have been developed,
the most popular ones being the Schr\"odinger Functional~\cite{Luscher:1992an,Sint:1993un} and
Rome-Southampton~\cite{Martinelli:1994ty} frameworks.
In this work we follow the latter. 
In both cases, the strategy comprises two distinct steps:
First the renormalisation factors are obtained non-perturbatively from lattice
simulations in a given intermediate scheme, hence the name
Non-Perturbative Renormalisation (NPR).
Then these factors are converted to $\MSbar$
providing us with a bridge between low and high energies.
This second step is done perturbatively, but in the framework
of the Rome-Southampton method, it can be done in standard continuum perturbation
theory so that we can take advantage of the multi-loop computation
available in the literature. 
In principle the method is relatively simple: one makes lattice measurements
of Green's functions with bilinear operators inserted into quark
$2$-point functions (for example), 
and compares the lattice results with continuum perturbative results.
This method can also be applied to more complicated Green functions,
higher twist bilinears, four-quark operators, etc. 
It is worth noting that although traditionally we work in massless limit,
a massive momentum scheme has been developed in~\cite{Boyle:2016wis}.
\\

In early lattice studies the operator insertion was at 
zero momentum transfer, see for example~\cite{Martinelli:1994ty,Franco:1998bm}
for the early direction of this approach.
This choice of kinematics corresponds to an exceptional momentum 
configuration and is subject to potential infrared issues.
The origin of the problem and potential solutions were already discussed in detail
in~\cite{Aoki:2007xm}. 
These infrared problems can become even more severe when the chiral limit is taken,
and had lead to significant discrepancies in the case of neutral meson
mixing~\cite{Garron:2016mva,Boyle:2017skn}.
With the improvement in lattice analyses as well as the use of dynamical fermions, the 
refinement of the non-perturbative structure that is of phenomenological 
interest became suspect due to the emergence of the latent infrared issues.\\

Therefore in~\cite{Sturm:2009kb,Gorbahn:2010bf,Almeida:2010ns,Gracey2011},
an extension to the zero momentum insertion 
for this class of Green's functions was developed. In particular the Green's 
functions were computed with a non-zero momentum flowing through the operator 
and external quark legs at what is now termed the fully Symmetric MOMentum 
(SMOM) subtraction point\footnote{To ease the notation we drop the `RI' in
RI/SMOM.}. This circumvents any potential ambiguity in assessing
infrared effects and in particular in the chiral limit. 
In either scenario of zero or non-zero 
operator momentum flow lattice measurements were assisted by matching to the 
continuum limit. By this we mean that one can compute the same Green's function
for any of the above operators in the continuum field theory at several loop 
orders in the $\MSbar$ scheme. In the zero momentum case this has been carried 
out to three loops, but for the non-zero momentum case this has
only been carried out to two loops, \cite{Sturm:2009kb,Gorbahn:2010bf,Almeida:2010ns}
through there has been progress towards the next loop order in \cite{Bednyakov:2020,Kniehl:2020}.
Such continuum  matching allows one to accurately tune to
the ultraviolet before extrapolating to the non-perturbative or low energy 
region where measurements are made and errors estimated.  \\
 
The effect of using a momentum configuration which is infrared secure
can be seen for example
in~\cite{Arthur:2010ht,Aoki:2010dy,Aoki:2010pe,Boyle:2017skn,Bi:2017ybi,Chang:2018uxx}
where the renormalisation factors for the various flavour non-singlet
quark bilinear or four-quark operators were calculated for both momentum configurations.
A quantity which can be used as a benchmark is the ratio of the vector and 
axial vector operators. On symmetry grounds this has to be unity 
 for Domain-Wall fermions
and thus provides a stringent test of determining which momentum configuration
was more reliable. It is evident from the analysis of,
for example, \cite{Arthur:2010ht,Bi:2017ybi, Chang:2018uxx} that the SMOM
ratio is unity over virtually the full momentum range and well into the deep 
infrared in marked contrast to the zero momentum case. While this augurs well 
for significant improvement in lattice measurements and errors, one question
naturally arises which will form the main focus of our article. \\

Now that it is established that the SMOM non-exceptional momentum configuration gives a 
clear improvement, it is only one of many possible configurations that can be 
studied. Indeed, for example in \cite{Sturm:2009kb,Gorbahn:2010bf}, a more general off-shell
case was developed which is termed IMOM to denote the interpolating momentum subtraction 
configuration. In the SMOM definition the momentum squared of the three 
external momenta were all equal. 
By contrast in the IMOM case the magnitude of 
the operator momentum flow is allowed to vary and this is governed by  
 an additional kinematic
parameter $\omega$. For example the SMOM case corresponds to $\omega$~$=$~$1$.
Varying $\omega$ between $0$ and $4$, which are the respective values giving 
infrared or collinear singularities, allows one to search for a value which can
improve the matching to the continuum, the non-perturbative behaviour
or maybe even reduce the lattice artefacts.
This therefore leads to our 
investigation to see if there is an optimal value for $\omega$, possibly 
different from one, which minimises errors and improves benchmarks such as that
studied in \cite{Arthur:2010ht,Bi:2017ybi, Chang:2018uxx}.
This would be the appropriate next stage after the 
success of SMOM lattice studies especially in the chiral limit. For instance 
the ratio of vector and axial vector renormalisation constants can be examined 
in the chiral limit to gauge pion mass effects. 
While $\omega$ will be our main
parameter the overall mass scale $\mu$ at which measurements are made will also
be important. Therefore to have a thorough investigation we have performed a 
lattice computation of various operators on several different lattices for half
integer values of $\omega$ between $0$ and $4$ for various values of $\mu$. 
This has allowed us to identify an optimal area of the $(\omega,\mu^2)$-plane 
where there is an effective plateau upon which measurements are reliably 
accurate. Moreover the picture appears to be consistent for the operators we 
consider, though the actual locations are not necessarily the same point for 
each case. It suggests that
 there should be a similar qualitative behaviour
 for operators with covariant derivatives, which we do not consider here.\\

Although the article 
primarily concentrates on the lattice computations and analysis we were 
indebted throughout to the various two loop $\MSbar$ continuum results 
accumulated over the last decade,
\cite{Sturm:2009kb,Gorbahn:2010bf,Almeida:2010ns,Bi:2017ybi,Bell:2016nar,Gracey2011}, to 
facilitate the continuum matching. Moreover the measurement of quark bilinear 
currents in the IMOM configuration will give crucial direction to $3$-point 
quark Green's functions such as those relevant to kaon mixing for example, 
\cite{Garron:2016mva,Boyle:2017skn}. In that scenario the effective $4$-point functions can be 
mimicked by products of the Green's functions we consider here.
It is particularly important in this case, as it has been argued that the choice
of kinematics was responsible for the discrepancy observed in the literature.

The paper is organised as follows: In Section \ref{sec:kinematics} we outline the kinematics
setup, explaining in detail how our incoming and outgoing momenta are chosen to lead
to a given value of $\omega$ for a given value of $\mu$. In Section~\ref{sec:implementation}
we explain how we obtain the $Z$-factors, and discuss our choice of projectors.
In Section~\ref{sec:results}, we present our results for the $Z-$factors
and their scale-evolutions.
We conclude in Section~\ref{sec:conclusions}.
Appendix~\ref{app:PT} shows some perturbative results,
such as the values of the conversion factors for matching from IMOM to $\MSbar$,
computed at Next-to-Next-To-Leading Order (NNLO)
More details about the lattice simulations can be found in Appendix~\ref{lattice}.

\sect{Kinematics}\label{sec:kinematics}

 We consider the interaction in the diagram below, where the incoming momentum is defined as
$p_1$, the outgoing momentum as $p_2$, and the momentum transfer as $q$.

\begin{figure}[H]
\bc
\includegraphics[scale=1]{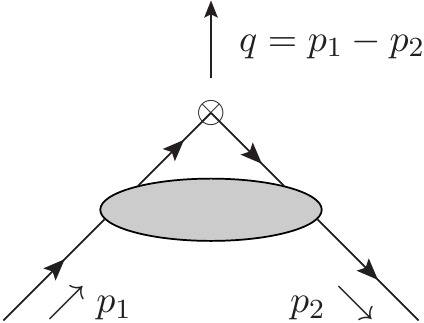}
\ec
\caption{We consider an incoming momentum $p_1$, outgoing momentum $p_2$ and a momentum transfer $q=p_1-p_2$.}
\label{fig:bil_kin}
\end{figure}

The method is based on the Rome-Southampton regularisation independent momentum scheme:
the fermions are off-shell,  we work in the chiral limit and the renormalisation scale
is given by the choice of external momenta.
In the original $\mathrm{RI/MOM}$ scheme (or $\mathrm{RI{}^\prime /MOM}$)~\cite{Martinelli:1994ty},
the incoming and outgoing momenta are equal,  $p_1 = p_2$, so that there is no momentum transfer, $q=0$.
The renormalisation scale $\mu$ is defined by $\mu^2=p_1^2=p_2^2$. In contrast, in the  $\mathrm{SMOM}$
schemes, the incoming and outgoing momenta are different: they are chosen
such that $q^2=p_1^2=p_2^2$; naturally the renormalisation scale is given by $\mu=\sqrt{q^2}$.
The advantage is to avoid exceptional kinematics which behave badly in the infrared
regime~\cite{Sturm:2009kb}. \\

\begin{figure}[H]
\begin{center}
\includegraphics[scale=1]{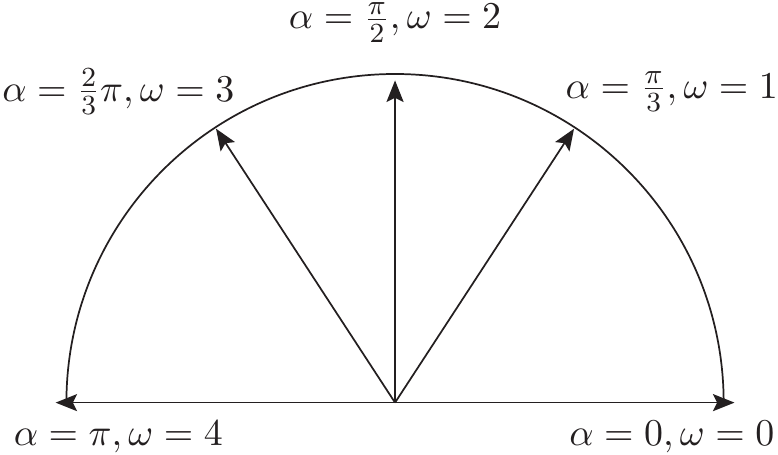}
\caption{The parameter $\omega$ is dependent upon the value of the angle $\alpha$ between the incoming and outgoing momenta.}
\label{fig:om_diagram}
\end{center}
\end{figure}

The above schemes can be generalised further.
We retain the condition $p_1^2 = p_2^2 = \mu^2$ but drop the requirement $q^2 = \mu^2$.
We define the variable $\omega = q^2/\mu^2$ to parameterise the kinematics. $\omega = 0$ corresponds to the
original RI/MOM and RI'/MOM schemes, $\omega=1$ corresponds to the SMOM scheme. The possible values of $\omega$
range between 0 and 4. Let us denote by $\alpha$ the angle between the incoming and outgoing momenta.
We choose our frame such that the momenta take the following form
(the time component is represented by the last entry):       
\bea 
\label{eq:p1}
p_1 &=& \mu \left[ 1, 0, 0, 0 \right],\\
\label{eq:p2}
p_2 &=& \mu \left[\cos \alpha, \sin \alpha, 0, 0\right].
\eea
This gives
\begin{equation}
q=p_1 - p_2 = \mu \left[1- \cos \alpha, -\sin \alpha, 0, 0\right] \;,
\end{equation}  
and
\begin{eqnarray}
q^2 &=& \mu^2\left(  (1-\cos \alpha)^2 + \sin^2 \alpha)\right) \;  \\
    &=& 2 \mu^2\left(  1 -\cos \alpha \right) \; \\
    &\equiv& \omega \mu^2 \; ,
\end{eqnarray}
where we have defined  $\omega = 2 (1-\cos\alpha)$. 
Clearly the parameter $\omega$ can take any real value between 0 and 4, and the value of $\omega$
is directly related to the value of $\alpha$ as shown in Fig~\ref{fig:om_diagram}.
We can check that the extreme cases $\omega=0$ and $\omega=4$
correspond to $p_1=p_2$ and $p_1=-p_2$, respectively,
and therefore these situations suffer from infrared and collinear singularities.
We also note that $\omega=0$ is the choice made in the original (exceptional)
kinematics of a standard RI/MOM scheme while $\omega=1$ corresponds
to the SMOM kinematic of~\cite{Sturm:2009kb}.
This has been known for a while, see for example \cite{Sturm:2009kb, Gorbahn:2010bf}.
Different choices of $\omega$ are referred to as {\em interpolating momentum}
schemes~\cite{Almeida:2010ns,Bell:2016nar,Gracey:2018fkg}. However, to our knowledge, there is no numerical
study or practical implementation of such kinematics. 
\\

Let us now turn to the lattice implementation. If $L$ is the physical spatial extent
of the lattice (we do not consider the temporal extent here for simplicity), we have 
\begin{eqnarray}
p_1 &=& \frac{2\pi}{L} \left[ l, 0, 0, 0 \right] \;,\\
p_2 &=& \frac{2\pi}{L} \left[ m, n, 0, 0 \right] \;,
\end{eqnarray}
where $l,m$ and $n$ are dimensionless (they are the Fourier modes
if they are chosen to be integers).
In order to obtain any desired value of $\mu$ and $\omega$ we take advantage of (partially)
twisted boundary conditions, which allow $l,m$ and $n$ to be any real numbers.
For any pair $(\mu,\omega)$, the numbers $(l,m,n)$ can be obtained from the kinematic
constraints $p_1^2=p_2^2=\mu^2$ and $q^2 = \omega \mu^2$
by solving
\begin{eqnarray}
       \mu &=& 2\pi l/L \;,\\
       l^2 &=& m^2+n^2 \;, \\
\omega l^2 &=& (l-m)^2+n^2  \;.
\end{eqnarray}
In order to show a concrete example, let us anticipate and consider
one of the ensembles used in this work, with spatial extent of $L/a=32$ and
inverse lattice spacing of $a^{-1}=2.383(9)$ GeV.
In that case, we can see that the following choice: 
\begin{eqnarray}
p_1 &=& \frac{2\pi}{L} \left[ 2.1372, 0, 0 , 0 \right],\\
p_2 &=& \frac{2\pi}{L} \left[ 1.60291, 1.41363, 0, 0 \right],
\end{eqnarray}
corresponds to $\mu=1$ GeV
and $\omega= 0.5$ 
with an uncertainty at the permille level. \\

It is worth noting another choice of kinematics given by
\footnote{Here we assume for simplicity that T=L, but in general the time component should
be rescaled if this is not the case.}
\begin{eqnarray}
p_1 &=& \frac{2\pi}{L} \left[ m, m, m, m \right],\\
p_2 &=& \frac{2\pi}{L} \left[ \pm m, \pm m, \pm m, \pm m \right].
\end{eqnarray}
Depending on the number of minus signs in $p_2$, we
can obtain any integer value of $\omega$ between $0$ and $4$.
With this choice we also expect the quark wave renormalisation
factor so to have similar lattice artefacts, whereas in the setup mentioned
above they will have different discretisation effects (except in some special cases).
In that case, one can expect the discretisation effects to be $\omega$-independent.
We note that the same procedure with fewer than four components can lead to non-integer
value of $\omega$: if the last component is fixed to zero, we can achieve the values
$\omega = 0, 4/3, 8/3, 4$. \\

In this work we only consider the setup given in Eqs.~(\ref{eq:p1})-(\ref{eq:p2})
as we also want to investigate arbitrary non-integer values of $\omega$.
We chose to vary $\omega$ between 0.5 and 4 with a step of 1/2.
In addition, we can use the $\omega$ dependence of the discretisation effects to obtain
an additional handle to estimate the systematic  uncertainties.

\section{Implementation} \label{sec:implementation}

\subsection{Lattice implementation}

For the bilinears, omitting flavour and spin-colour indices for clarity (we only consider
flavour non-singlet operators) we define $O_\Gamma = \bar \psi \Gamma \psi$,
with $\Gamma = \mathbb{1}, \gamma_\mu,  \gamma_\mu \gamma_5, \gamma_5$
(S,V,A,P).
We sketch here the computation of the propagators and of the bilinears
needed to renormalise the quark mass and the quark wave function.
We drop the volume factors and set $a=1$ for simplicity.\\

Following~\cite{Gockeler:1998ye}, the momentum source propagators are computed
by first solving
\be
\label{eq:Dirac_mom}
\sum_x D(y,x) \tilde{G}_x(p) = e^{ip.y} \mathbb{1}\;, 
\ee
for a given momentum $p$, on Landau-gauged fixed configurations.
In the previous equation, $D$ denotes a generic Dirac matrix in spin-colour space,
regardless of the lattice discretisation.
Then we obtain the momentum source propagators
by multiplying by the corresponding phase factor
\be
G_x(p) = \tilde{G}_x(p)e^{-ip.x} = \sum_z D^{-1}(x,z)e^{ip.(z-x)}. \label{npr4}
\ee
So within our conventions, $G_x(p)$ is an incoming propagator (with respect to $x$)
with momentum $p$. After gauge average, the outgoing one is then given by
$G_x(-p) = \gamma_5 G_x(p)^\dagger \gamma_5$.
The advantage of the method proposed in~\cite{Gockeler:1998ye}
compared to a traditional Fourier transom is that there sum over $z$ in Eq.(\ref{npr4}) 
 significantly reduces the statistical uncertainties. 
For the bilinears, we consider the diagram shown in Fig~\ref{fig:bil_kin} 
and compute ($\langle ... \rangle$ denotes the gauge average)
\bea
\label{Vgamma}
V_\Gamma(p_2,p_1) &=& \la \psi(p_2)  O_\Gamma \bar \psi (p_1)\ra \;, \\ 
&=& \sum_x \la G_x(-p_2) \Gamma G_x(p_1) \ra \;.
\eea
Defining $G(p) = \sum_x G_x(p)$, we then amputate the external legs
\be
\Pi_\Gamma =
\langle G^{-1}(-p_2)\rangle V_\Gamma(p_2,p_1) \langle G^{-1}(p_1)\rangle.
\ee
This quantity $\Pi_\Gamma$ is obviously a matrix in spin-colour space, 
the renormalisation factors $Z_\Gamma$ are then defined
by projecting it to the corresponding tree level value.
A possible choice (which defines a so-called $\gamma_\mu$-scheme) is to use the same $\Gamma$ matrix as in the vertex
\be
\Lambda_\Gamma = \frac{1}{F} {\cal{P}}[\Pi_\Gamma] = \frac{1}{F}\text{Tr}[\Gamma \Pi_\Gamma],
 \label{Born_def} \ee
where the trace is taken in colour space and $F$ is the corresponding tree level value.

\subsection{Z-factors}
\label{subsec:Zfactors}

Let us now specify our conventions for the renormalisation factors. We denote
the renormalised quantities with an index $R$, whereas the bare quantities
do not carry extra indices.
The quark wave function renormalisation factor $Z_q$ is defined by $\psi_R = \sqrt{Z_q} \psi$,
and $ G_R = Z_q G$ represents the renormalised fermion propagator.
For the bilinears defined above, the renormalised quantities are
denoted by $O_\Gamma^R = Z_\Gamma O_\Gamma$.
Note that for the corresponding Green's functions, we have as usual
$\Pi_\Gamma^R = Z_\Gamma/Z_q \times \Pi_\Gamma$. \\

The renormalisation factors for the bilinears are then obtained by imposing 
\be
\label{eq:impose}
\lim_{m\to 0}
\frac{Z_\Gamma (\mu,\omega)}{Z_q(\mu)} \times \Lambda_\Gamma
|_{(\mu^2=p_1^2=p_2^2, (p_1-p_2)^2=\omega\mu^2)} = 1,
\ee
where $m$ is the (bare) quark mass (before taking the chiral limit,
we work with degenerate quark masses for simplicity). Note that
we could take also the chiral limit by sending the renormalised mass to zero,
here we assume that there is no additive mass 
renormalisation~\footnote{With Domain-Wall
fermions, a small additive renormalisation of the quark mass is also
necessary. For simplicity we also assume the residual mass has already been included;
here the quark mass renormalises multiplicatively.}.
There are several possible ways to define the quark field renormalisation constant $Z_q$.
For example in $\mathrm{RI{}^\prime /MOM}$, one imposes a condition
directly on the fermionic propagator
\be
  Z_q(\mu) = -i \lim_{m\to0} \left[ \frac{1}{12p^2}  \text{Tr}
  \left[
    \slashed{p}  G^{-1}(p) \right]
    \right]_{\mu^2 = p^2}  \;.
\ee
\\
 
We now turn to the quark mass;
The choice of the field renormalisation fixes the 
renormalisation factor $Z_m$, which relates the bare and renormalised masses as
$m_R = Z_m m$. 
To be specific, we write in Fourier space the inverse bare fermion propagator
\be
G^{-1}(p) = i\slashed{p} + m + \Sigma =  i\slashed{p}(1+\Sigma^V) + m (1 +\Sigma^S) \;.
\ee
 and similarly, having a generic MOM scheme in mind, 
 we write for the renormalised propagator,
\be
G_R^{-1}(p) = i\slashed{p} + m_R + \Sigma_R =  i\slashed{p}(1+\Sigma_R^V) + m_R (1 +\Sigma_R^S) \;.
\ee
Since $ G_R = Z_q G$, we have
\bea
(1+\Sigma_R^V) &=& Z_q^{-1}(1+\Sigma^V) \;, \\
(1+\Sigma_R^S) &=& Z_q^{-1} Z_m^{-1}(1+\Sigma^S)\;. 
\eea
Traditionally in a MOM scheme, one defines the renormalisation factors $Z_m$ and $Z_q$ 
by imposing that the renormalised parts are finite in the chiral limit at a certain renormalisation
point. For example the renormalisation factor $Z_m$ can then be extracted from
\be
Z_m(\mu) = \frac{1}{Z_q(\mu)} \lim_{m\to0} \left[
  \frac{1}{12m} \text{Tr} \left[ G^{-1}(p) \right]
  \right]_{\mu^2 = p^2}  \;,
\ee
which is equivalent to
\be
  \lim_{m\to0} \left[
  \frac{1}{12m_R} \text{Tr} \left[ G_R^{-1}(p) \right]
  \right]_{\mu^2 = p^2} =1 \;.
\ee
In~\cite{Sturm:2009kb}, it was suggested to replace this condition by
\be
\lim_{m\to0} \left\{
\frac{1}{12m_R} \left[
  \text{Tr} \left[ G_R^{-1}(p) \right]_{\mu^2 = p^2}  
  -\frac{i}{2} \text{Tr} \left[ q_\mu \Pi^{\mu}_{A,R} \gamma_5\right]_{\mu^2 = p_1^2=p_2^2=(p_1-p_2)^2}
  \right]
  \right\}=1 \;.
  \ee
The motivation for this definition of $Z_m$ is that it leads to
$Z_m = 1/Z_P$
if $Z_P$ is defined from Eq.(\ref{Born_def}) at the same kinematic 
point. 
This follows from requiring that the 
Ward-Takahashi identities (WTI) given below are satisfied
both for the bare and renormalised quantities
(for degenerate quark masses)
    \bea
  \label{eq:VWI}
  q_\mu \Pi_{V^\mu}(p_1,p_2) &=& -i(G^{-1}(p_2) - G^{-1}(p_1) )\;,\\
  q_\mu \Pi_{A^\mu}(p_1,p_2) &=& 2im \Pi_P(p_1,p_1)
  - i\left( \gamma_5 G^{-1}(p_2) + G^{-1}(p_1) \gamma_5 \right)\;,
  \eea 
in the case of degenerate quark masses. 
Imposing this condition with the SMOM kinematics
leads to maintaining the WTI on the renormalised quantities. In particular
the procedure leads to $Z_A=Z_V$.\\

\subsection{Our definitions}

We deviate from the choice made in~\cite{Sturm:2009kb} for several reasons.
Firstly, some derivations in~\cite{Sturm:2009kb} rely on the SMOM kinematics,
which we want to generalise here. 
Secondly, rather than imposing the renormalisation condition directly on the quark
propagators, we take advantage of the good chiral properties of the Domain-Wall discretisation
and extract the renormalisation factors from the bilinears.
Thirdly, in our framework, we use the local vector and axial vector current which are
related to the conserved (or partially conserved) ones by the finite ``renormalisation''
factors $Z_V$ and $Z_A$ respectively.
In this paper we are using Domain Wall fermions,
so we have $Z_A = Z_V$ up to small artefacts. However the SMOM
scheme can also be used for other lattice fermions, with
$Z_A \ne Z_V$ so we do not want to restrict ourselves to the case
$Z_A=Z_V$.In this paper we are using Domain Wall fermions,
so we have $Z_A = Z_V$ up to small artefacts. However the SMOM
scheme can also be used for other lattice fermions, with
$Z_A \ne Z_V$ so we do not want to restrict ourselves to the case
$Z_A=Z_V $.
Our values of $Z_A=Z_V$ are known from previous studies
on the same lattice, and are 
given in Appendix~\ref{lattice} for convenience.
Additionally, within our framework we define $Z_m$ directly as
\be
Z_m = 1/Z_S \;\;
\ee
and we also define 
\bea
\label{eq:ZV}
\frac{Z_V}{Z_q(\mu,\omega)} \lim_{m\to0} \left[\Lambda_V\right]_{\text{IMOM}} =1\;,\\
\label{eq:ZS}
\frac{Z_S(\mu,\omega)}{Z_q(\mu,\omega)} \lim_{m\to0}\left[\Lambda_S\right]_{\text{IMOM}} =1\;.
\eea
where the shorthand IMOM stands for $\mu^2=p_1^2=p_2^2, (p_1-p_2)^2=\omega\mu^2$.
(These equations are used for the quark wave function and the mass renormalisation
factors.) 
More explicitly, we define
\bea
\label{eq:Zq}
Z_q(\mu,\omega) &=&  Z_V \lim_{m\to0} \left[\Lambda_V\right]_{\text{IMOM}} \;,\\
\label{eq:Zm}
Z_m(\mu,\omega)  &=& \frac{1}{Z_V}  \lim_{m\to0} \left[\frac{\Lambda_S}{\Lambda_V}\right]_{\text{IMOM}}   \;,
\eea
where $\Lambda_{V,S}$ are the bare (and local) vertices defined above.
Note that both $Z_m$ and $Z_q$ acquire an $\omega$-dependence through these definitions.
This dependence is discussed in section~\ref{sec:results}. 
The main advantage of this approach is that the Green's functions of the bilinears
have a much safer infrared behaviour than the quark propagators.\\

In order to completely fix the definition of $Z_q$ and $Z_m$ in
Eqs. (\ref{eq:Zq}) and (\ref{eq:Zm})
we need to specify our the choice of projector ${\cal P}_\mu$ for the vector Green's function, 
\be
\Lambda_V = \frac{1}{F} {\cal P}_\mu[ \Pi_{V^{\mu}} ]\;.
\ee
We implement the so-called $\gamma_\mu$ and $\slashed{q}$-projectors
(the trace is taken over both Dirac and colour indices) 
\bea
\label{eq:gammamuProj}
\Lambda_V^{(\gamma_\mu)} &=& \frac{1}{48} \text{Tr}[ \gamma_\mu \Pi_{V^{\mu}} ]\;, \\
\label{eq:qslashProj}
\Lambda_V^{(\slashed{q})} &=& \frac{q^{\mu}}{12 q^2} \text{Tr}[ \slashed{q} \Pi_{V^{\mu}}\, ]\;,
\eea
where obviously $q=p_1-p_2$ is the momentum transfer of the IMOM kinematic
defined above, $\mu^2=p_1^2=p_2^2, (p_1-p_2)^2=\omega\mu^2$.
Plugging these definitions in Eqs (\ref{eq:Zq}) and (\ref{eq:Zm}),
we see that we have defined two kinds of IMOM schemes
similarly to was done in the SMOM case.
In order to keep track of what projector was used, we introduce the respective  notations
$
Z_m^{(\gamma_\mu)} (\mu,\omega)$ and $Z_m^{(\slashed{q})} (\mu,\omega)$ for the renormalisation factors.
\\

The renormalisation factors for the other bilinears are defined exactly
in the same way as Eqs~\ref{eq:ZV},\ref{eq:ZS}. 
This, together with a lattice formulation which preserves chiral symmetry,
leads to $Z_V=Z_A$ and $Z_S=Z_P$, apart from the physical effects of 
spontaneous chiral symmetry breaking, which may become important
at low $p^2$ and $q^2$. 
This is discussed in~Section~\ref{sec:results}.
Naturally for a lattice formulation which break explicitly chiral symmetry,
one can follow~\cite{Sturm:2009kb} with the modification $q^2=\omega \mu^2$.

\subsection{Nomenclature}

To define unambiguously a momentum scheme for a composite operator
one needs to specify the kinematics (the choice of momenta)
and the choice of projectors, not only for the Green's function
corresponding to the operator in question, but also for the quark 
 wave function.\footnote{
 Strictly speaking we also need to specify the gauge, but
it is standard to assume that the computation is performed in the Landau gauge. }
Historically, when the Rome-Southampton method was introduced
in~\cite{Martinelli:1994ty},
the kinematics was the exceptional case, $p_1=p_2$ and  $Z_q$
was defined through a $\gamma_\mu$-projector, i.e. Eq.~(\ref{eq:gammamuProj}).
Although the authors of~\cite{Martinelli:1994ty} mention explicitly
that other choices were possible, by convention ``RI/MOM scheme''
refers to this specific choice of kinematics and wave function.
Similarly, it also became standard to call ``RI${}^\prime$/MOM scheme''
a scheme in which the kinematics is still the exceptional case
but the quark wave function is renormalised through
a $\slashed{q}$-projector, ie Eq.~(\ref{eq:qslashProj}).
(In \cite{Martinelli:1994ty} the definition of the projector
differs from the one given here, but both definitions are equivalent
up to lattice artefacts).\\

We turn now to the SMOM schemes. 
By definition, the ``S'' of ``SMOM'' refers to the situation
where $p_1^2=p_2^2=(p_1-p_2)^2$, ie $\omega=1$,
or ``Symmetric''~\cite{Sturm:2009kb}. 
For the quark wave function, the convention in~\cite{Sturm:2009kb}
was that ``RI/SMOM'' refer to a $\slashed{q}$-projector and ``RI/SMOM$_{\gamma_\mu}$"
to a $\gamma_\mu$-projector.
Strictly speaking there is also a subtlety in the definition
of $Z_m$, see section~\ref{subsec:Zfactors}.
\\

Here, following the conventions adopted by RBC-UKQCD (see for example~\cite{Boyle:2017skn}),
we use the notations $Z^{(X)}(\mu, \omega)$, where $X\in(\gamma_\mu, \slashed{q})$
indicates the choice of the projector for the wave function.
Obviously the choice of kinematics is explicitly given by $\omega$. 
When needed, we also refer to these schemes as IMOM$^{(\gamma_\mu)}$ and 
IMOM$^{(\slashed{q})}$.

\begin{table}[th]
  \begin{center}
  \begin{tabular}{c|c|c}
    \hline 
    Name & Kinematics & Projector for $Z_q$ \\
    \hline
    RI/MOM & $p_1=p_2$ & $\gamma_\mu$ \\
    RI${}^\prime$/MOM & $p_1=p_2$ & $\slashed{q}$ \\
    RI/SMOM$_{\gamma_\mu}$ & $p_1^2=p_2^2=(p_1-p_2)^2$ & $\gamma_\mu$ \\
    RI/SMOM & $p_1^2=p_2^2=(p_1-p_2)^2$ & $\slashed{q}$ \\
    \hline
    IMOM$^{(\gamma_\mu)}$  & $p_1^2=p_2^2=\mu^2, \; (p_1-p_2)^2=\omega \mu^2 $ & $\gamma_\mu$ \\
    IMOM$^{(\slashed{q})}$  & $p_1^2=p_2^2=\mu^2, \; (p_1-p_2)^2=\omega \mu^2 $ & $\slashed{q}$ \\    
  \end{tabular}
  \end{center}
  \end{table}

\subsection{Non-perturbative running}
To compute the non-perturbative scale evolution of $Z_m$:
we define $\Sigma_{m}$ as
\be
\label{eq:sigma_def}
\Sigma^{(X)}_{m}(a,\mu,\mu_0,\omega, \omega_0) = \lim_{m\rightarrow 0}\frac{Z^{(X)}_m(a,\mu,\omega)}{Z^{(X)}_m(a,\mu_0,\omega_0)}\;,
\ee
where $X$ can either be $\gamma_\mu$ or $\slashed{q}$. 
 We can then take the continuum limit $a^2 \rightarrow 0$ and define
\be
\sigma^{(X)}_{m}(\mu,\mu_0,\omega, \omega_0) = \lim_{a^2 \rightarrow 0} \Sigma^{(X)}_{m}(a,\mu,\mu_0,\omega, \omega_0) \;.
\ee
Similarly, for the quark wave function we define $\sigma_{q}$, by just substituting
$Z_q$ for $Z_m$ in the above definitions.
Since in practice we have only two lattice spacings, we add a systematic
error as an estimate of the residual discretisation effects.
This error is obtained by adding half the difference beween the extrapolated value
and the value on the the finest lattice spacing (see Appendix~\ref{app:CL}).
We now have for each value of $\mu, \mu_0, \omega, \omega_0$ a central value of $\sigma_{m,q}$,
some statistical error and a systematic error from the continuum extrapolation
In the following we will always sum the systematic and statistical uncertainties in
quadrature except when stated otherwise.
We found that the chiral extrapolations are well under control and that an associated systematic 
error would be negligible compared to the other sources of error.
The main motivation for looking at these quantities rather than
the $Z$-factors themselves is the existence of the continuum limit.
We can also compare the non-perturbative running obtained from our
lattice simulation with the perturbative prediction.

\section{Results} \label{sec:results}

 To determine the relationship between lattice results and
$\overline{\mathrm{MS}}$ results we need to compare lattice and
perturbation theory in a kinematic region where both perturbation
theory and lattice results are reliable. In this region the
perturbative and lattice results should run in the same way, so the
ratio should be constant. We will see deviations from constancy in
regions where perturbation theory converges slowly (for example at
small $\mu$ or small $q^2 = \omega \mu^2$. We will also see
non-constancy if lattice artefacts are important, which is likely to
happen at large $\mu^2$ and large $\omega \mu^2$.
We also note that our procedure to estimate the residual
discretisation errors could potentially underestimate them
in the large $\mu^2$ and large $\omega \mu^2$ regions. \\

In the following sections we will first keep $\omega$ constant and
compare the $\mu$ dependence of the lattice results and the continuum.
An advantage of IMOM is that we have two parameters, $\mu^2$ and
$\omega$, so we can better test that we are working in a region where
it is sensible to trust both perturbative and lattice
results. Accordingly, we next will vary both $\mu$ and $\omega$ and
study if a plateau is emerging where we can reliably determine $Z$.

\subsection{Quark mass renormalisation factor}

\begin{figure}[t]
\bc
\includegraphics[scale=.82]{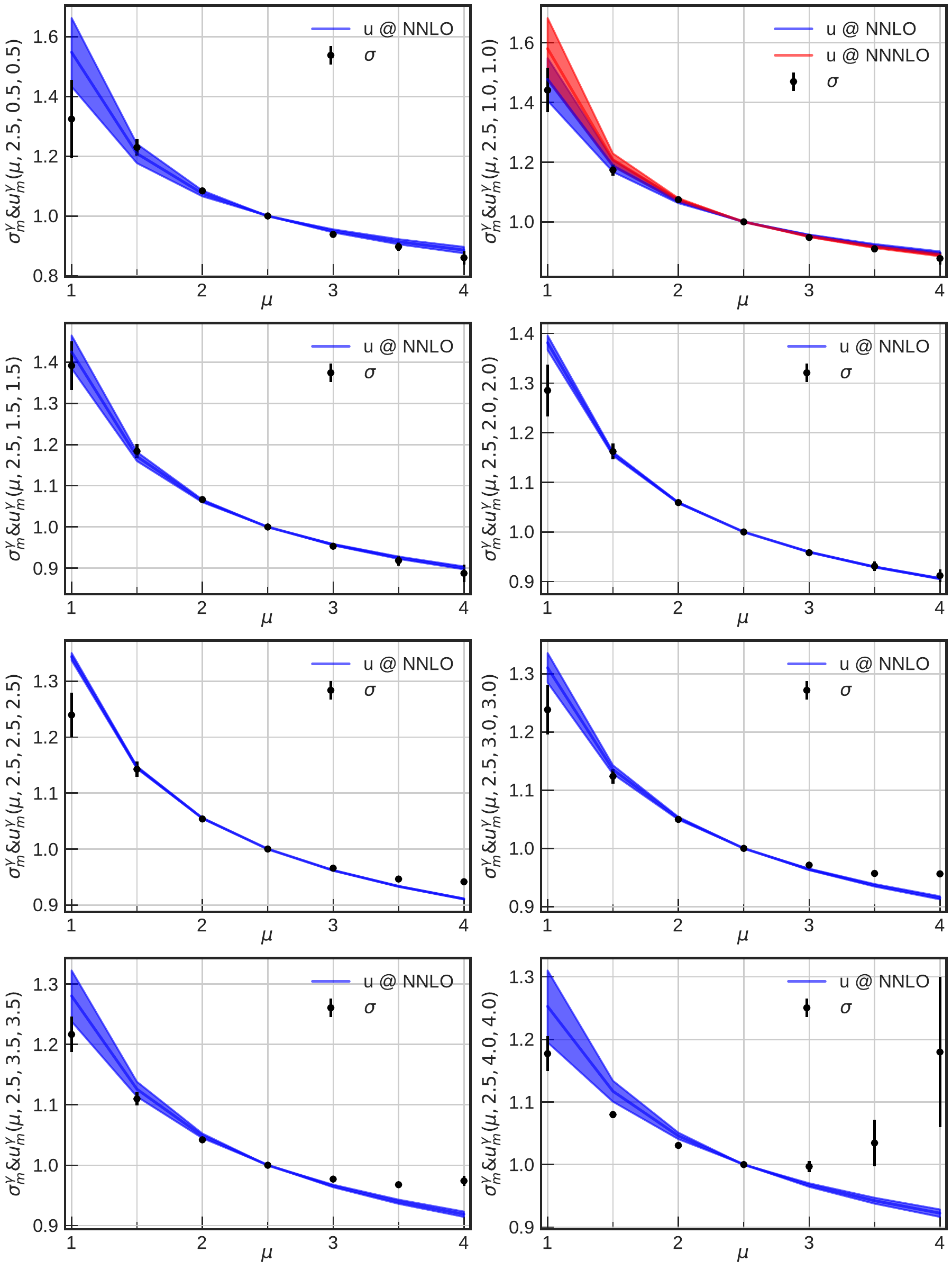}
\ec
\caption{Comparison of the non-perturbative and perturbative running
  for $Z_m^{(\gamma_\mu)}$. 
Note that for $\omega=1$ the perturbative running is known at N$^3$LO.}
\label{fig:Compare_m_gamma_mu0_25}
\end{figure}

\begin{figure}[t]
\bc
\includegraphics[scale=.82]{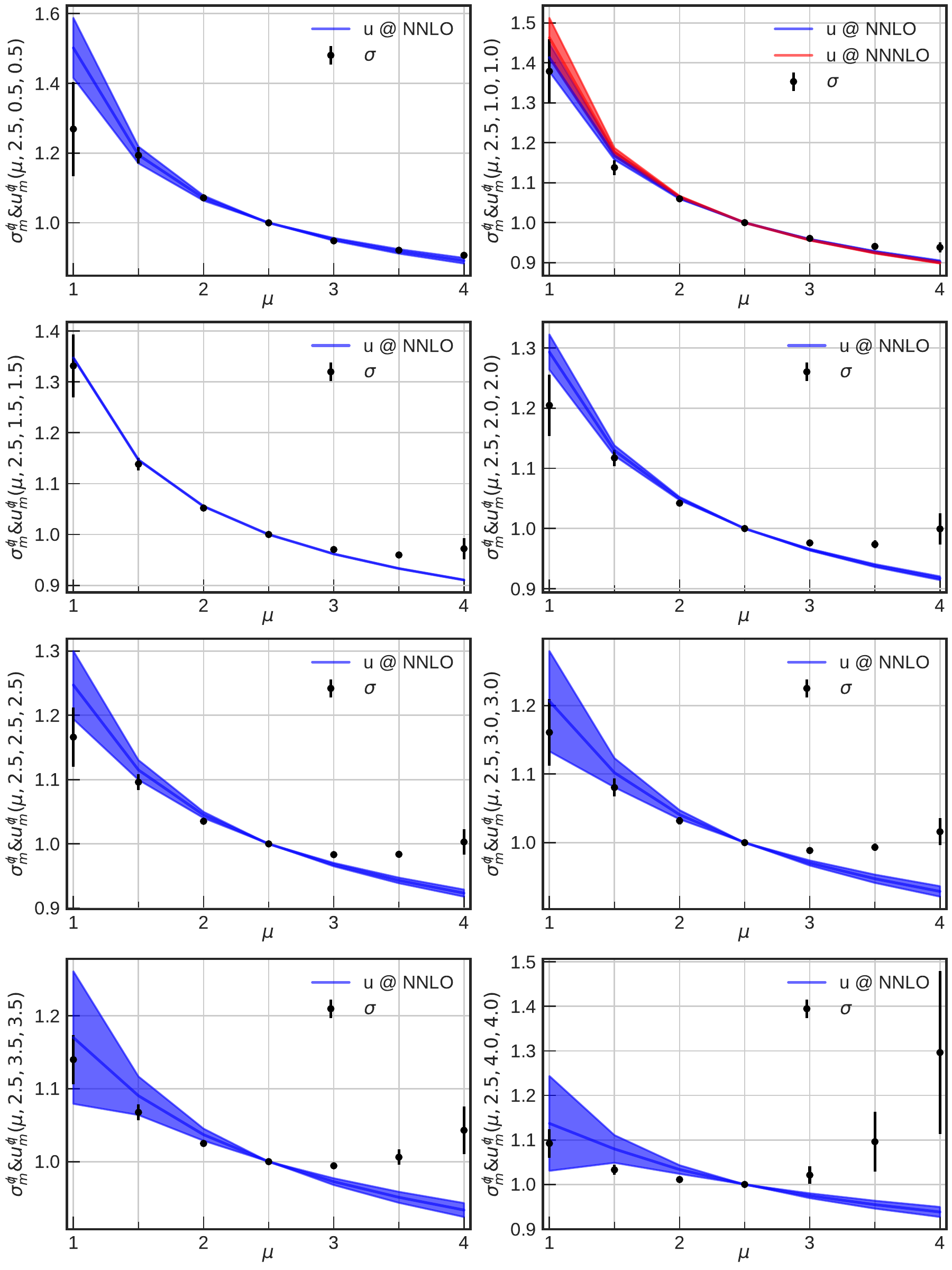}
\ec
\caption{Same as Fig.~\ref{fig:Compare_m_gamma_mu0_25} for  $Z_m^{(\slashed{q})}$. }
\label{fig:Compare_m_qslash_mu0_25}
\end{figure}

We now apply this scheme to determine $Z$ for the mass operator and quark
wave-function.
The running of the quark mass is shown in Fig~\ref{fig:Compare_m_gamma_mu0_25}
for the $\gamma_\mu$-scheme and Fig.~\ref{fig:Compare_m_qslash_mu0_25} for the
$\qslash$-scheme. In these plots we show both the non non-perturbative 
scale evolution $\sigma^{(\gamma_\mu)}_{m}(\mu,\mu_0,\omega, \omega_0)$
and the perturbative prediction $u^{(\gamma_\mu)}_{m}(\mu,\mu_0,\omega, \omega_0)$.
In order to  study the $\mu$-evolution for fixed values of $\omega$, we set
$\omega=\omega_0 =0.5, 1.0, 1.5, \ldots, 4.0$ and let $\mu$ vary between 1 and 4 GeV.
We find a good agreement for intermediate values of $\mu$ and $\omega$, where both
perturbation theory and lattice artefacts are expected to be under good control.
In fact perturbation theory works surprisingly well even for small values of $\mu$,
where significant discrepancies with the lattice results only emerge at $\mu \simeq
1$GeV.
Out of the two schemes, perturbation theory and lattice results agree best in the
$\gamma_\mu$-scheme.
The onset of lattice artefacts for large values of $\mu$ and $\omega$ becomes only
relevant for $q^2 \gtrsim 25 \mathrm{GeV}^2$.
This becomes particularly visible for large values of $\omega = 4$, where
perturbation theory also becomes less reliable.
The discrepancy of lattice results and perturbation theory occurs already at values
of $q^2 \gtrsim 10 \mathrm{GeV}^2$ in the $\qslash$-scheme.

\begin{figure}[t]
\bc
\includegraphics[width=\textwidth]{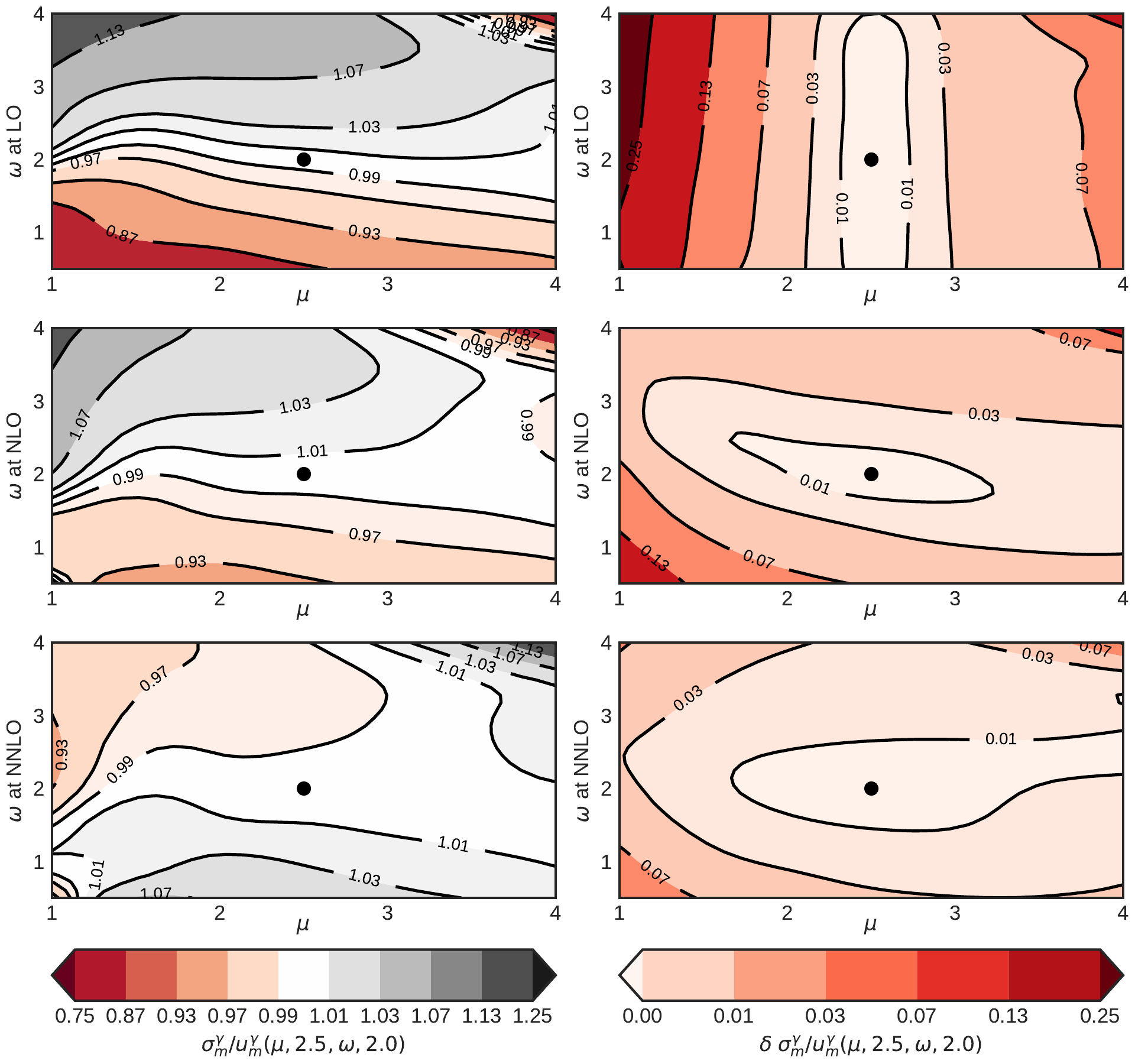}
\ec
\caption{Ratio of the non-perturbative running over the perturbative
    prediction for $Z_m^{(\gamma_\mu)}$. The central value is shown on the left
 and the error on the right. This ratio is exactly one (by definition) for the point
    ($\mu=2.5$ GeV, $\omega=2$). }
\label{fig:ratio_gamma}
\end{figure}

\begin{figure}[t]
\bc
\includegraphics[width=\textwidth]{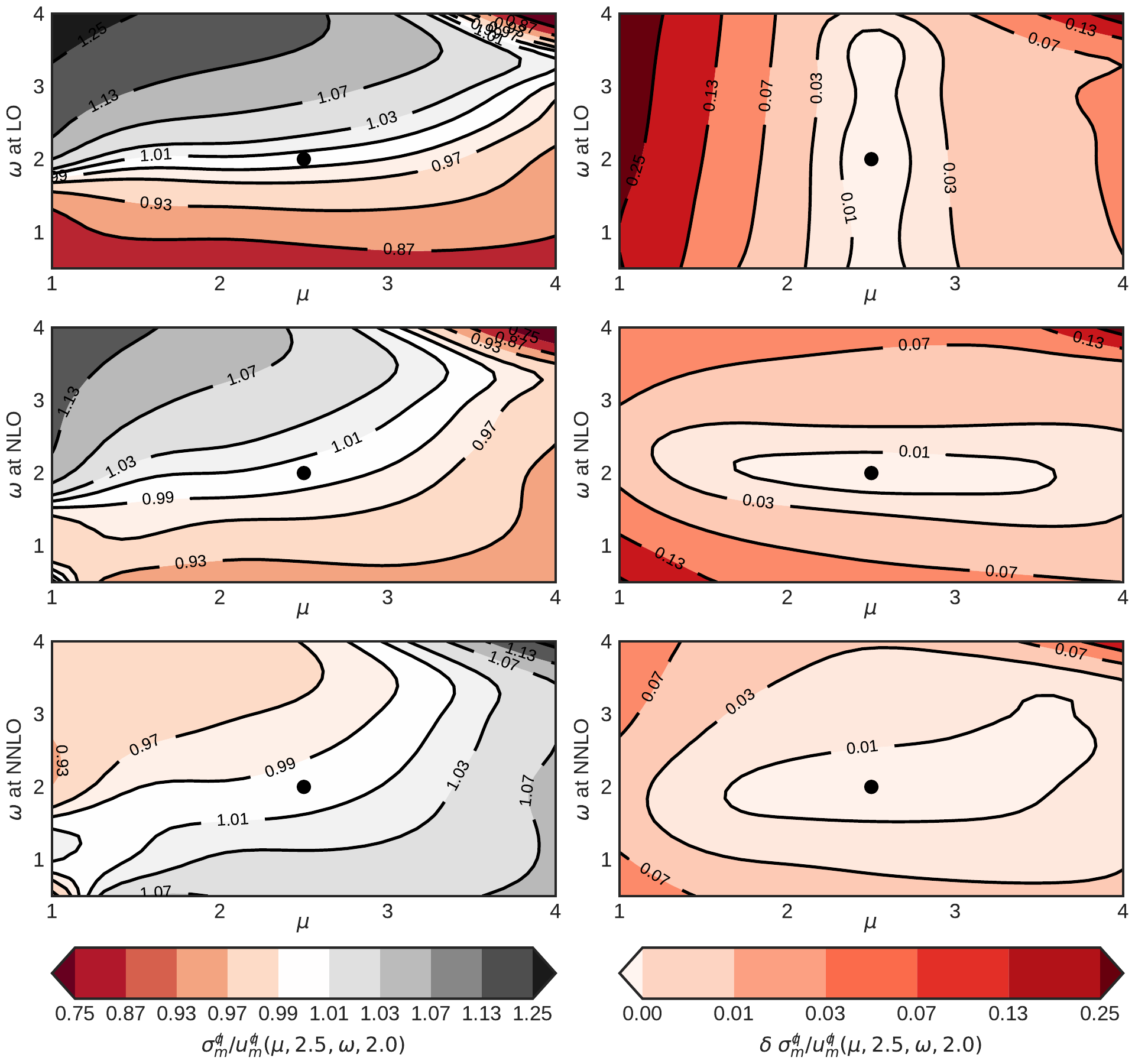}
\ec
\caption{Same as Fig.~\ref{fig:ratio_gamma} for 
$Z_m^{(\slashed{q})}$. }
\label{fig:ratio_qslash}
\end{figure}

We show our results for the scale evolution
of the quark mass in the $(\mu,\omega)$ plane in Fig.~\ref{fig:ratio_gamma}
for the $\gamma_\mu$-scheme and in   Fig.~\ref{fig:ratio_qslash}
we show the same quantity for the $\slashed{q}$-scheme.
Here we allow $\omega\ne \omega_0$ and show the ratio of the non-perturbative
running divided by the perturbative prediction as a function of $\mu$ and $\omega$
at LO, NLO and NNLO, while fixing $\omega_0=2$ and $\mu_0=2.5$ GeV.
The discrepancy from one can then be compared with the combined lattice and
perturbative uncertainties that have been added in quadrature.


As expected we observe that increasing the order of the perturbative expansion
improves the agreement with the non-perturbative evolution.
The corner of the planes are affected by larger systematic errors,
in particular where $\omega=4$ and/or $\mu > 3.5$ GeV where the discretisation effects
become more sizeable. However our data agree with NNLO at a few percent level for
a large part of the $(\mu,\omega)$plane: approximately for
$1.5 \mbox{ GeV} \le \mu \le 3.5 \mbox{ GeV}$ and $1 \le \omega \le 3$. \\

\clearpage
For completeness, we study the next order in perturbation theory for the case $\omega=1$, 
as the three loop matching has been computed in~\cite{Bednyakov:2020, Kniehl:2020}
and the $\MSbar$ four-loop anomalous dimension in can be found in~\cite{Chetyrkin:1999pq}.
We take $\mu_0 = 2$ GeV and  $\mu= 2.5$ GeV and compute the running in different schemes,
see Table~\ref{table:sigmam_NNNLO}. In Table~\ref{table:sigmam_conv_NNNLO}, we show the
contributions of the higher orders. Although the corrections to the leading order
contributions are rather small, we see that the convergence in $\MSbar$ seems to be much
better than in the MOM schemes, in the sense the $(i+1)$-th correction  is much smaller
than the  $i$-th order. However, the important observations  here are that the leading order
gives by far the main contribution and that the perturbative series seems to converge nicely
to the non-perturbative values~\footnote{The attentive reader would have noticed that in
  Table~\ref{table:sigmam_NNNLO} The NLO for $\gamma_\mu$ and $\qslash$  are identical.
  This seems to be nothing but a numerical accident: adding more significant figures lead to
  0.941112 for $\gamma_\mu$ and  0.944132 for $\qslash$.}.

\begin{table}[htb]
  \begin{center}
    \begin{tabular} {|c| c         c        c       c        c      |}
      \hline
      Scheme        & LO        & NLO       & NNLO      & NNNLO     & NP    \\
      \hline
      $\MSbar$      &  0.9537   &  0.9456   &  0.9441   &  0.9439  &       \\
      $\MSbar\leftarrow \gamma_\mu $ & 0.9537 & 0.9350 &  0.9389 &  0.9426 &  \\
      $\MSbar\leftarrow \qslash $ & 0.9537 & 0.9451 &  0.9462 & 0.9475 & \\
      $\gamma_\mu$   &  0.9537   &  0.9411   &  0.9357  &  0.9318  &  0.9307(62) \\
      $\qslash$     &  0.9537   &  0.9441   &  0.9415   &  0.9400  & 0.9436(46)\\
      \hline
    \end{tabular}
  \end{center}
  \caption{Running between 2 and 2.5 GeV for the quark mass.
    First we perform the computation directly in $\MSbar$
    then and in the SMOM ($\omega=1$) schemes $\gamma_\mu$ and  $\qslash$.
    In this case the running is known at NNNLO. 
    The lattice results are denoted by NP (non-perturbative).
    We also show the  running computed first non-perturbatively then converted to
    $\MSbar$, we denote them by $\MSbar\leftarrow \gamma_\mu $ and  $\MSbar\leftarrow \qslash$,
    respectively.
   }
\label{table:sigmam_NNNLO}
\end{table}

\begin{table}[htb]
\begin{center}
  \begin{tabular} { |c| c         c        c       |}
    \hline
    Scheme       &  NLO-LO  &  NNLO-NLO  &  NNNLO-NNLO   \\
    \hline
   $\MSbar$     &   -0.0081 & -0.0015  &  -0.0002  \\
   $\gamma_\mu$  &   -0.0126 & -0.0054  &  -0.0040 \\
    $\qslash$   &   -0.0096 & -0.0026  &  -0.0017 \\
    \hline
\end{tabular}
  \end{center}
\caption{
  Study of the convergence of the perturbative series for running of the quark mass between 2 and 2.5 GeV
  in $\MSbar$,  SMOM-$\gamma_\mu$  and $\qslash$. }
\label{table:sigmam_conv_NNNLO}
\end{table}

\clearpage
Finally for completeness, in Tables~\ref{tab:sigma_m_gamma} and~\ref{tab:sigma_q_gamma} we give our results
for the non-perturbative running of the quark mass, with $\mu_0 =2$ GeV
and $\omega=\omega_0=0.5, 1.0, \ldots, 4.0$. In Appendix~\ref{app:extensive}, we
show more numerical results, where we vary our parameters $\omega,\omega_0, \mu$
and $\mu_0$ in various ways. Our conclusion remains that the
non-perturbative results agree very well with the perturbative ones
as long as we stay from the corner of the $(\omega,\mu)$ plane.

\begin{table}[t]
\bc
\begin{tabular}{ |c| c  c  c  c  c  c  | }
  \hline
  $\omega / \mu=$
      & $1.0$           & $1.5$            & $2.5$           & $3.0$           & $3.5$          & $4.0$             \\
  \hline
 $0.5$  &  $1.219(123)$ &  $1.131(21)$  &    $0.922( 8)$  &  $0.866(13)$  &  $0.827(18)$  &  $0.793(26)$ \\
 $1.0$  &  $1.343(64)$  &  $1.092(16)$  &    $0.931( 6)$  &  $0.882(11)$  &  $0.846(17)$  &  $0.816(24)$ \\
 $1.5$  &  $1.306(52)$  &  $1.110(12)$  &    $0.937( 5)$  &  $0.894(11)$  &  $0.861(17)$  &  $0.831(25)$ \\
 $2.0$  &  $1.214(49)$  &  $1.099(11)$  &    $0.944( 5)$  &  $0.904(10)$  &  $0.878(14)$  &  $0.860(17)$ \\
 $2.5$  &  $1.176(37)$  &  $1.085( 9)$  &    $0.949( 5)$  &  $0.916( 9)$  &  $0.898(10)$  &  $0.893( 7)$ \\
 $3.0$  &  $1.180(39)$  &  $1.070( 7)$  &    $0.952( 6)$  &  $0.925( 8)$  &  $0.911( 8)$  &  $0.910( 4)$ \\
 $3.5$  &  $1.169(27)$  &  $1.065( 6)$  &    $0.959( 5)$  &  $0.937( 6)$  &  $0.928( 4)$  &  $0.935( 5)$ \\
 $4.0$  &  $1.142(27)$  &  $1.049( 5)$  &    $0.970( 2)$  &  $0.968(10)$  &  $1.004(36)$  &  $1.144(116)$ \\
\hline
\end{tabular}
\caption{Non-perturbative running for the quark mass in the $\gamma_\mu$ scheme. }
\label{tab:sigma_m_gamma}
\ec
\end{table}

\begin{table}[t]
\bc
\begin{tabular}{ |c| c  c  c  c  c  c  | } 
\hline
  $\omega / \mu=$
& $1.0$           & $1.5$            & $2.5$           & $3.0$           & $3.5$          & $4.0$             \\
\hline
 $0.5$  &  $1.182(128)$ &  $1.110(21)$  &   $0.933( 8)$  &  $0.886( 9)$  &  $0.860( 8)$  &  $0.846( 8)$ \\
 $1.0$  &  $1.302(73)$  &  $1.074(18)$  &   $0.944( 5)$  &  $0.907( 5)$  &  $0.888( 4)$  &  $0.885(11)$ \\
 $1.5$  &  $1.266(57)$  &  $1.082(10)$  &   $0.951( 3)$  &  $0.922( 3)$  &  $0.912( 3)$  &  $0.924(17)$ \\
 $2.0$  &  $1.157(49)$  &  $1.074(11)$  &   $0.959( 4)$  &  $0.936( 5)$  &  $0.933( 4)$  &  $0.958(22)$ \\
 $2.5$  &  $1.127(43)$  &  $1.060(10)$  &   $0.966( 5)$  &  $0.950( 5)$  &  $0.950( 3)$  &  $0.969(16)$ \\
 $3.0$  &  $1.126(45)$  &  $1.047( 8)$  &   $0.969( 5)$  &  $0.958( 6)$  &  $0.962( 3)$  &  $0.985(15)$ \\
 $3.5$  &  $1.114(32)$  &  $1.042( 7)$  &   $0.976( 4)$  &  $0.970( 4)$  &  $0.982( 7)$  &  $1.018(28)$ \\
 $4.0$  &  $1.080(30)$  &  $1.023( 7)$  &   $0.989( 6)$  &  $1.009(25)$  &  $1.081(71)$  &  $1.274(182)$ \\
\hline
\end{tabular}
\ec
\caption{Non-perturbative running for the quark mass in the $\qslash$ scheme.}
\label{tab:sigma_m_qslash}
\end{table}

\clearpage
\subsection{Quark wave function renormalisation factor}

We also study the convergence for the quark wave function,  $Z_q^{(X)}$,
see Table~\ref{table:sigmaq_NNNLO} and Table~\ref{table:sigmaq_conv_NNNLO}.
Here the situation is very different from the quark mass mainly as there is
no contribution at leading order (in the Landau gauge).
In the $\qslash$-scheme, the non-perturbative results differ significantly
from the perturbative predictions; however one should keep in mind that the
running is very small in magnitude. It is also interesting to have a close
look at the $\qslash$ scheme where there the perturbative prediction is known at N$^3$LO.
As we can see the perturbative series to converge very poorly in
the sense that the relative difference decrease very slowly as
we increase the order. The difference between the non-perturbative
result and the N$^3$LO, namely $\sim 1.0195 - 1.0113 \sim 0.0082$,
could be explained by higher order in the perturbative series. \\


\begin{table}[t]
  \begin{center}
    \begin{tabular} {|c| c         c        c       c        c      |}
      \hline
      Scheme      & LO        & NLO       & NNLO      & NNNLO     & NP    \\
      \hline
      $\MSbar$     &  1.0      &  1.0048    &  1.0062  &  1.0064   &       \\
      $\MSbar\leftarrow \gamma_\mu $ &  1.0   &   1.0069  & 1.0078 & N.A. & \\
      $\MSbar\leftarrow \qslash $ &  1.0      &  1.0195   &  1.0175   &  1.0146 &  \\   
      $\gamma_\mu$  & 1.0       &  1.0017    &  1.0020  &  N.A      &  1.0037(20)\\
      $\qslash$    &  1.0      &  1.0048    &  1.0081  &  1.0113   &  1.0195(25) \\
      \hline
    \end{tabular}
  \end{center}
  \caption{Running between 2 and 2.5 GeV for the quark wave function in $\MSbar$
    and in the SMOM schemes $\gamma_\mu$($\omega=1$)  and  $\qslash$.
    In this case the running is known at NNNLO. }
\label{table:sigmaq_NNNLO}
\end{table}

\begin{table}[t]
\begin{center}
  \begin{tabular} { |c| c         c        c       |}
    \hline
    Scheme       &  NLO-LO  &  NNLO-NLO  &  NNNLO-NNLO   \\
    \hline
   $\MSbar$     &   0.0048 &  0.0013   & 0.0003  \\
   $\gamma_\mu$  &   0.0017 &  0.0003   &         \\
    $\qslash$   &   0.0048 &  0.0033   & 0.0032  \\
    \hline
\end{tabular}
  \end{center}
\caption{Study of the convergence of the perturbative series for running of the quark wave function
  between 2 and 2.5 GeV in $\MSbar$,  SMOM-$\gamma_\mu$  and $\qslash$. }
%
\label{table:sigmaq_conv_NNNLO}
\end{table}

%

\begin{figure}[t]
\bc
\includegraphics[width=\textwidth]{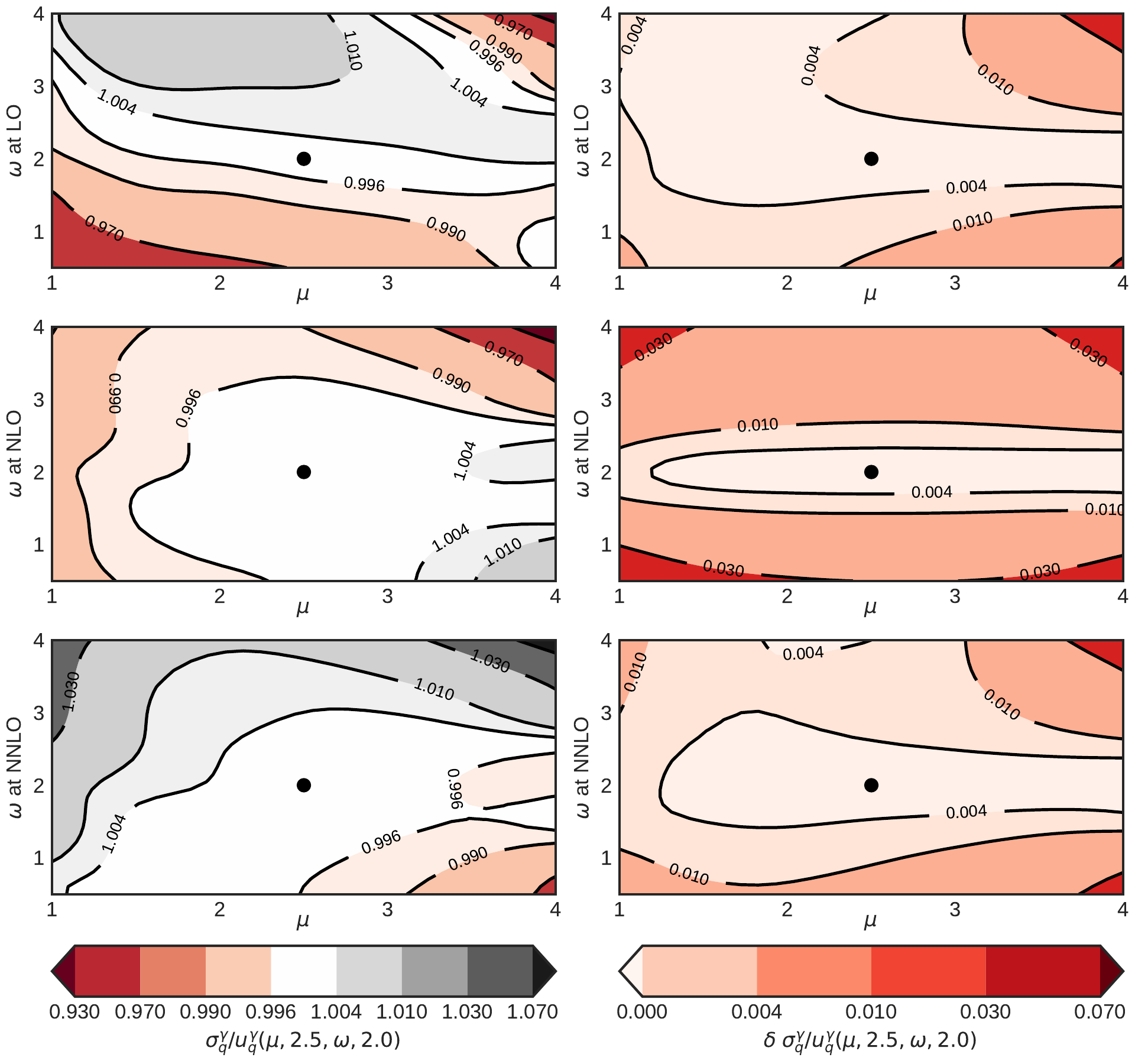}
\ec
\caption{Same as Fig.~\ref{fig:ratio_gamma} for  $Z_q^{(\gamma_\mu)}$}
\label{fig:ratio_q_gamma}
\end{figure}

\begin{figure}[t]
\bc
\includegraphics[width=\textwidth]{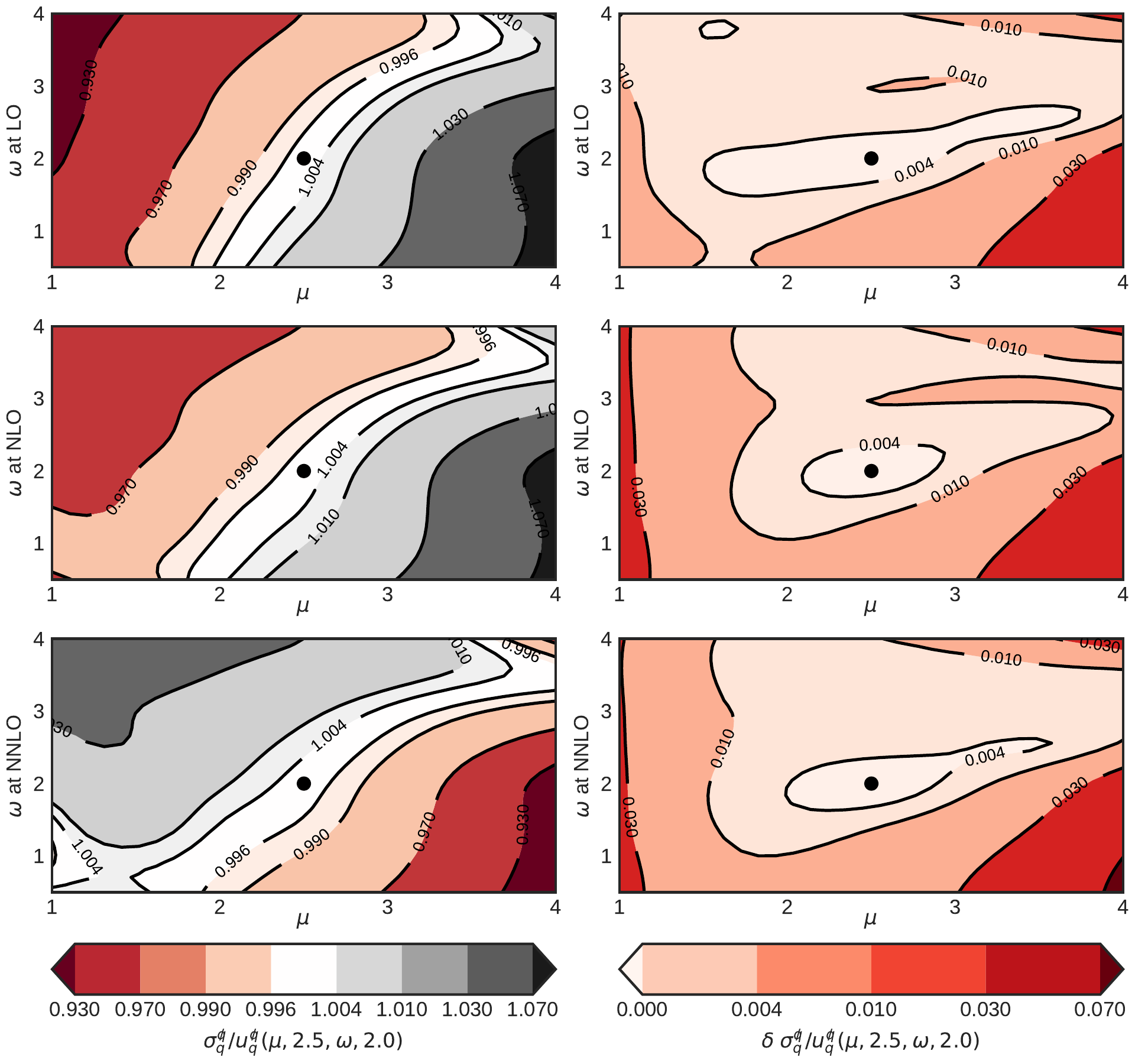}
\ec
\caption{Same as Fig.~\ref{fig:ratio_gamma} for  $Z_q^{(\slashed{q})}$.}
\label{fig:ratio_q_qslash}
\end{figure}

Keeping these facts in mind, we show our results in Fig.~\ref{fig:ratio_q_gamma}
in \ref{fig:ratio_q_qslash} for $X=\gamma_\mu$ and $X=\slashed{q}$ respectively.
Once again we divide the non-perturbative running by the perturbative prediction,
we fix $\omega_0=2$ and $\mu_0=2.5$ GeV and let $\omega$ and $\mu$ vary.

\begin{figure}[t]
\bc
\includegraphics[scale=.82]{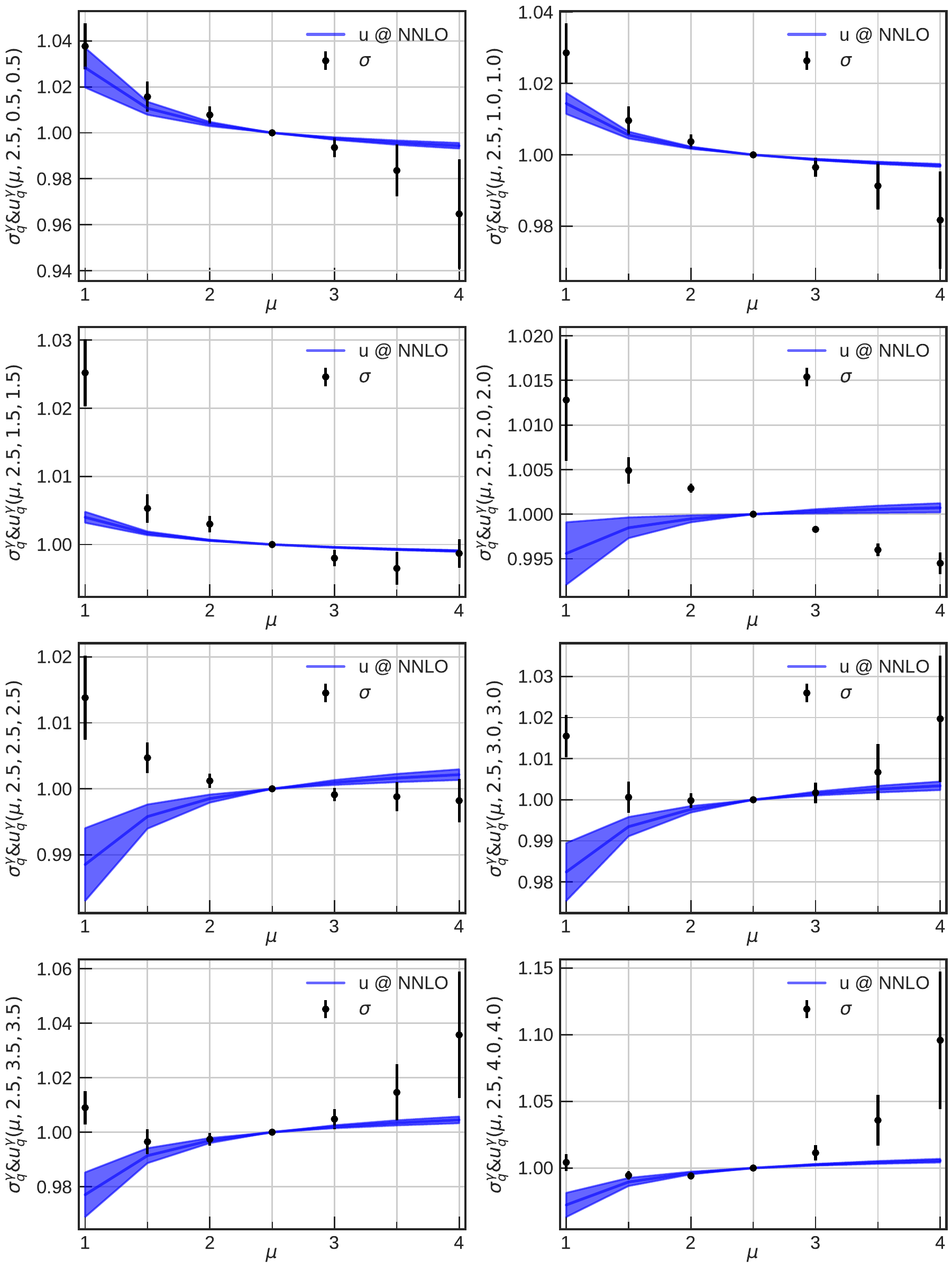}
\ec
\caption{Same as Fig.~\ref{fig:Compare_m_gamma_mu0_25} for  $Z_q^{(\gamma_\mu)}$}
\label{fig:Compare_q_gamma_mu0_25}
\end{figure}

\begin{figure}[t]
\bc
\includegraphics[scale=.82]{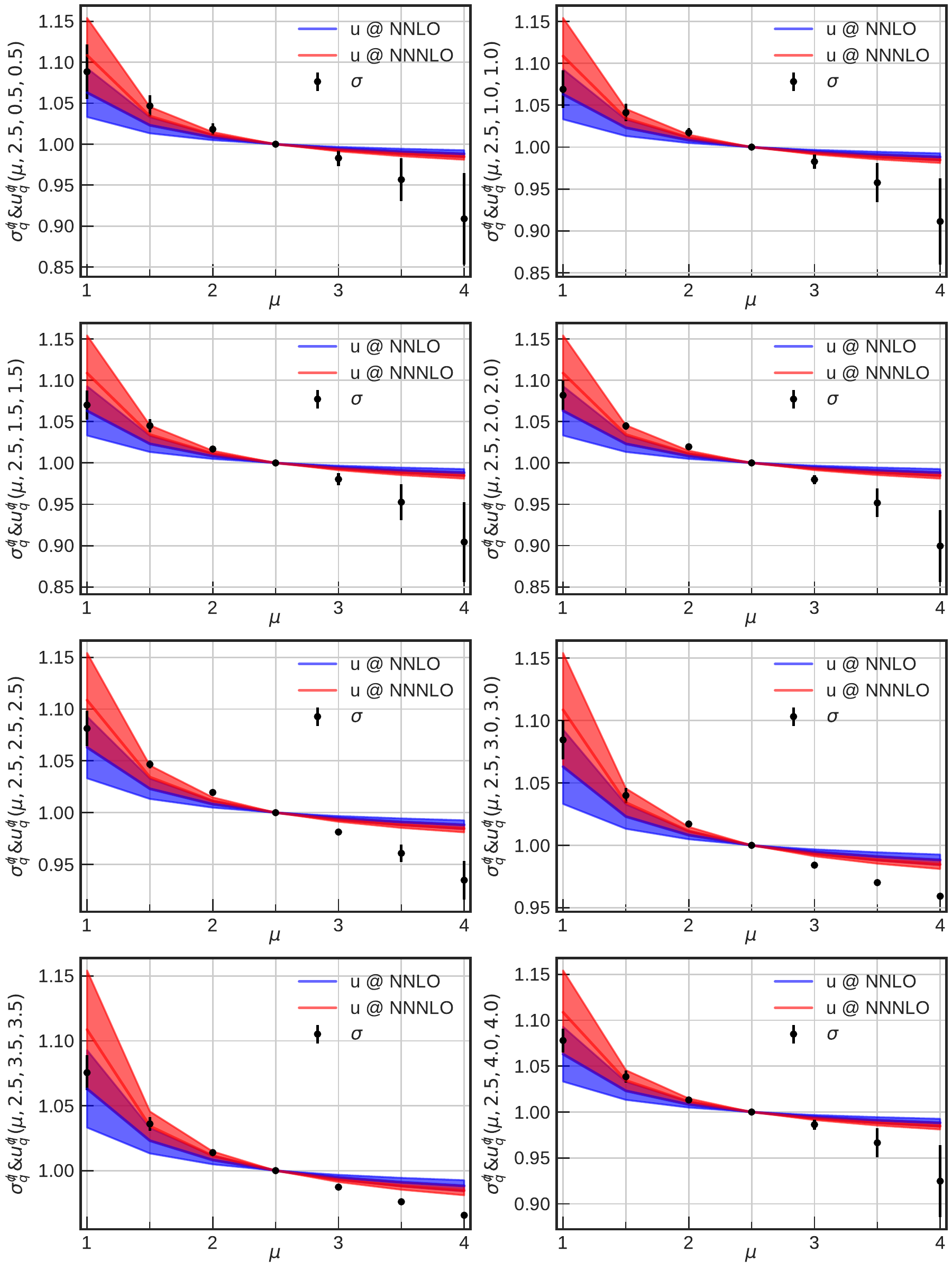}
\ec
\caption{Same as Fig.~\ref{fig:Compare_m_gamma_mu0_25} for  $Z_q^{(\slashed{q})}$. }
\label{fig:Compare_q_qslash_mu0_25}
\end{figure}

\clearpage
\subsection{Study of the $\omega$-(in)dependence of $\sigma_q$ }

\begin{figure}[htb]
  \bc
  \begin{tabular}{cc}
    \includegraphics[width=0.5\textwidth]{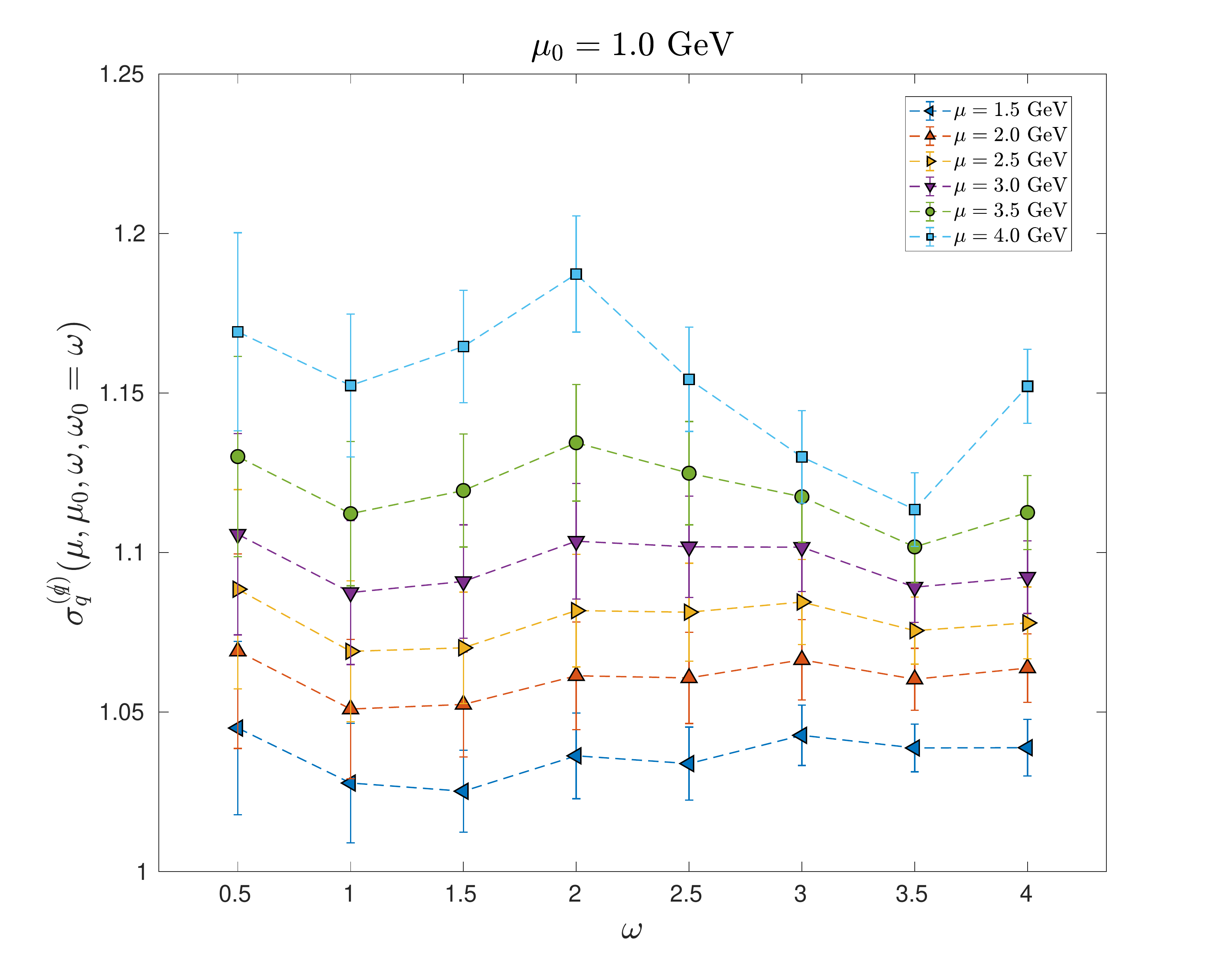} &
    \includegraphics[width=0.5\textwidth]{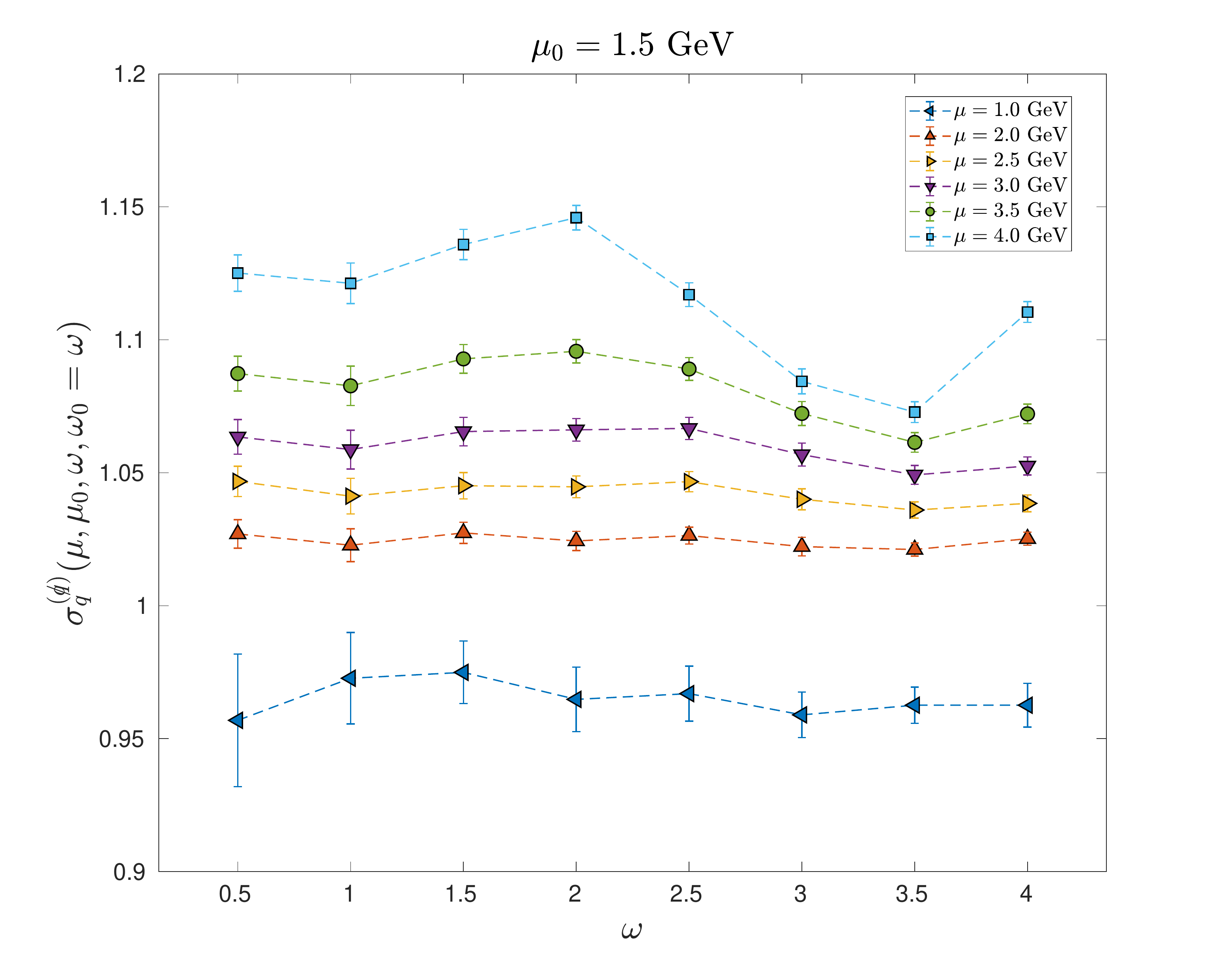} \\
    \includegraphics[width=0.5\textwidth]{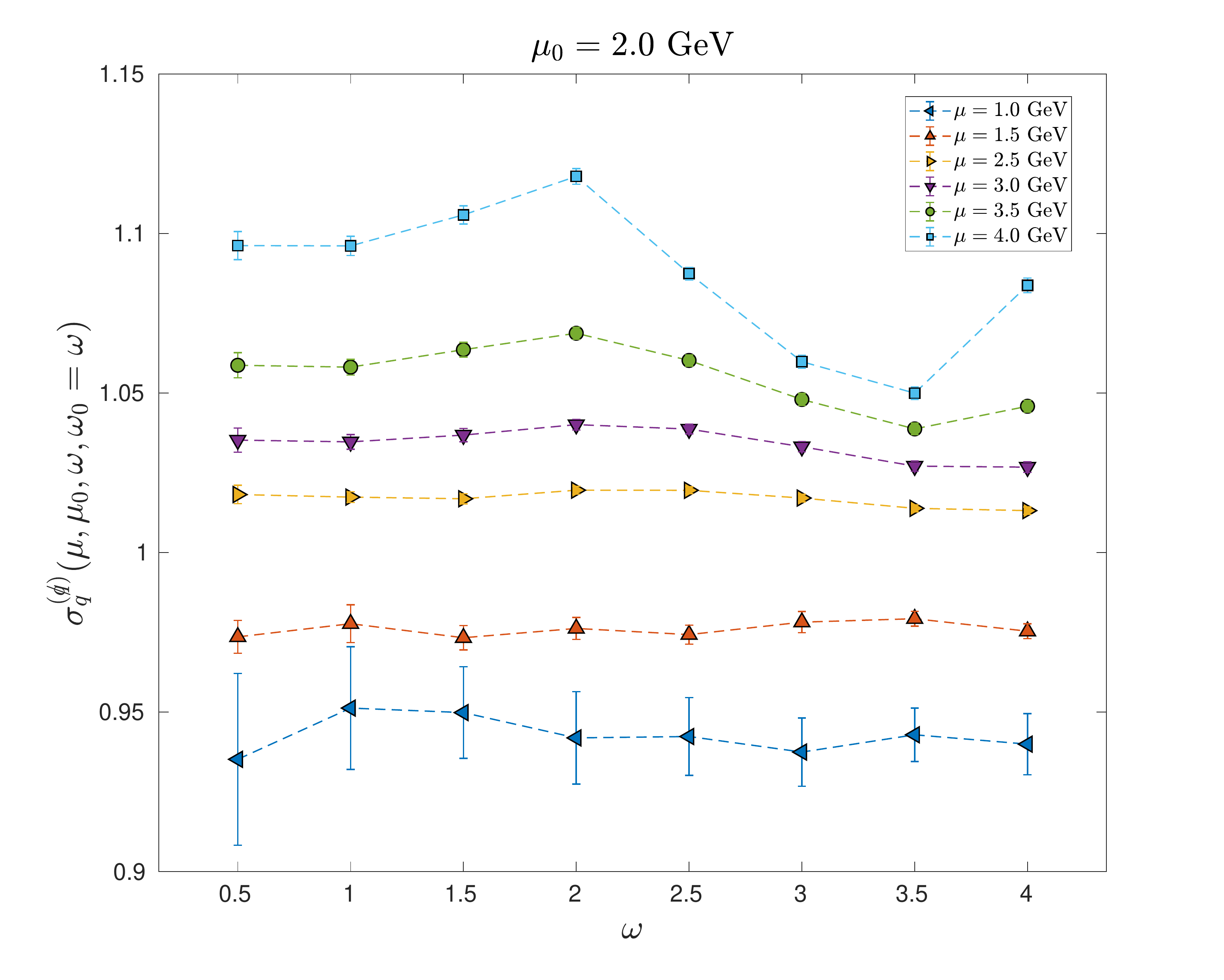} &
    \includegraphics[width=0.5\textwidth]{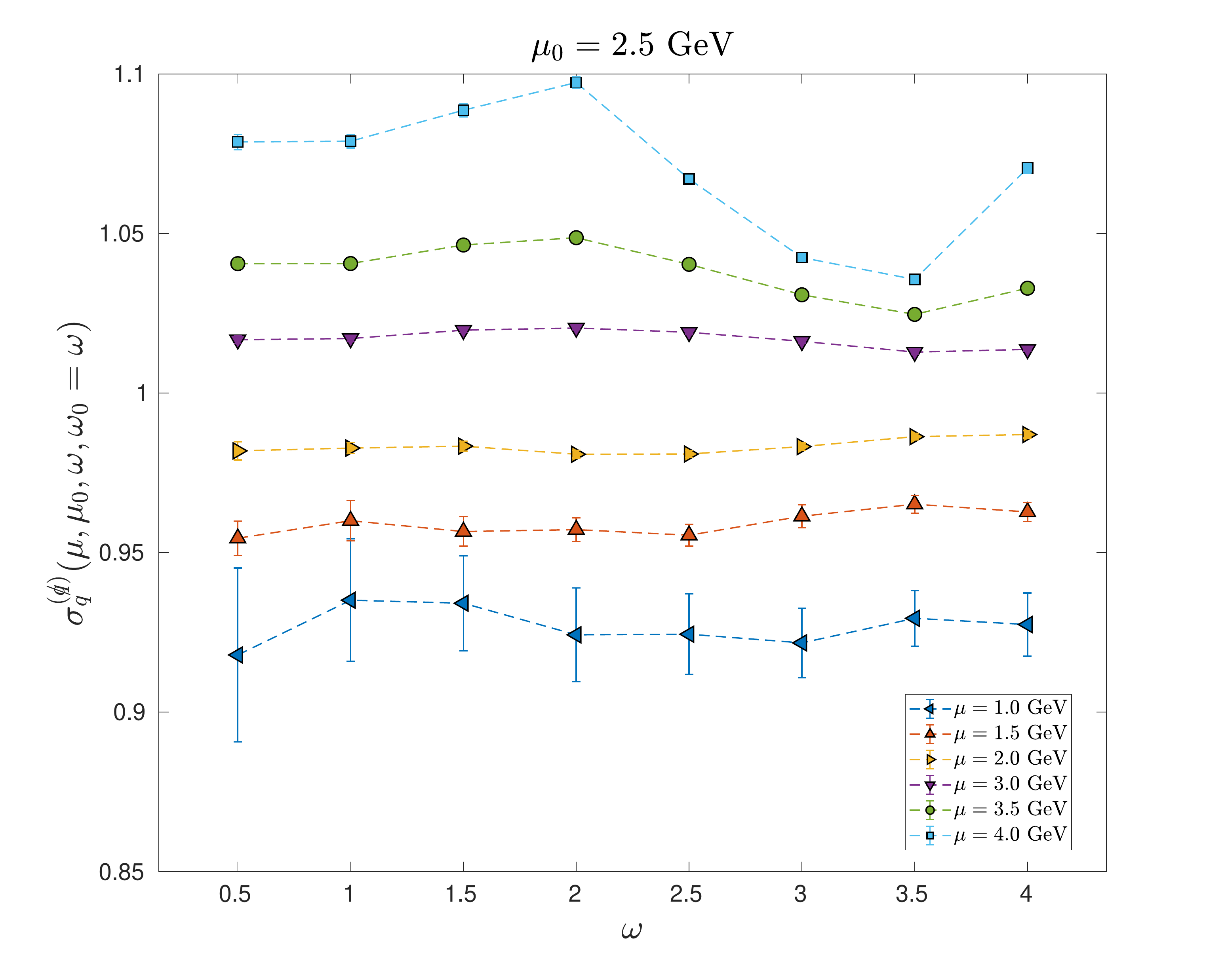} \\
    \includegraphics[width=0.5\textwidth]{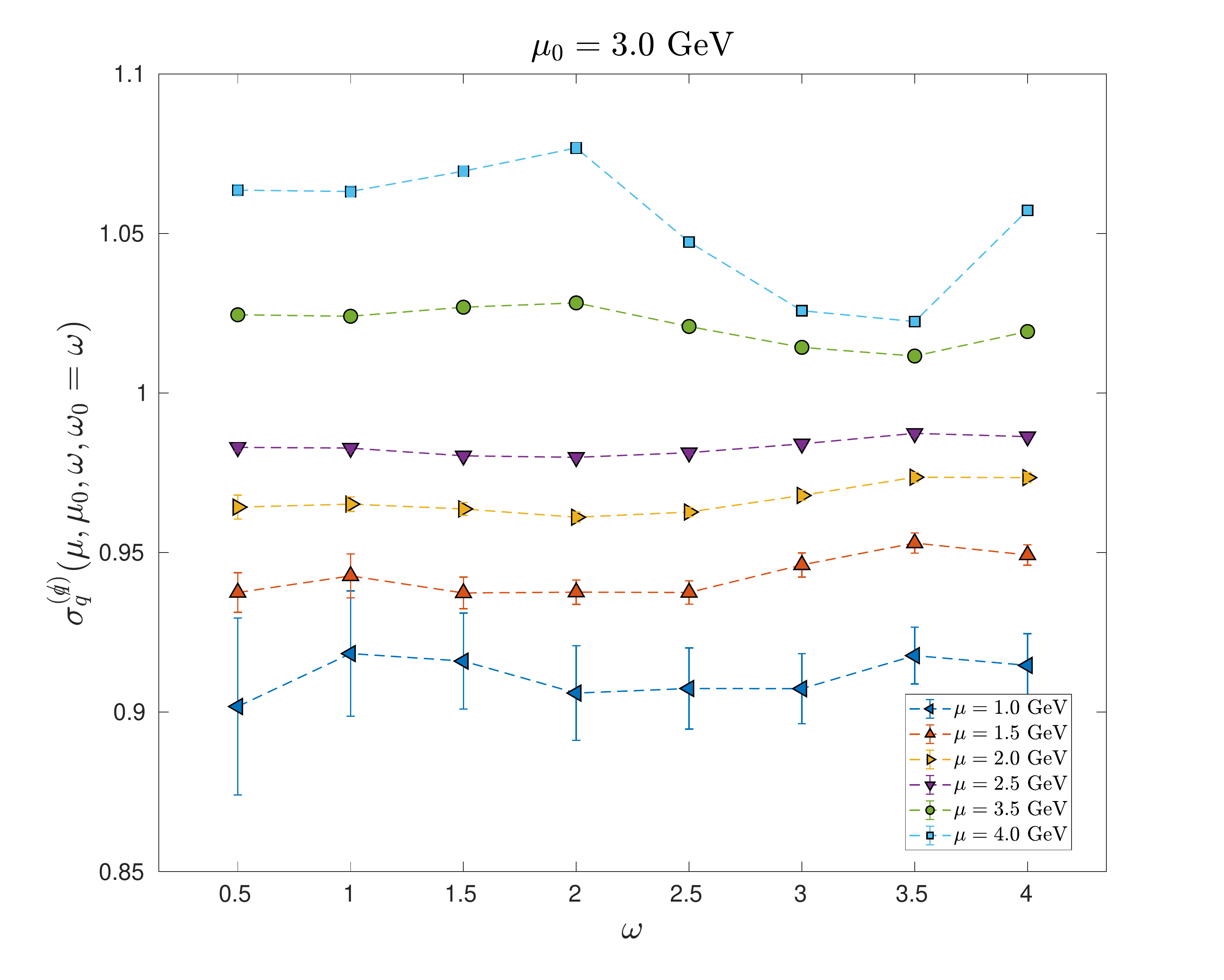} &
    \includegraphics[width=0.5\textwidth]{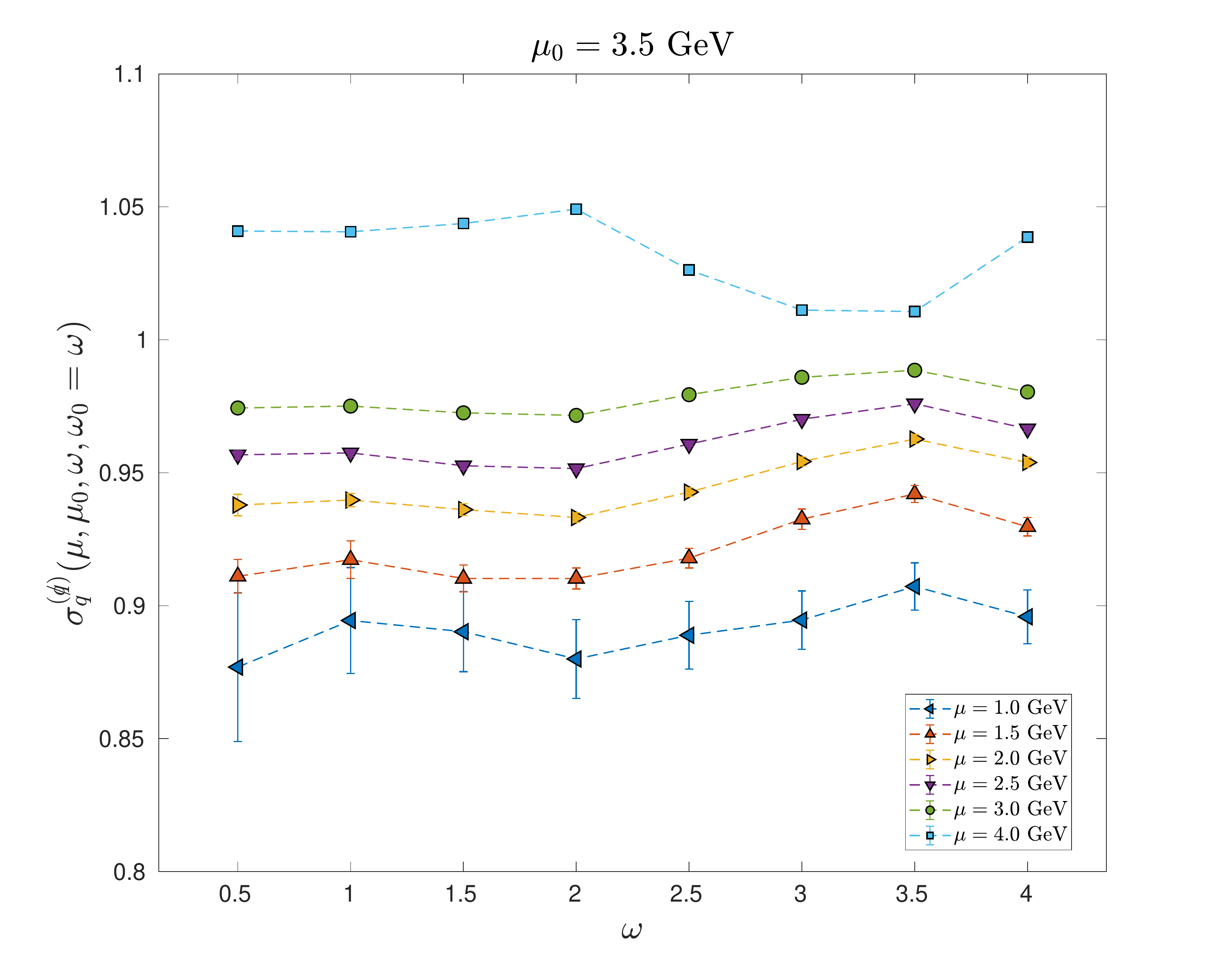} \\
    \end{tabular}
\ec
\caption{
    $\sigma_q^{(\qslash)}(\mu,\mu_0, \omega, \omega_0) $  for $\omega_0=\omega$, 
    statistical error only. 
    }
\label{fig:sigma_q_vs_w}
\end{figure}

\begin{figure}[htb]
  \bc
  \begin{tabular}{cc}
    \includegraphics[width=0.5\textwidth]{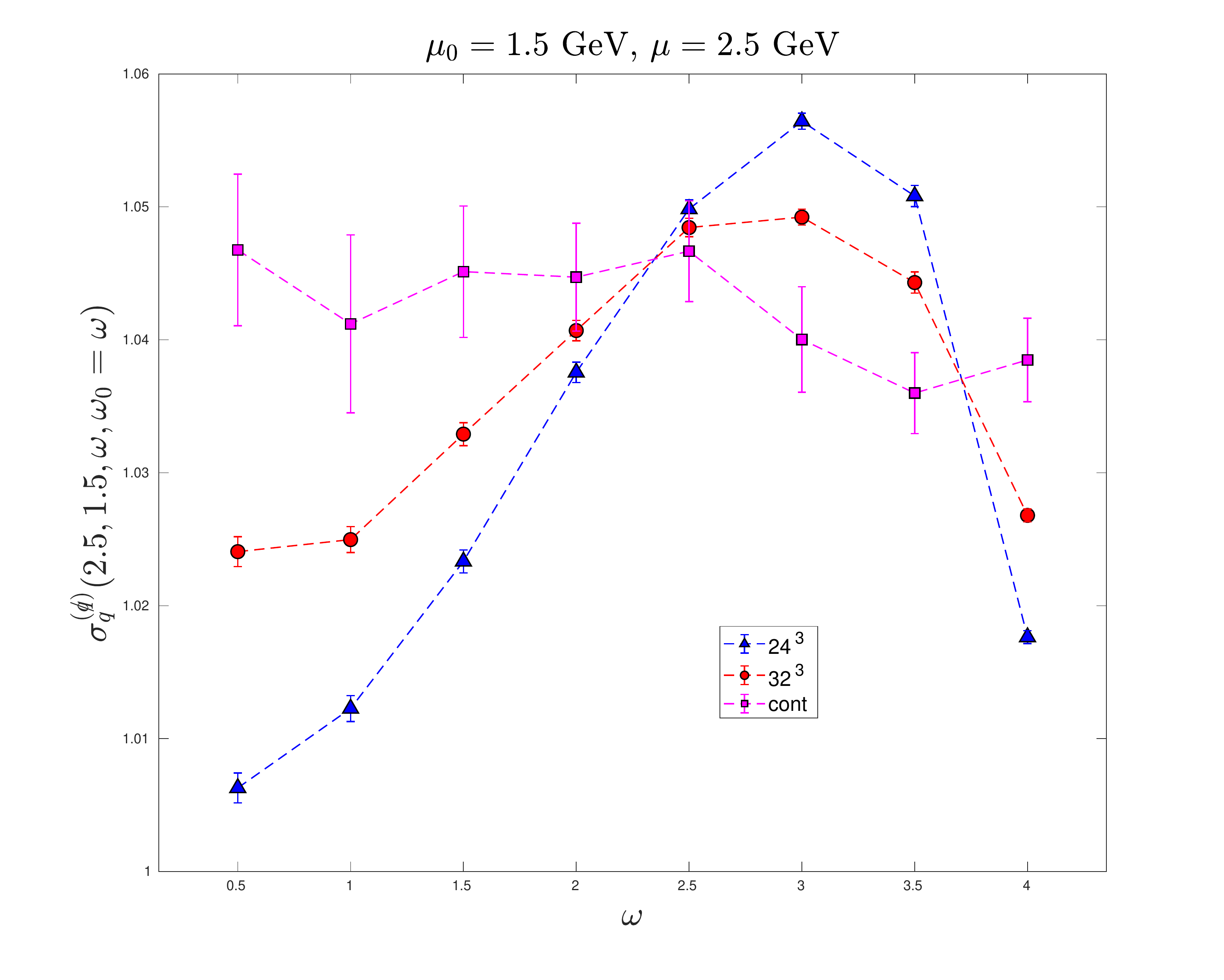} &
    \includegraphics[width=0.5\textwidth]{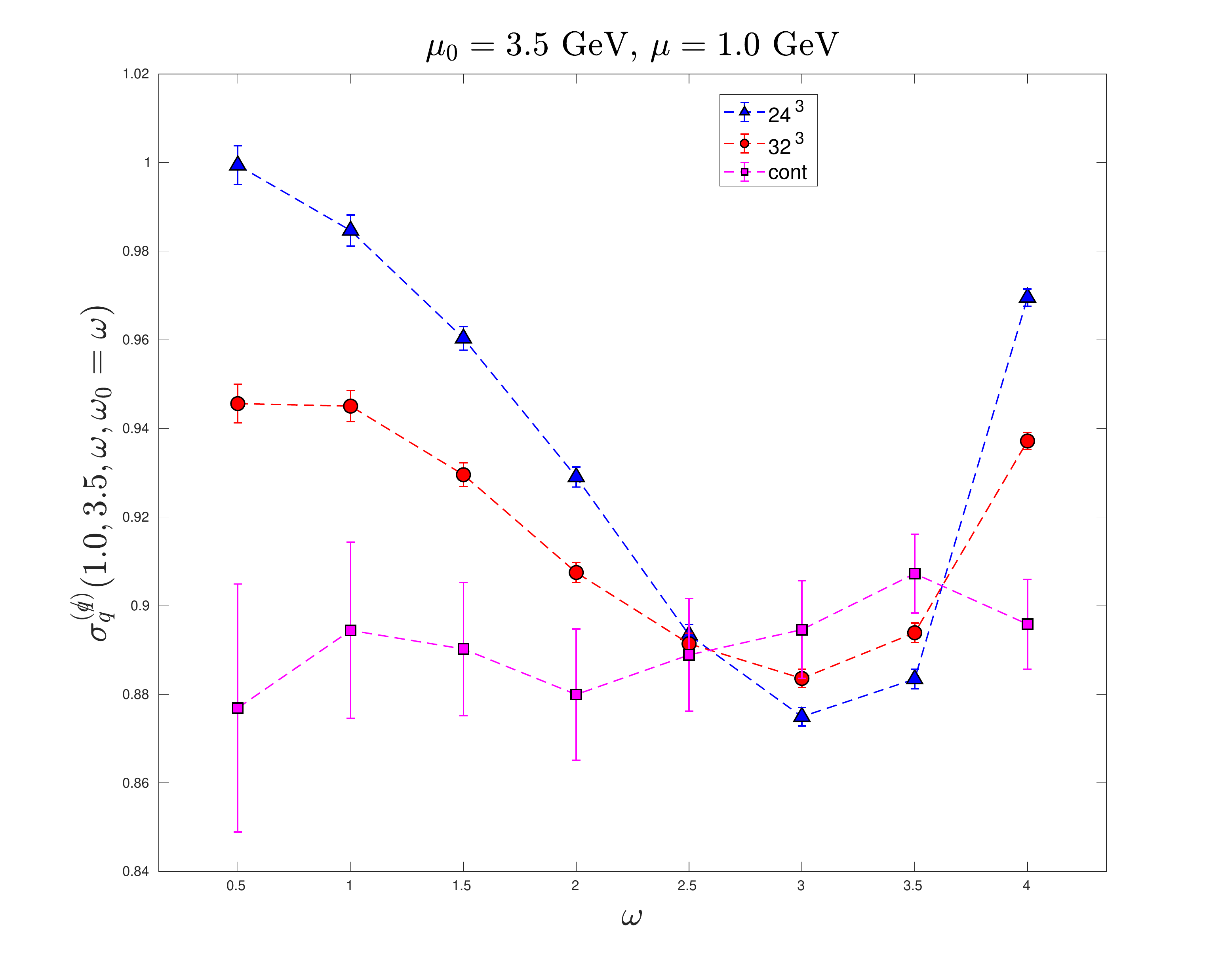} \\
    \end{tabular}
\ec
\caption{Example of continuum extrapolations for  $\sigma_q^{(\qslash)}(\mu,\mu_0, \omega, \omega_0) $. 
}
\label{fig:cl_sigma_q_vs_w}
\end{figure}

Due to the vector WTI, Eq~\ref{eq:VWI}, we might expect $Z_q^{(\qslash)}$ to be
$\omega$-independent up to lattice artefacts. This is illustrated in
Fig.~\ref{fig:sigma_q_vs_w} where we show $\sigma_m^{(\qslash)}(\mu,\mu_0, \omega, \omega_0) $
as a function of $\omega=\omega_0$ for different values of $\mu$ and $\mu_0$.
We can see that this property is rather well satisfied as long as the scales
remain moderate, say $\mu, \mu_0 \le 3$  GeV.
However it is worth noting that this invariance is only true in the continuum,
as can be seen in Fig.~\ref{fig:cl_sigma_q_vs_w}. At finite lattice spacing,
within our small statistical errors, the lattice artefacts are visible
and $Z_q^{\qslash}$ clearly show a $\omega$ dependence.
These figures also suggest that the lattice artefacts are well under control,
with two lattice spacings, as long we do not go too high in energy.
Note that in these plots, the systematic errors due to the continuum extrapolations
are not included. The non perturbative-scale evolution of $Z_q$, $\sigma_q(\mu,\mu_0)$
with $\mu_0=2$ GeV and $\omega=\omega_0$ for various values of $\mu$ and $\omega$ can be
found in Tables~{\ref{tab:sigma_q_gamma} and  \ref{tab:sigma_q_qslash}}.

Also it is interesting to notice in Fig~\ref{fig:cl_sigma_q_vs_w} how the discretisation
effects depend on $\omega$. Clearly for $\omega=2.0$ and $2.5$
the lattice artefacts are much smaller than for $\omega=1.0$.
Of course this effect is quantity-dependent but it this could be useful
in future computations. 

\begin{table}[t]
\bc
\begin{tabular}{ |c| c  c  c  c  c  c  | }
  \hline
  $\omega / \mu=$
      & $1.0$           & $1.5$            & $2.5$           & $3.0$           & $3.5$          & $4.0$             \\
  \hline
$0.5$ & $0.972(   8)$   & $0.993(   4)$    & $1.008(   4)$   & $1.014(   8)$   & $1.023(  14)$   & $1.040(  26)$     \\ 
$1.0$ & $0.976(   8)$   & $0.994(   3)$    & $1.004(   2)$   & $1.007(   5)$   & $1.012(   8)$   & $1.021(  15)$     \\ 
$1.5$ & $0.978(   4)$   & $0.998(   2)$    & $1.003(   1)$   & $1.005(   2)$   & $1.006(   3)$   & $1.004(   3)$     \\ 
$2.0$ & $0.990(   7)$   & $0.998(   2)$    & $1.003(   0)$   & $1.005(   1)$   & $1.007(   1)$   & $1.008(   1)$     \\ 
$2.5$ & $0.987(   5)$   & $0.997(   2)$    & $1.001(   1)$   & $1.002(   2)$   & $1.002(   3)$   & $1.003(   4)$     \\ 
$3.0$ & $0.985(   4)$   & $0.999(   2)$    & $1.000(   2)$   & $0.998(   4)$   & $0.993(   9)$   & $0.978(  18)$     \\ 
$3.5$ & $0.989(   5)$   & $1.001(   2)$    & $0.997(   2)$   & $0.993(   6)$   & $0.982(  13)$   & $0.959(  27)$     \\ 
$4.0$ & $0.990(   5)$   & $0.999(   1)$    & $0.994(   3)$   & $0.983(   8)$   & $0.957(  22)$   & $0.887(  60)$     \\ 
\hline
\end{tabular}
\caption{Non-perturbative running for the quark wave function in the $\gamma_\mu$ scheme. }
\label{tab:sigma_q_gamma}
\ec
\end{table}

\begin{table}[t]
\bc
\begin{tabular}{ |c| c  c  c  c  c  c  | } 
\hline
  $\omega / \mu=$
& $1.0$           & $1.5$            & $2.5$           & $3.0$           & $3.5$          & $4.0$             \\
\hline
$0.5$ & $0.935(  27)$   & $0.974(   7)$    & $1.018(   7)$   & $1.035(  16)$   & $1.059(  31)$   & $1.096(  56)$     \\ 
$1.0$ & $0.951(  19)$   & $0.978(   6)$    & $1.017(   6)$   & $1.035(  14)$   & $1.058(  28)$   & $1.096(  51)$     \\ 
$1.5$ & $0.950(  15)$   & $0.973(   4)$    & $1.017(   4)$   & $1.037(  11)$   & $1.064(  25)$   & $1.106(  49)$     \\ 
$2.0$ & $0.942(  15)$   & $0.976(   3)$    & $1.020(   3)$   & $1.040(   8)$   & $1.069(  20)$   & $1.118(  45)$     \\ 
$2.5$ & $0.942(  14)$   & $0.974(   3)$    & $1.019(   1)$   & $1.039(   3)$   & $1.060(   9)$   & $1.087(  20)$     \\ 
$3.0$ & $0.937(  12)$   & $0.978(   4)$    & $1.017(   2)$   & $1.033(   2)$   & $1.048(   2)$   & $1.060(   2)$     \\ 
$3.5$ & $0.943(  10)$   & $0.979(   3)$    & $1.014(   2)$   & $1.027(   3)$   & $1.039(   3)$   & $1.050(   2)$     \\ 
$4.0$ & $0.940(  10)$   & $0.975(   4)$    & $1.013(   3)$   & $1.027(   8)$   & $1.046(  18)$   & $1.084(  40)$     \\
\hline
\end{tabular}
\ec
\caption{Non-perturbative running for the quark wave function in $\qslash$ scheme. Here we expect the results
  to be $\omega$-independent.}
\label{tab:sigma_q_qslash}
\end{table}

\clearpage
\subsection{Study of the chiral symmetry breaking effects}

One of the original arguments to motivate the RI/SMOM schemes is a drastic reduction
 in the effects of spontaneous chiral symmetry. Even though these effects are physical,
 they can prevent a clean determination of the renormalisation factors
 because they are absent from the perturbative calculations. This is particularly true
for quantities like $\Pi_P$ and $\Sigma^S$
 where the presence of pseudo-Goldstone poles can
completely dominate the signal. In practice, the vertex functions from which
we want to extract the Z-factors are 'polluted' by negative powers of the quark mass
or the momentum scale. Of course these chiral symmetry breaking effects are non-perturbative
and disappear for high-momentum. However, in practice we want to keep the Rome-Southampton windows
as open as possible, so it is always desirable to reduce these infrared contamination
or at least keeping them under control.
\\

\begin{figure}[t]
  \bc
  \begin{tabular}{cc}
    \includegraphics[width=0.5\textwidth]{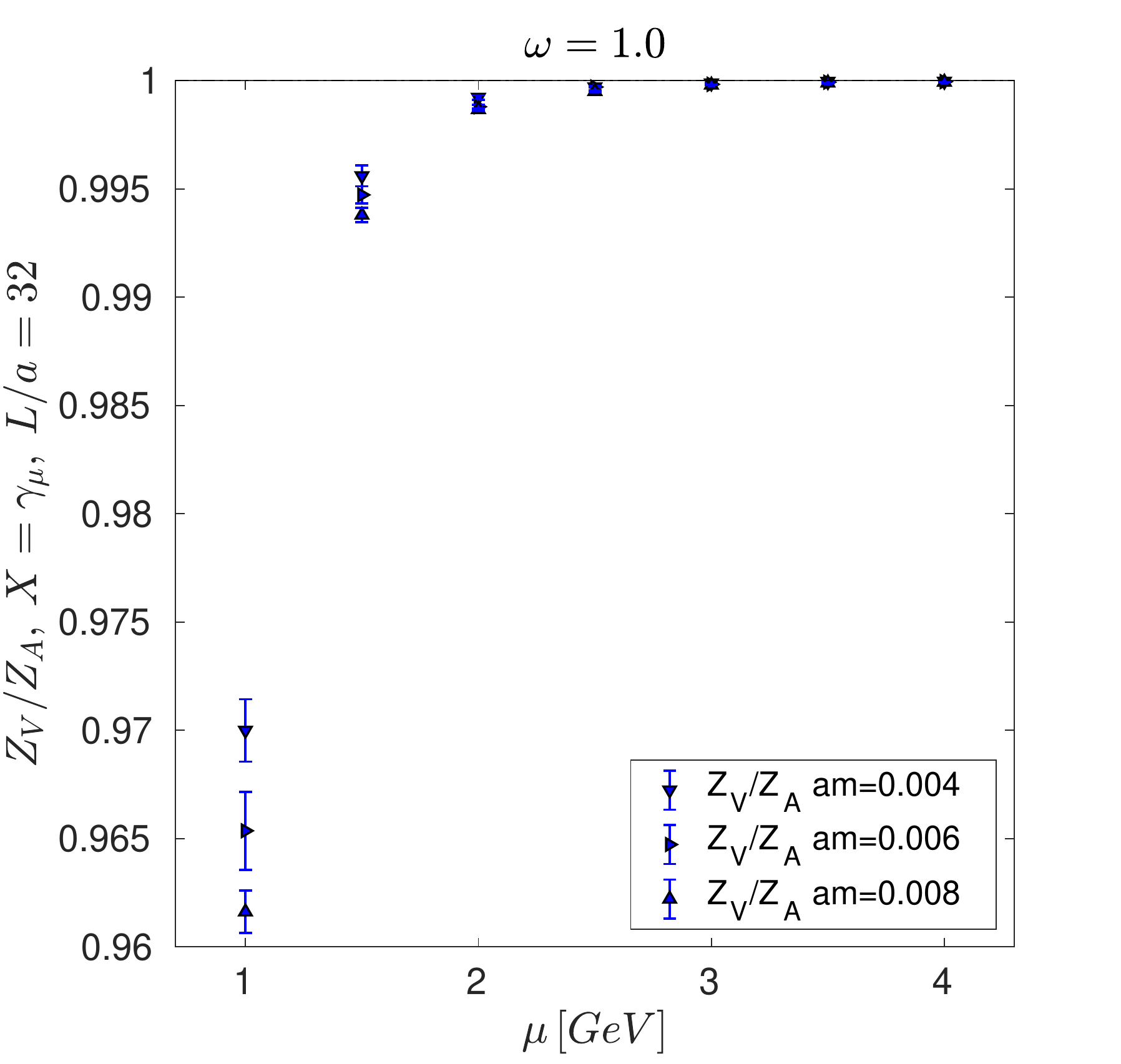} &
    \includegraphics[width=0.5\textwidth]{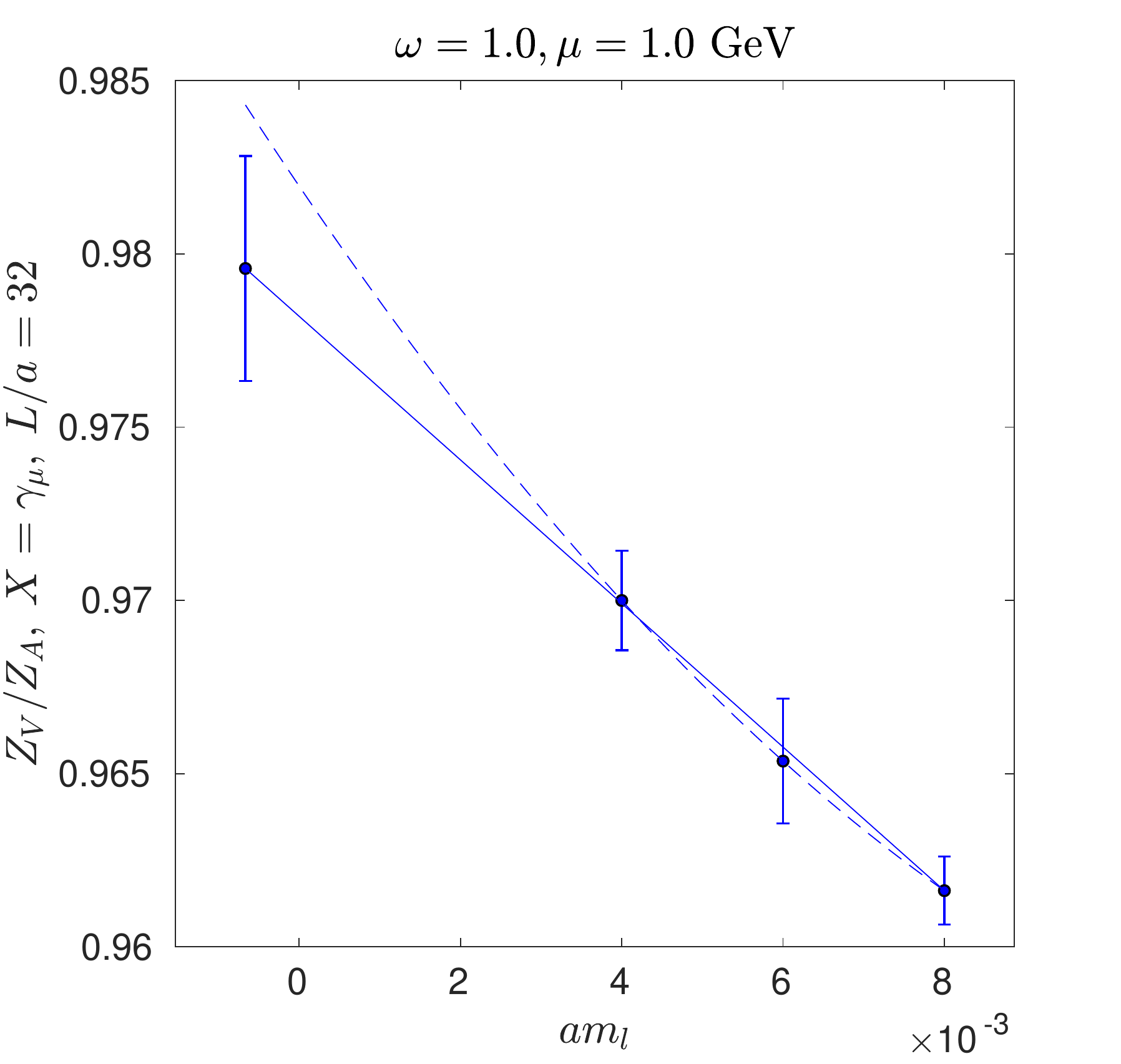} 
    \end{tabular}
\ec
\caption{$Z_V/Z_A$ for the ${\gamma_\mu}$-projector as a function of $\mu$ for $\omega=1$.
    On the left panel we show the results at finite quark masses as a function of the energy scale.
    As an example of chiral extrapolations, on the right panel we show our results for $\mu =1 $ GeV,
    $\omega=1$. We perform a linear extrapolation (solid line), a quadratic extrapolation
    is shown as a dash line for illustration.
}
\label{fig:ZVoA}
\end{figure}

If chiral symmetry is exactly realised then we should find that $Z_S=S_P$ and $Z_V = Z_A$.
Therefore, in order to study these chiral symmetry breaking affects as a
function of $\omega$,
we study the deviation of $Z_V$ from $Z_A$ and $Z_S$ from $Z_P$.
We start by the analysis of the ratio $Z_V/Z_A$, see Fig.~\ref{fig:ZVoA}
for $\omega=1$ using the  the ${\gamma_\mu}$-projector.
This ratio collapses quickly to one as the energy scale increases:
at $\mu=2$ GeV, the deviation from one is at most at the per-mille level. 
We also find the quark mass dependence to be linear within our statistical error -
as expected - therefore we extrapolate to the chiral limit using a straight line.
Our results for the $\qslash$-projector are very similar.
\clearpage
In  Fig.~\ref{fig:ZVoA} we show our results in the chiral limit for the
different values of $\omega$ and find that $Z_V/Z_A$ is always very close to one.
At our lowest energy point $\mu=1$ GeV, we can observe a deviation from one
at the order of a percent. Our results seem to indicate a trend
that $Z_V/Z_A$ increases toward one when $\omega$ increases between 0.5 and 2
($Z_V/Z_A \sim 0.98$ at $\omega=1$ and  $Z_V/Z_A \sim 0.99$ at $\omega=2$),
but this could well be a statistical effect. 
All together we find that $Z_V/Z_A =1 $ to a very good approximation
for all values of $\omega$. This true for both values of the lattice spacing and
both projectors $X\in (\gamma_\mu, \qslash)$.
\begin{figure}[t]
  \bc
  \begin{tabular}{cc}
    \includegraphics[width=0.5\textwidth]{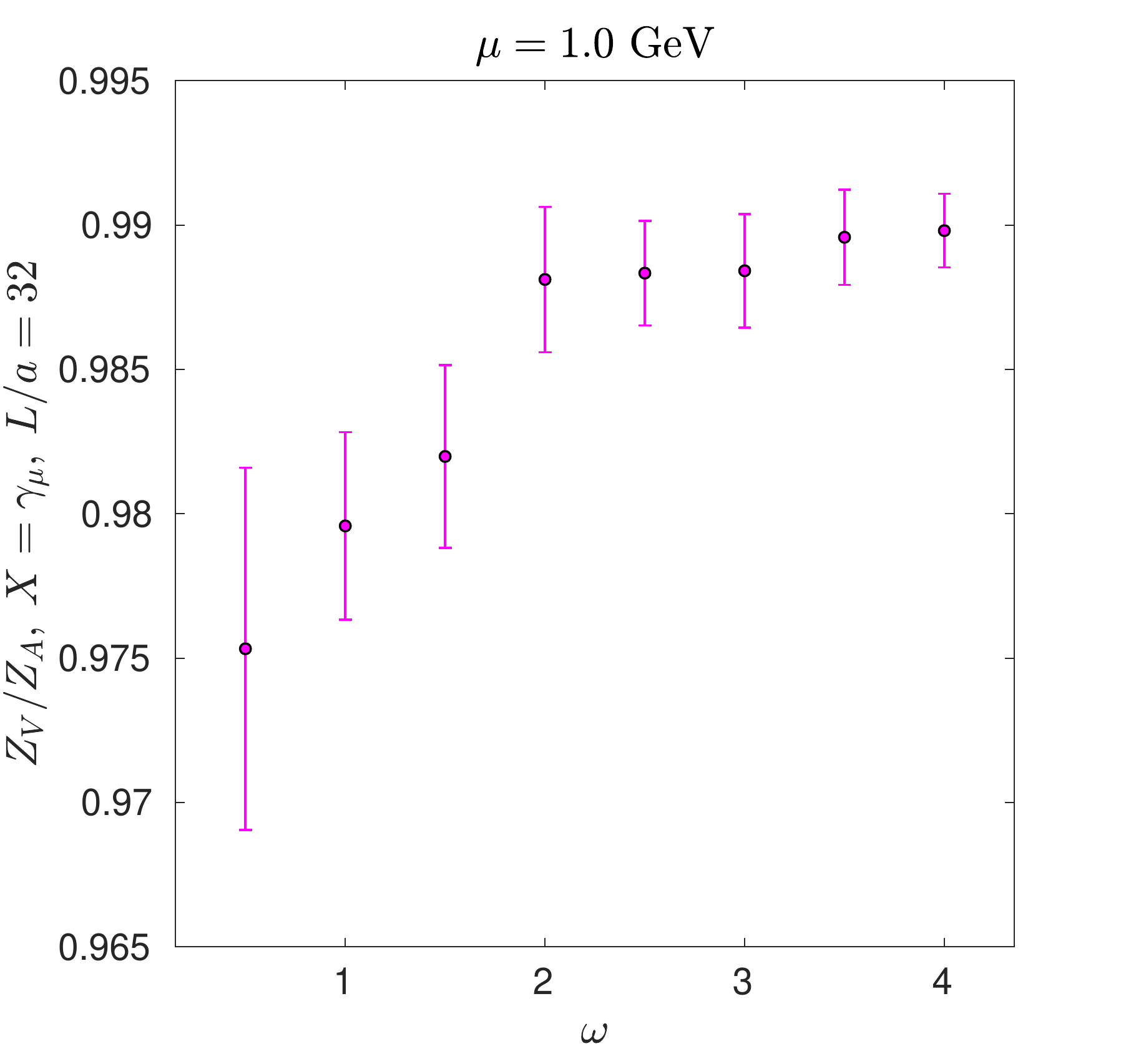} &
    \includegraphics[width=0.5\textwidth]{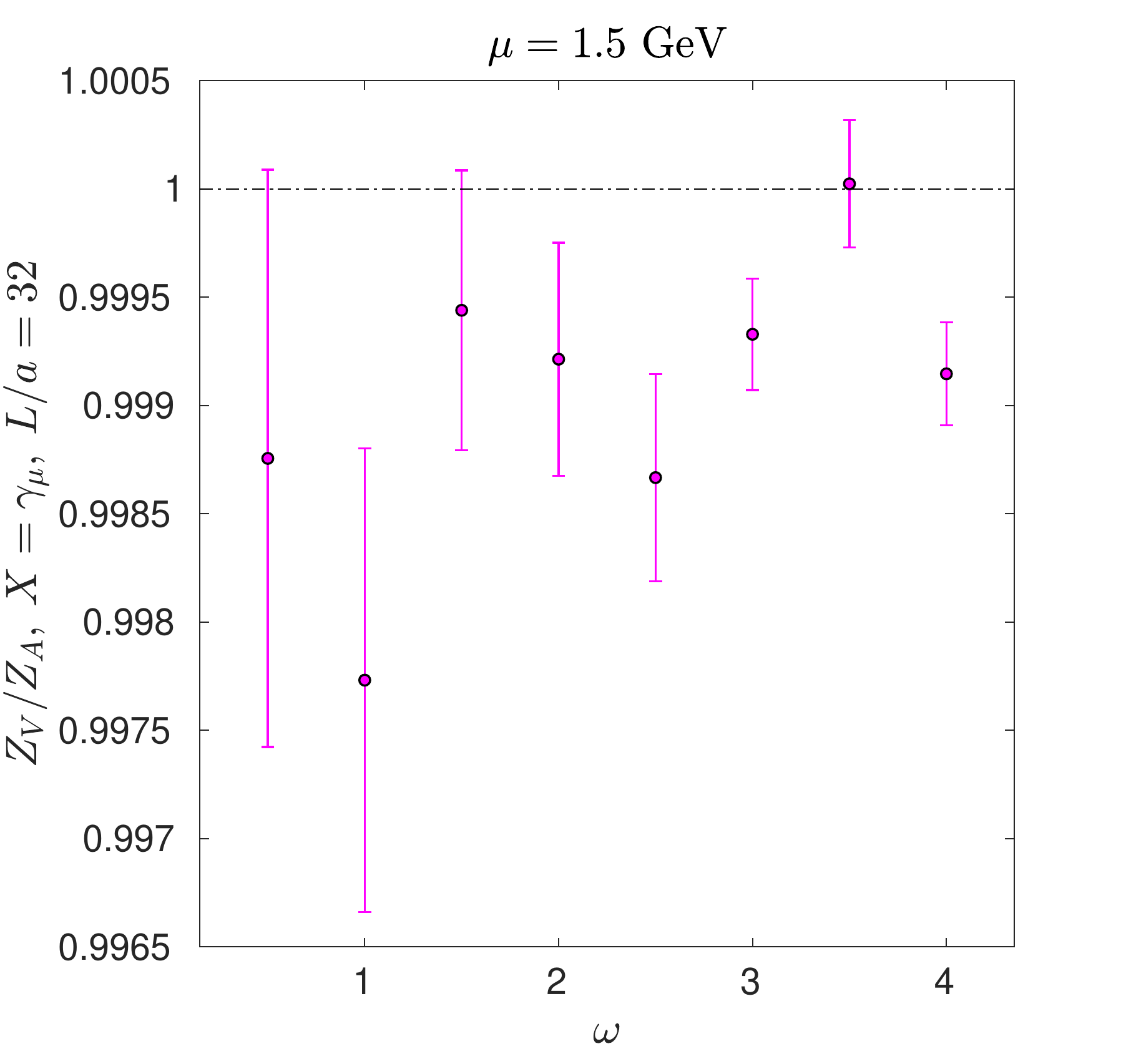} \\
    \includegraphics[width=0.5\textwidth]{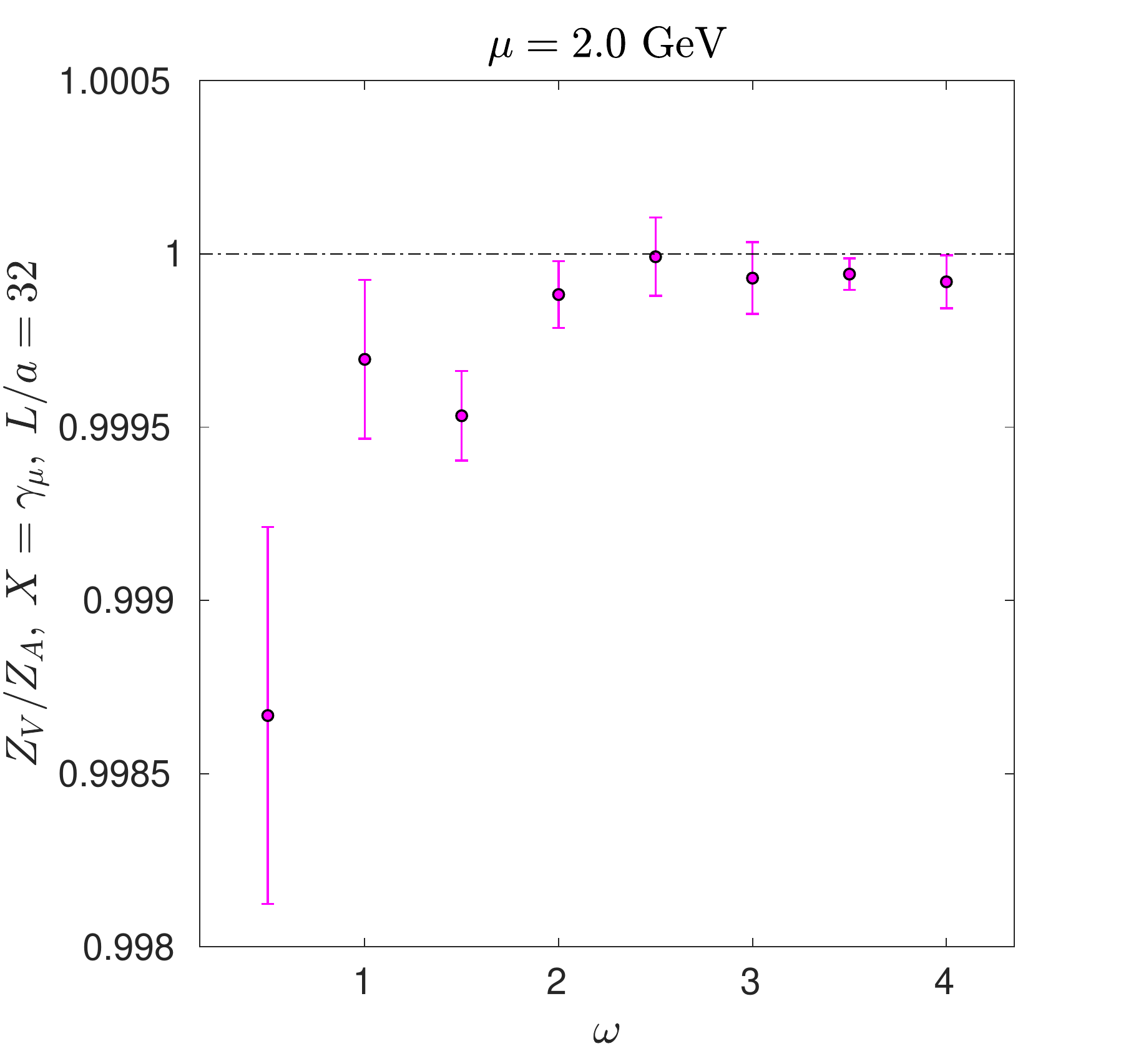} &
    \includegraphics[width=0.5\textwidth]{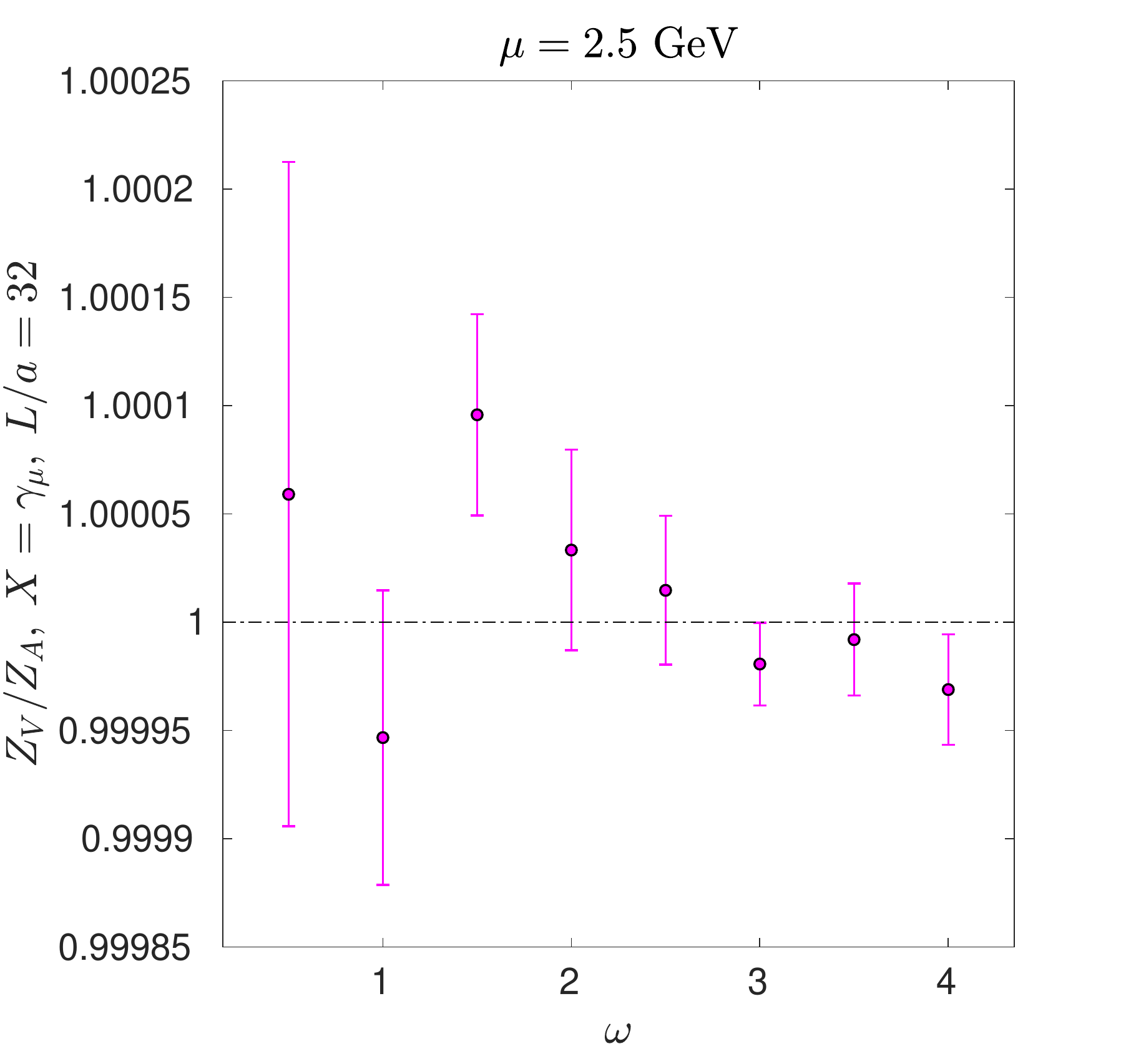} \\
    \includegraphics[width=0.5\textwidth]{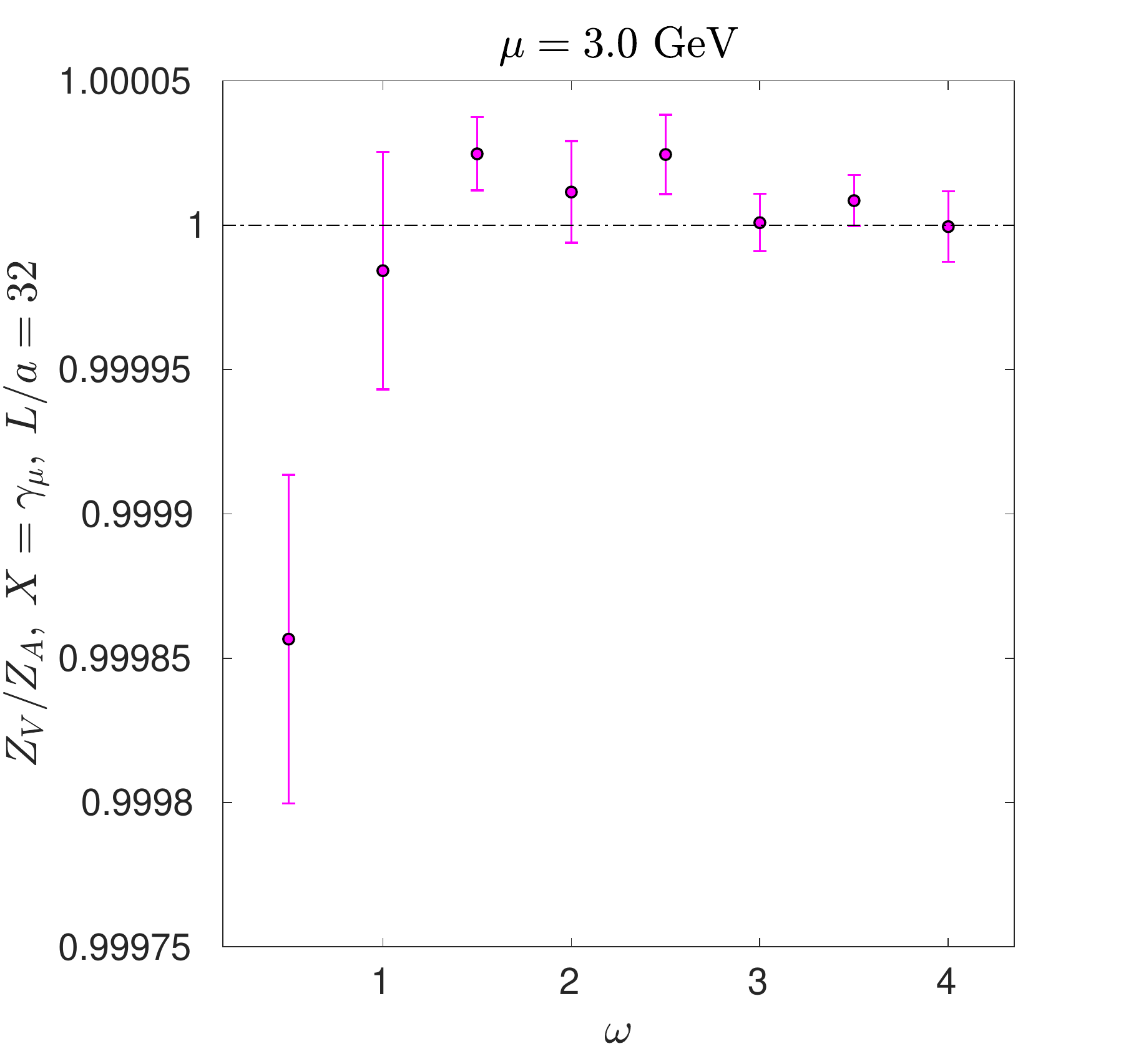} &
    \includegraphics[width=0.5\textwidth]{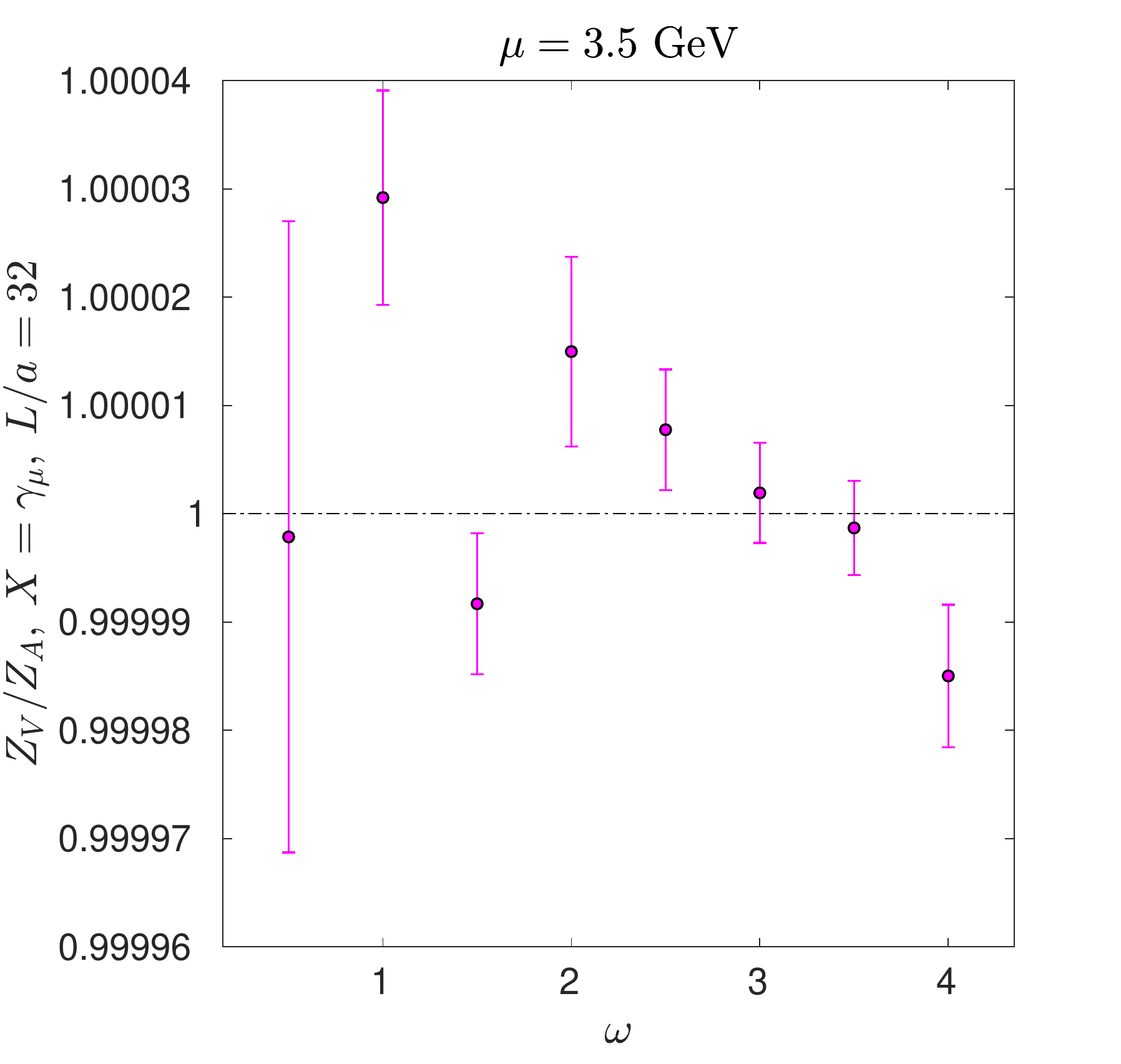} 
    \end{tabular}
\ec
\caption{\footnotesize{$Z_V/Z_A$ for the ${\gamma_\mu}$ projector as a function of $\omega$ for various
    values of $\mu$.}}
\label{fig:ZVoA_vs_mu}
\end{figure}

\begin{figure}[t]
  \bc
  \begin{tabular}{cc}
    \includegraphics[width=0.5\textwidth]{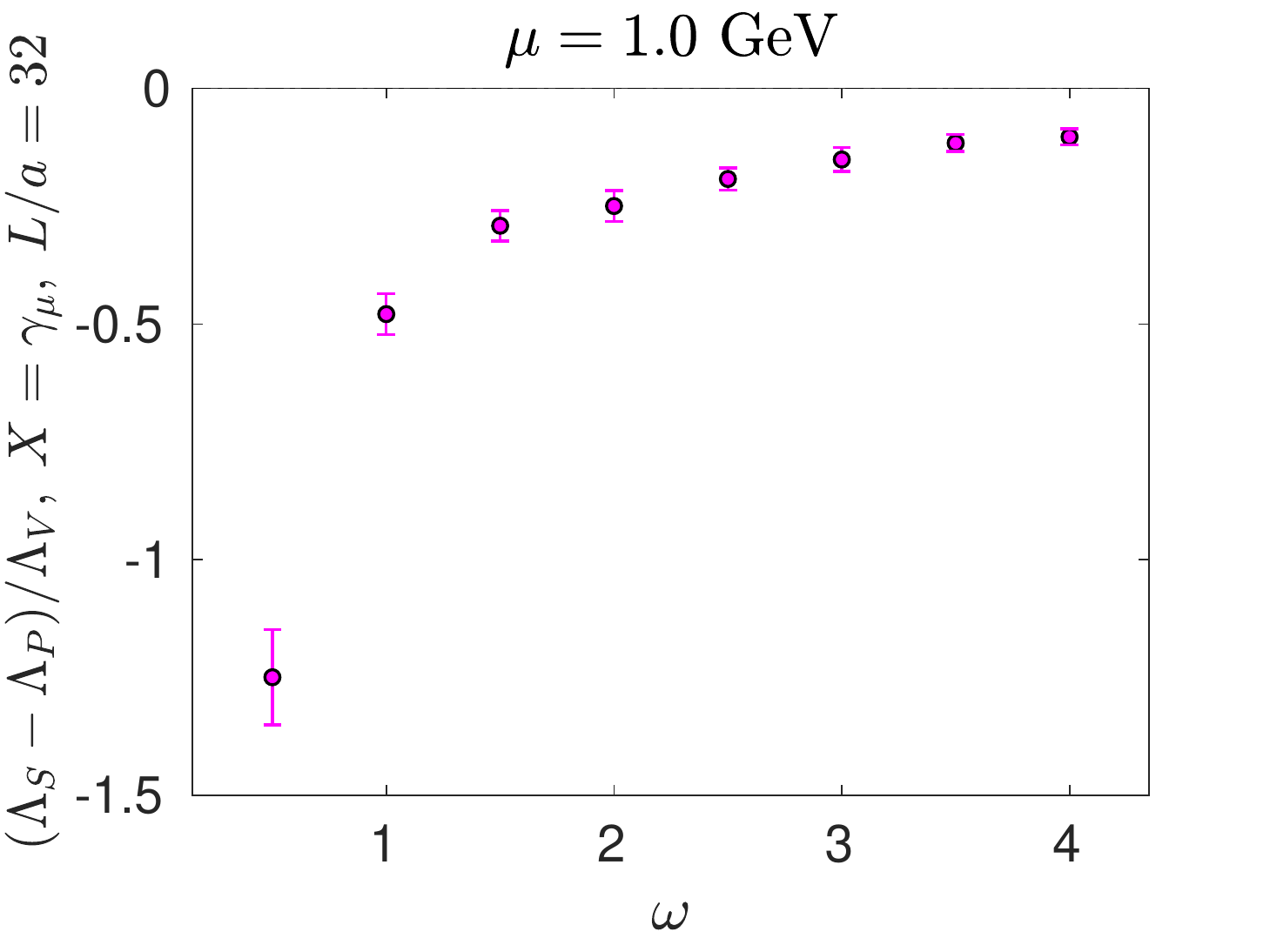} &
    \includegraphics[width=0.5\textwidth]{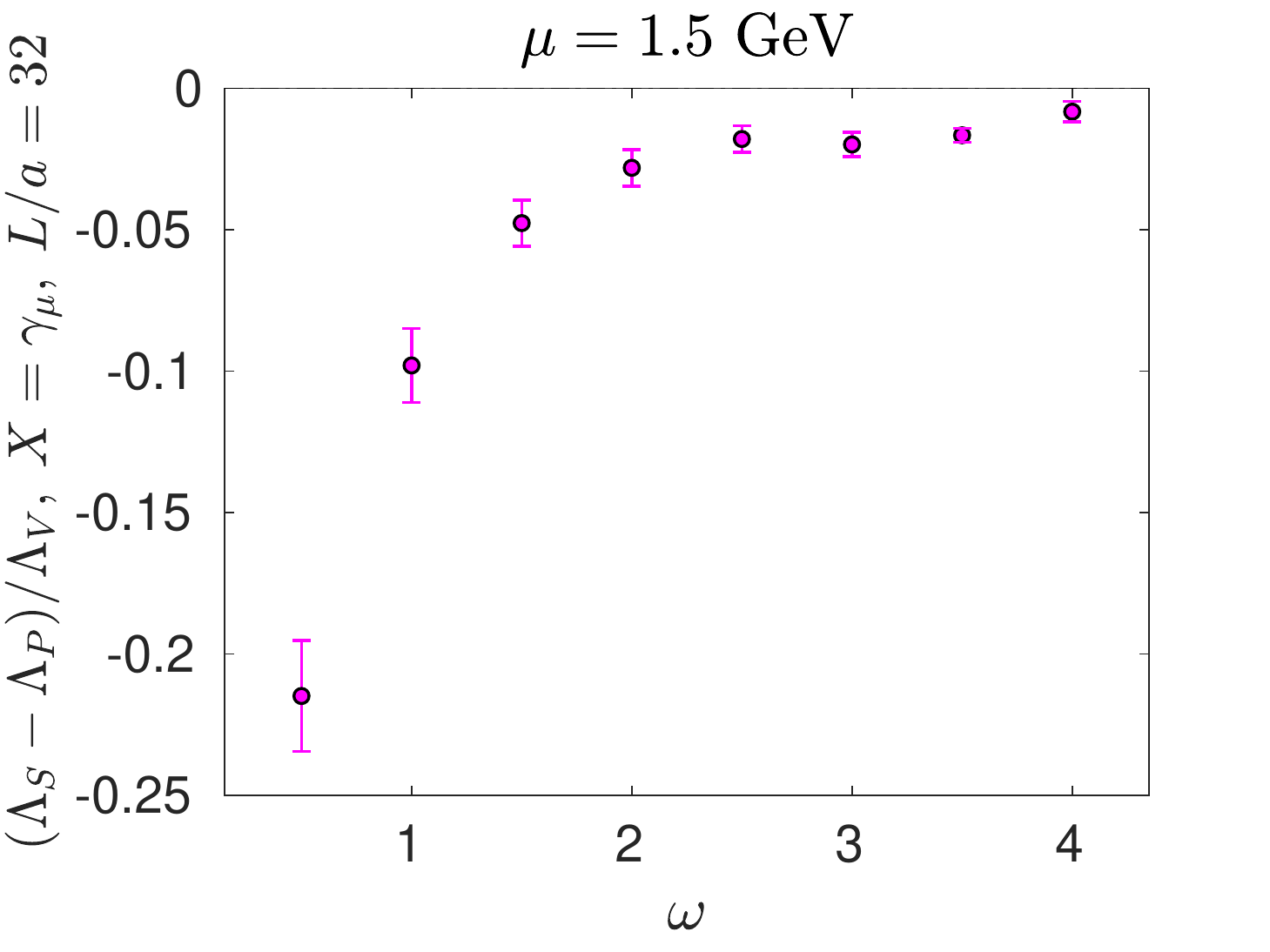} \\
    \includegraphics[width=0.5\textwidth]{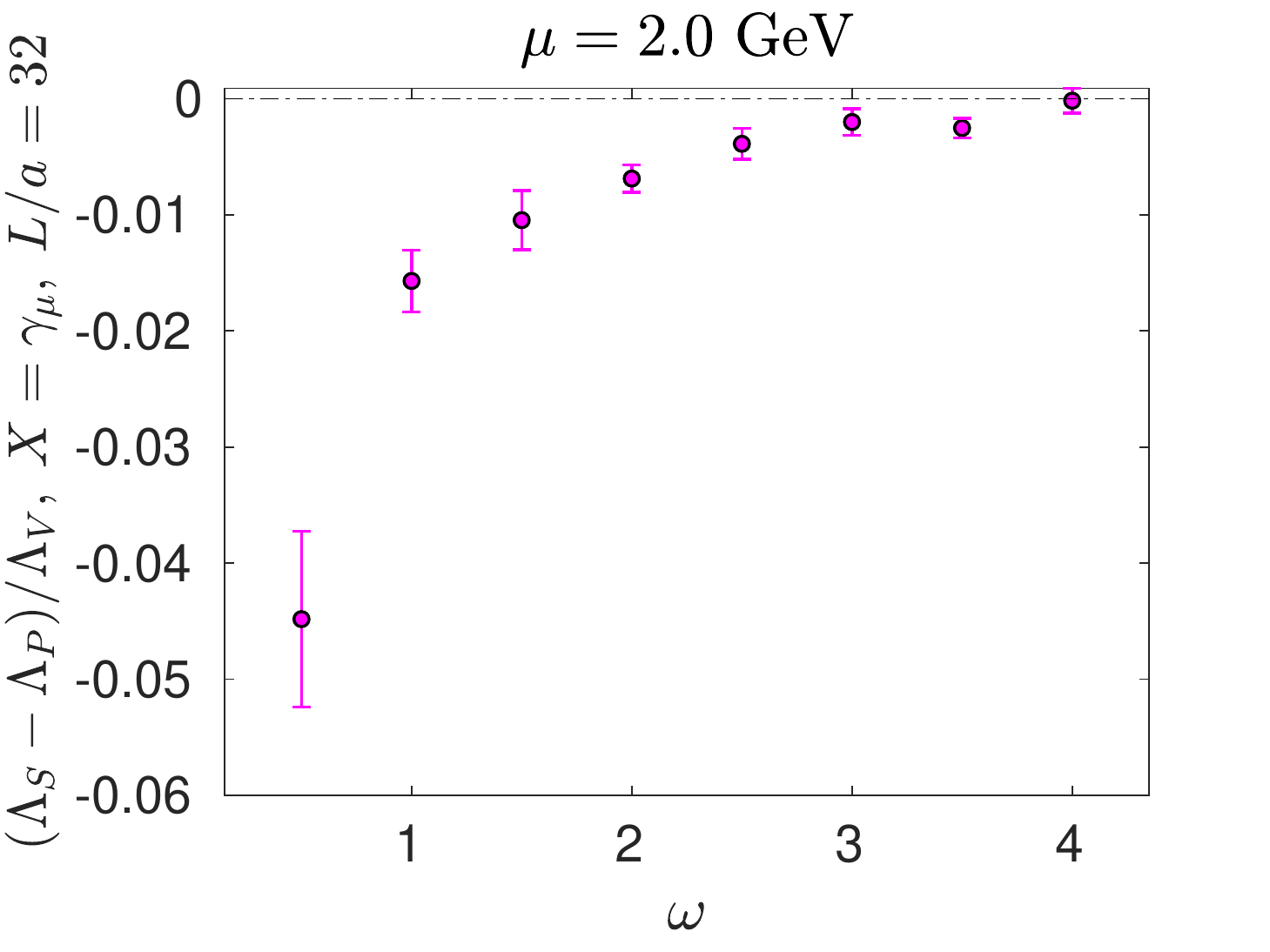} &
    \includegraphics[width=0.5\textwidth]{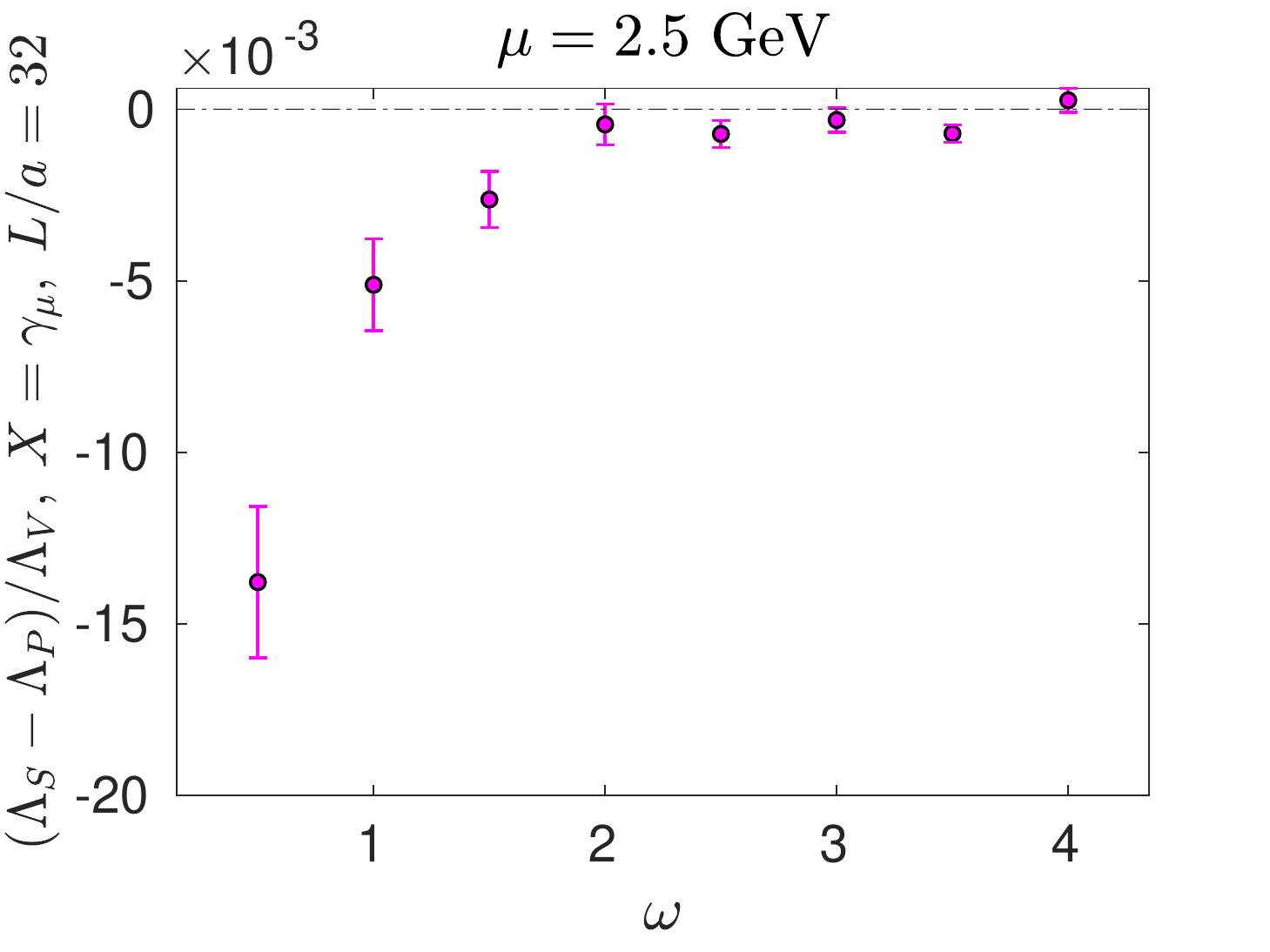} \\
    \includegraphics[width=0.5\textwidth]{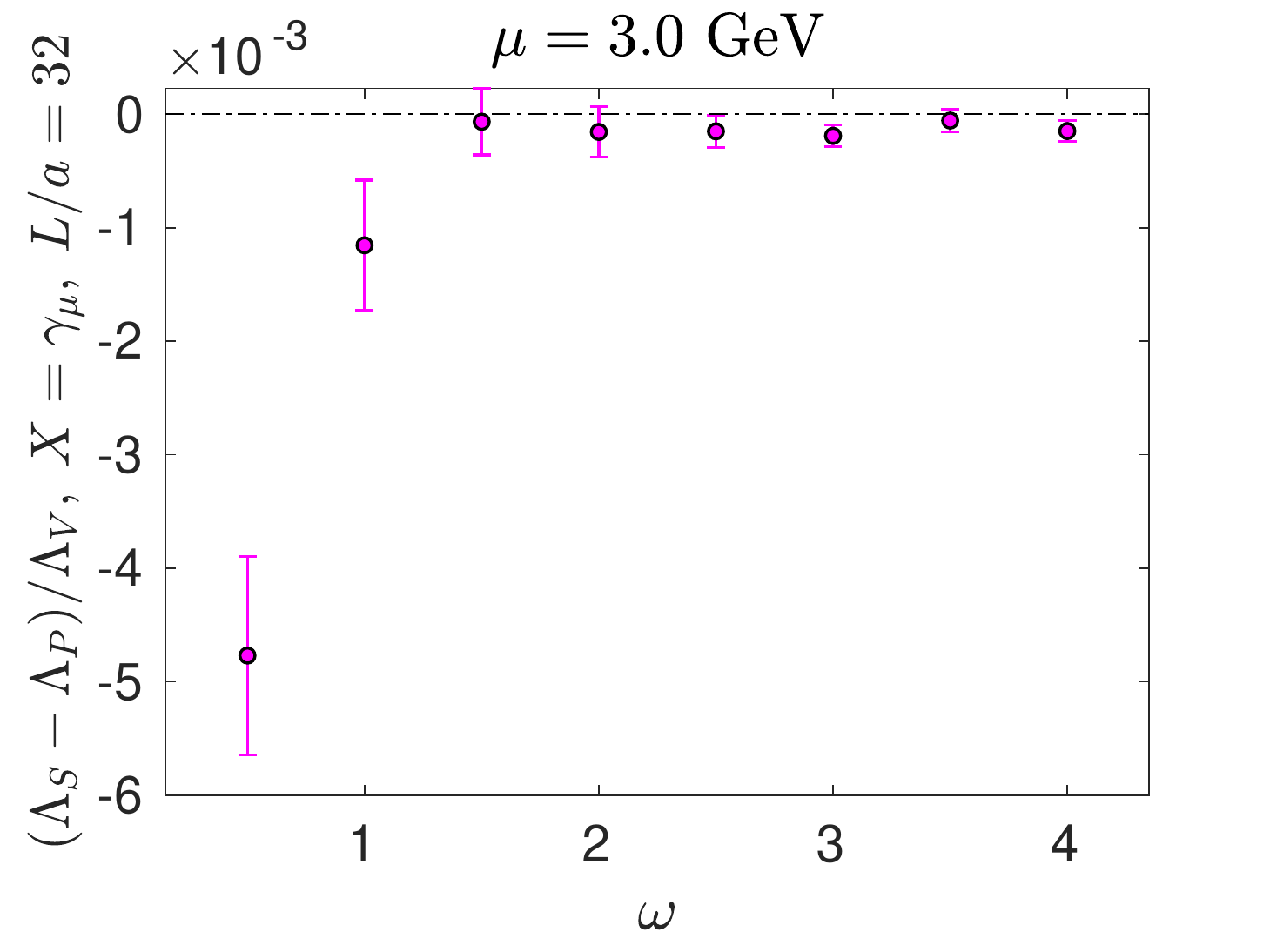} &
    \includegraphics[width=0.5\textwidth]{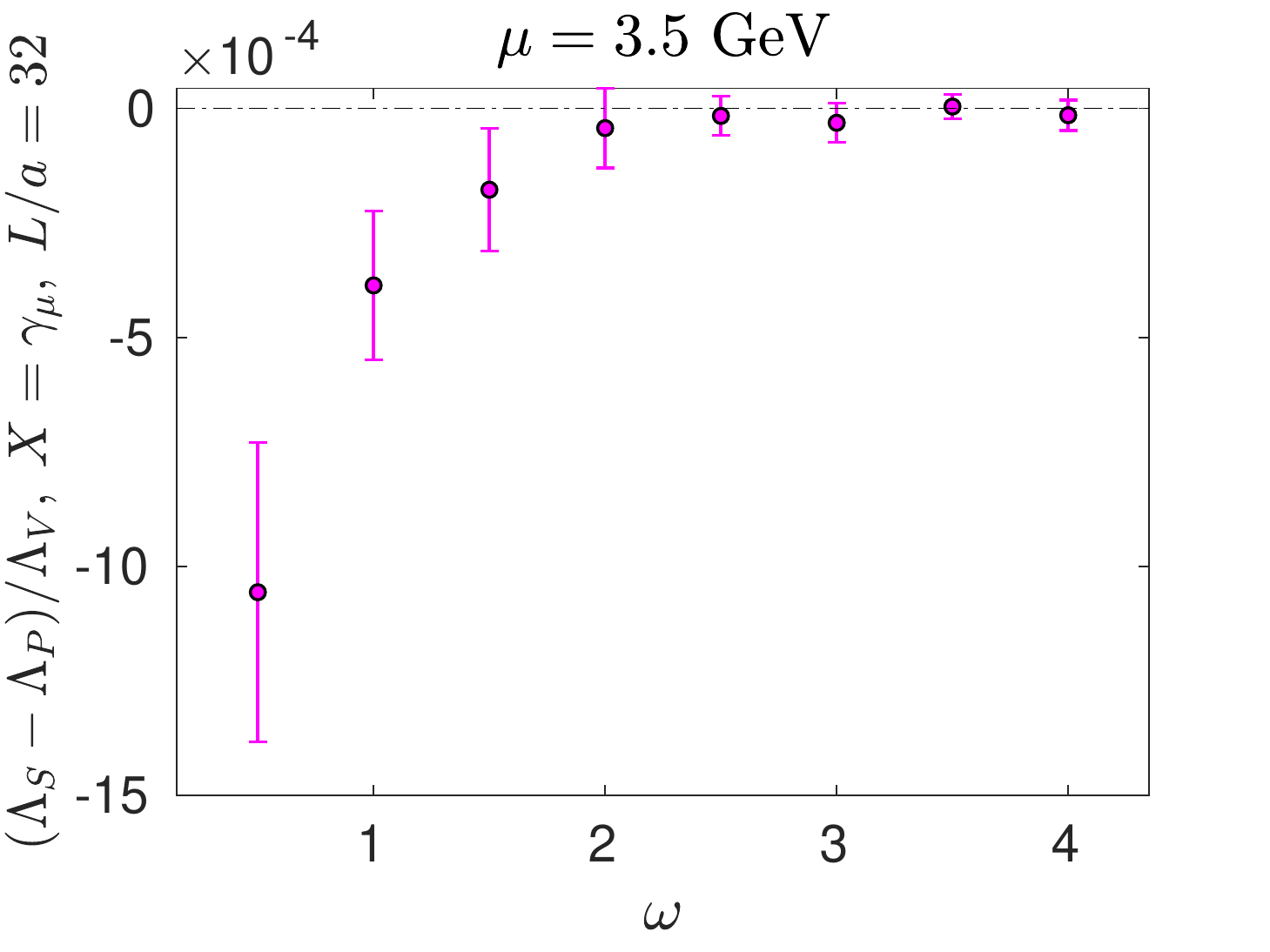} 
    \end{tabular}
\ec
\caption{\footnotesize{$(\Lambda_S-\Lambda_P) / \Lambda_V$ for the ${\gamma_\mu}$ projector as a function of $\omega$ for various
    values of $\mu$.}}
\label{fig:LSmP_vs_w}
\end{figure}

We also study $(\Lambda_S - \Lambda_P)/\Lambda_V$, which is proportional to $Z_V (1/Z_S -1/Z_P)$.
Our results are shown in \ref{fig:LSmP_vs_w}. Here it is clear that
the IR effects due to chiral symmetry breaking are affected by the
value of $\omega$. We can observe that at $\mu=1.5$ GeV,
$(\Lambda_S - \Lambda_P)/\Lambda_V \sim 0.1$. The same quantity reduces to $\sim 0.03$
for $\omega=2.0$. It will be interesting to perform a similar study on
four-quark operators, especially those where the IR contamination
can be important and a source of disagreement 
(see for example the section on BSM kaon mixing in FLAG~\cite{Aoki:2021kgd}
and~\cite{Boyle:2017skn}).

\begin{figure}[t]
  \bc
  \includegraphics[width=0.5\textwidth]{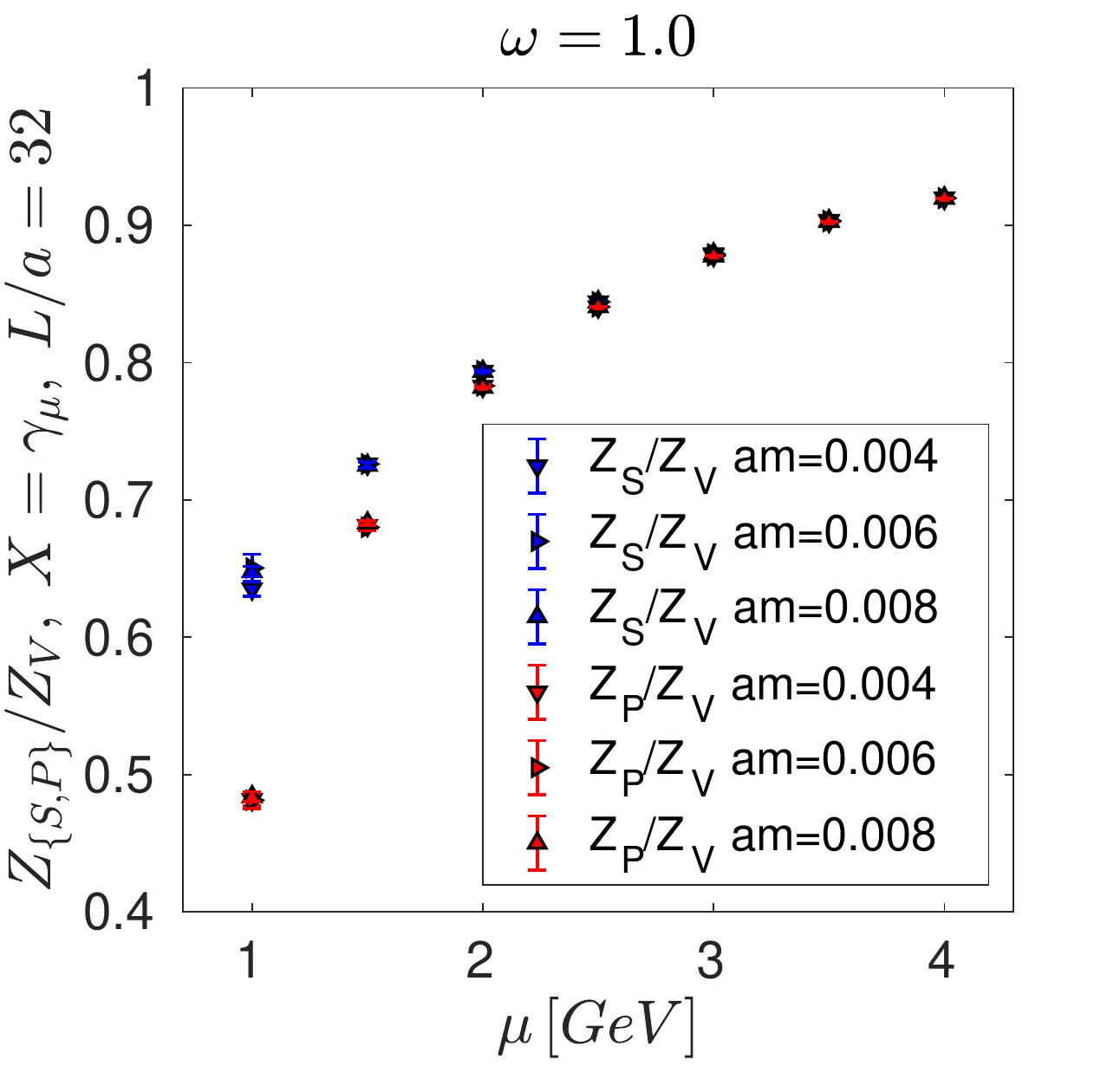}
  \ec
  \caption{
    $Z_S$ and $Z_P$ at finite quark masses versus the renormalisation
    scale for $\omega=1$. 
  }
    \label{fig:ZS_vs_mu}
\end{figure}

\section{Conclusions and outlook}
\label{sec:conclusions}

As a proof of concept we have implemented several IMOM schemes,
defined via two different projectors, and for various kinematics. In
this framework we have determined the renormalisation factors and
non-perturbative scale evolution functions of the quark mass and of
the quark wave function. To compare our lattice results with 
NNLO results in continuum perturbation theory we also present the
numerical form of the perturbative scheme conversion factors for these
schemes in these general kinematics.

We find that the non-pertubative and perturbative results agree very
well as long as we stay from the corner of the $\omega,\mu$ plane,
with one exception, namely $Z_q^{(\qslash)}$. Where, we argued that
the reason for this relatively bad agreement is the poor convergence
of the perturbative expansion.  Clearly having several values of
$\omega$ help to have a better handle on the systematic errors coming
from the NPR procedure.  As an application, we have seen a example
where the discretisation effects depend significantly on $\omega$ (see
$\sigma_q^{(\qslash)}$ where the discretisation effects for $\omega=2,2.5$
are much smaller than for $\omega=1$) as well as the perturbative convergence. In
this proof of concept study only two lattice spacings have been
used. Adding a finer lattice could potentially allow us to test the
agreement of the perturbative and non-perturbative window even
further.

Generally speaking, it is well-known that the SMOM kinematics ($\omega=1$)  leads to
much cleaner determinations of the renormalisation factors than the $\omega=0$ case.
In this work, we have shown that increasing the value of $\omega$
leads to smaller contributions of the pseudo-Goldstone poles contamination
in $\Lambda_P-\Lambda_S$. It will be interesting to extend this study to the case of
four-quark operators where the infrared contaminations due to chiral symmetry
breaking are significantly more sizeable. 

\section{Acknowledgements} \label{ackn}
This work was supported by the Consolidated Grant ST/T000988/1 and
the work of JAG by a DFG Mercator Fellowship. The quark propagators
were computed on the DiRAC Blue Gene Q Shared Petaflop system at
the University of Edinburgh, operated by the Edinburgh Parallel
Computing Centre on behalf of the STFC DiRAC HPC Facility
(www.dirac.ac.uk). This equipment was funded by BIS National
E-infrastructure capital grant ST/K000411/1, STFC capital grant
ST/H008845/1, and STFC DiRAC Operations grants ST/K005804/1 and
ST/K005790/1. DiRAC is part of the National E-Infrastructure.  \\

We warmly thank our colleagues of the RBC and UKQCD collaborations.
We are particularly indebted to Peter Boyle,  Andreas J\"uttner,
J~Tobias Tsang for many interesting discussions. We also thank Peter
Boyle for his help with the UKQCD hadron software. We wish to thank
Holger Perlt for his early contribution in this area. N.G. thanks
his collaborators from the California Lattice (CalLat) collaboration
and in particular those working on NPR: David Brantley,  Henry
Monge-Camacho, Amy Nicholson and Andr\'e Walker-Loud. \\


\clearpage
\begin{appendices}

  \section{Perturbation Theory}
\label{app:PT}

In order to make contact to phenomenology, results obtained from
lattice simulations need to be converted to a scheme that is used in
continuum perturbation theory such as $\MSbar$. The non-perturbative
renormalisation schemes used in this work serve as an intermediate
scheme that can be defined on the Lattice and in the continuum. This
enables us to compare the non-perturbative with the perturbative
change of the renormalisation scheme parameters $\mu$ and $\omega$.
This appendix provides the ingredients for the perturbative scheme
changes at NNLO.

The dependence of the strong coupling constant on the dimensional
regularisation scale $\nu$ in the $\MSbar$ scheme is governed by the
$\beta$-function
\begin{equation}
\label{eq:beta-function}
\nu^2 \frac{d}{d\nu^2} \left(\frac{\alpha_s}{4\pi}\right)
\equiv -\sum_{i=0}^{\infty} \beta_i   \left(\frac{\alpha_s}{4\pi}\right )^{i+2}\;.
\end{equation}
Although it is known at
five-loop~\cite{Baikov:2016tgj,Herzog:2017ohr,Luthe:2017ttg}, we will
only use the first three coefficients in the expansion to be
consistent with our NNLO analysis of the running. In the following we will need the first two coefficients
\begin{align}
\beta_0 &= 11 - \frac{2}{3}N_f \label{beta0} \;,\\
\beta_1 &= 102 - \frac{38}{3}N_f \;,
\end{align}
where we fixed the number of colours $N_c = 3$ in this and the
following expressions. Using $\alpha_S(M_Z) = 0.1179(10)$ and
$M_Z = 91.1876(21)$ GeV \cite{Zyla:2020zbs} we find the required
values of the strong coupling constant for $N_f = 3$ flavours given in
Table~\ref{Tab:alphaS}.
\begin{table}
\begin{center}
\begin{tabular} {|c| c         c        c       c        c        c         c |}
  \hline
  $\nu$ /GeV           & 1.0    &  1.5   &  2.0   & 2.5    & 3.0    & 3.5     & 4.0 \\ \hline
  $\alpha_S(\nu)$  &  0.4698 & 0.3467 & 0.2950 & 0.2652 & 0.2452 & 0.2307 &  0.2196 \\
\hline
\end{tabular}
\end{center}
\caption{Values of the strong running coupling as a function of the dimensional regularisation scheme $\nu$.
  They are obtained in the $\MSbar$ scheme in the three-flavour theory.}
\label{Tab:alphaS}
\end{table}
To arrive at three active flavours, we decoupled the bottom and charm
quark at $\nu_b = 5$ GeV and $\nu_c = 1.5$ GeV respectively using
central values $m_b(m_b) = 4.13$ GeV and $m_c(m_c) = 1.27$ GeV for the
quark masses in the $\MSbar$ scheme\cite{Zyla:2020zbs}. The relevant
threshold effects and running of the strong coupling constant are
implemented using RunDec\cite{Chetyrkin:2000yt}.

The fields and masses in the $\mathrm{IMOM}^{(X)}$ schemes are related
to the $\mathrm{\overline{MS}}$ scheme as
$\psi_{\MSbar}(\nu) = (Z_q^{\MSbar}/ Z_q^{(X)}) \psi_X(\mu, \omega)$
and
$m^{\mathrm{ \overline{MS}}}(\nu) = (Z_m^{\MSbar}/ Z_m^{(X)})
m_X(\mu,\omega)$,
where $X \in (\gamma_\mu, \slashed{q})$. The continuum perturbation
theory expansion of the conversion factors
\begin{equation}
  \label{eq:conversion-dim-reg}
   \frac{Z_i^{\MSbar}(\epsilon, \nu)}{Z_i^{(X)}(\epsilon, \mu,\omega)} 
 \overset{\epsilon \to 0}{= } 
 C_i^{(X)}(\nu,\mu,\omega) \,, \qquad i \in \{ q,m \} \;, 
\end{equation}
are known up to two loops \cite{Gorbahn:2010bf, Bell:2016nar} for the
IMOM scheme and up to three loops \cite{Bednyakov:2020,Kniehl:2020} in
the SMOM limit, i.e.\ where $\omega = 1$. Writing 
\begin{equation}
\label{eq:conversion-perturbative-expansion}
C_i^{(X)} = 1 + \frac{\alpha_s}{4 \pi} C_i^{(X,1)} + \frac{\alpha_s^2}{(4 \pi)^2} C_i^{(X,2)}
\end{equation}
and setting $\mu = \nu$ we find the numerical conversion factors as a
function of $\omega$ given in Table.~\ref{tab:cm-pt}.
\begin{table}[t]
\begin{center}
\begin{tabular}{|c|c|c|c|c|}
\hline
 $\omega $ & $C_m^{(\slashed{q},1)}$ & $C_m^{(\slashed{q},2)}$ & $C_m^{(\gamma,1)}$ & $C_m^{(\gamma,2)}$ \\
\hline
 $0.5$ & $-2.422$ & $-64.756+5.988 N_f$ & $-3.248$ & $-89.07+7.571 N_f$ \\
 $1.0$ & $-0.646$ & $-22.608+4.014 N_f$ & $-1.979$ & $-55.032+6.162 N_f$ \\
 $1.5$ & $0.778$ & $10.344 +2.432 N_f$ & $-0.964$ & $-28.916+5.035 N_f$ \\
 $2.0$ & $1.994$ & $38.567 +1.08 N_f$ & $-0.098$ & $-6.829+4.072 N_f$ \\
 $2.5$ & $3.071$ & $63.81 -0.115 N_f$ & $0.667$ & $12.741 +3.222 N_f$ \\
 $3.0$ & $4.042$ & $86.961 -1.195 N_f$ & $1.358$ & $30.56 +2.454 N_f$ \\
 $3.5$ & $4.933$ & $108.544 -2.184 N_f$ & $1.99$ & $47.076 +1.752 N_f$ \\
 $4.0$ & $5.757$ & $128.894 -3.1 N_f$ & $2.575$ & $62.576 +1.102 N_f$ \\ 
\hline \hline
 $\omega $ & $C_q^{(\slashed{q},1)}$ & $C_q^{(\slashed{q},2)}$ & $C_q^{(\gamma,1)}$ & $C_q^{(\gamma,2)}$ \\
\hline
 $0.5$ & $0$ & $-25.464+2.333 N_f$ & $0.825$ & $1.53 +0.75 N_f$ \\
 $1.0$ & $0$ & $-25.464+2.333 N_f$ & $1.333$ & $9.599 +0.185 N_f$ \\
 $1.5$ & $0$ & $-25.464+2.333 N_f$ & $1.742$ & $15.476 -0.269 N_f$ \\
 $2.0$ & $0$ & $-25.464+2.333 N_f$ & $2.093$ & $20.137 -0.658 N_f$ \\
 $2.5$ & $0$ & $-25.464+2.333 N_f$ & $2.403$ & $24.001 -1.004 N_f$ \\
 $3.0$ & $0$ & $-25.464+2.333 N_f$ & $2.684$ & $27.292 -1.316 N_f$ \\
 $3.5$ & $0$ & $-25.464+2.333 N_f$ & $2.942$ & $30.147 -1.603 N_f$ \\
 $4.0$ & $0$ & $-25.464+2.333 N_f$ & $3.182$ & $32.66 -1.869 N_f$ \\
\hline
\end{tabular}
\end{center}\caption[]{\label{tab:cm-pt} The perturbative expansion
  coefficients of the conversion factors $C_m^{(\gamma)}$,
  $C_m^{(\slashed{q})}$, $C_q^{(\gamma)}$ and
  $C_q^{(\slashed{q})}$ as a function of $\omega$. Note that $C_q^{(\slashed{q})}$ is independent of $\omega$.}
\vspace{4mm}
\end{table}
The scheme transformation can then be written as a product
\begin{equation}
U_i^{(X)} (\mu_1,\mu_0,\omega_1,\omega_0) = C_i^{(X)}(\nu_1,\mu_1,\omega_1) U_i^{\MSbar} (\nu_1,\nu_0) C_i^{(X)^{-1}}(\nu_0,\mu_0,\omega_0) \bigg|_{
\begin{subarray}{l}
\mu_0=\nu_0 \\ \mu_1=\nu_1
\end{subarray}}
\end{equation}
of the conversion factor, its inverse, and the and the $\MSbar$
evolution kernel $U_i^{\MSbar} (\nu_1,\nu_0)$. The evolution Kernel
fulfils the renormalisation group equation
\begin{equation}
\label{eq:u-rge}
\nu^2 \frac{d}{d \nu^2}U_i^{\MSbar} (\nu,\nu_0) = - \gamma_i^{\MSbar} U_i^{\MSbar} (\nu,\nu_0)
\end{equation}
and we expand the anomalous dimensions 
\begin{equation}
\label{eq:gamma-expansion}
\gamma_i^{(X)} \equiv \sum_{k=0}^{\infty} \gamma_i^{(X,k)}  \left(\frac{\alpha_s}{4\pi}\right )^{k+1}\; ,
\end{equation}
where $X$ denotes the renormalisation scheme.  Hence we can transform
for example a light quark masses renormalised in given scheme $X$ at
different kinematic points $(\mu_1,\omega_1)$ and $(\mu_0,\omega_0)$
via
\begin{equation}
\label{eq:mass-pt-trafo}
m^{(X)}(\mu_1,\omega_1) = U_m^{(X)}(\mu_1, \mu_0,\omega_1,\omega_0) m^{(X)}(\mu_0,\omega_0)\,,
\end{equation}
where the logarithms $\log \mu_1/\mu_0$ are summed using
renormalisation group improved perturbation theory. 
Explicitly, we find at NNLO 
\begin{equation*}
\begin{split}
U_i^{(X)} (\mu_1,\mu_0,\omega_1,\omega_0) = &
\left( 1 + \frac{\alpha_s(\mu_1)}{4 \pi} J^{(X,1)}_i(\omega_1) + 
           \frac{\alpha_s^2(\mu_1)}{(4 \pi)^2} J^{(X,2)}_{i}(\omega_1) \right)
\left( \frac{\alpha_s(\mu_0)}{\alpha_s(\mu_1)} \right)^{\gamma^{(0)}_i/\beta_0} \times \\
& \qquad \left( 1 - \frac{\alpha_s(\mu_0)}{4 \pi} J^{(X,1)}_i(\omega_0) + 
           \frac{\alpha_s^2(\mu_0)}{(4 \pi)^2} \left[(J^{(X,1)}_i(\omega_0))^2  - J^{(X,2)}_i(\omega_0) \right]  \right)\,,
\end{split}
\end{equation*}\label{Uexpanded}
where the $J_X^{(1)}$ and $J_X^{(2)}$ are given by
\begin{equation}
\label{eq:js-pt}
\begin{split}
J^{(X,1)}_i =& \frac{\gamma_i^{(X,1)}}{\beta_0} - \frac{\beta_1\gamma_i^{(0)}}{\beta_0^2}\;,\\
J^{(X,2)}_i =& \frac{1}{2} \left( {J^{(X,1)}}^2 + \frac{\gamma_i^{(X,2)}}{\beta_0} + \frac{\beta_1^2\gamma_i^{(0)}}{\beta_0^3}
- \frac{\beta_1\gamma_i^{(X,1)}}{\beta_0^2} - \frac{\beta_2\gamma_i^{(0)}}{\beta_0^2} \right) \;.
\end{split}
\end{equation}
The leading order anomalous dimensions
\begin{equation}
\label{eq:lo-adms}
\begin{split}
\gamma_q^{(0)} &= 0 \;, \\
\gamma_m^{(0)} &= 4 \;
\end{split}
\end{equation}
are scheme independent. The expressions for the NLO and NNLO 
anomalous dimensions are given in Table~\ref{tab:gam-pt} for different values of $\omega$ as a function of the number of flavours $N_f$.
\begin{table}[t]
\begin{center}
\begin{tabular}{|c|c|c|}
\hline
 $\omega$ & $\gamma _m^{(\slashed{q},1)}$ & $\gamma _m^{(\slashed{q},2)}$ \\ \hline
 $0.5$ & $93.981 -3.837 N_f$ & $2985.27 -398.857 N_f+6.256 N_f^2$ \\
 $1.0$ & $74.434 -2.653 N_f$ & $1816.8 -273.08 N_f+3.623 N_f^2$ \\
 $1.5$ & $58.78 -1.704 N_f$ & $948.781 -176.459 N_f+1.515 N_f^2$ \\
 $2.0$ & $45.395 -0.893 N_f$ & $240.849 -95.917 N_f-0.288 N_f^2$ \\
 $2.5$ & $33.558 -0.175 N_f$ & $-364.298-25.958 N_f-1.882 N_f^2$ \\
 $3.0$ & $22.868 +0.473 N_f$ & $-896.718+36.367 N_f-3.322 N_f^2$ \\
 $3.5$ & $13.075 +1.066 N_f$ & $-1374.46+92.858 N_f-4.641 N_f^2$ \\
 $4.0$ & $4.006 +1.616 N_f$ & $-1809.31+144.71 N_f-5.862 N_f^2$ \\ 
\hline \hline
 $\omega $ & $\gamma _m^{(\gamma ,1)}$ & $\gamma _m^{(\gamma ,2)}$ \\ \hline
 $0.5$ & $103.057 -4.387 N_f$ & $3655.83 -479.684 N_f+8.367 N_f^2$ \\
 $1.0$ & $89.101 -3.541 N_f$ & $2704.63 -382.794 N_f+6.487 N_f^2$ \\
 $1.5$ & $77.943 -2.865 N_f$ & $1993.76 -308.337 N_f+4.984 N_f^2$ \\
 $2.0$ & $68.413 -2.288 N_f$ & $1409.35 -246.123 N_f+3.701 N_f^2$ \\
 $2.5$ & $59.993 -1.777 N_f$ & $905.536 -191.915 N_f+2.567 N_f^2$ \\
 $3.0$ & $52.396 -1.317 N_f$ & $458.454 -143.458 N_f+1.544 N_f^2$ \\
 $3.5$ & $45.441 -0.895 N_f$ & $53.89 -99.383 N_f+0.607 N_f^2$ \\
 $4.0$ & $39.005 -0.505 N_f$ & $-317.4-58.784 N_f-0.26 N_f^2$ \\
\hline \hline
 $\omega $ & $\gamma _q^{(\gamma ,1)}$ & $\gamma _q^{(\gamma ,2)}$ \\ \hline
 $0.5$ & $13.257 -0.783 N_f$ & $417.983 -65.569 N_f+1.741 N_f^2$ \\
 $1.0$ & $7.667 -0.444 N_f$ & $200.702 -36.683 N_f+0.988 N_f^2$ \\
 $1.5$ & $3.171 -0.172 N_f$ & $43.556 -14.519 N_f+0.382 N_f^2$ \\
 $2.0$ & $-0.684+0.062 N_f$ & $-79.969+3.809 N_f-0.137 N_f^2$ \\
 $2.5$ & $-4.102+0.269 N_f$ & $-181.296+19.56 N_f-0.597 N_f^2$ \\
 $3.0$ & $-7.195+0.456 N_f$ & $-266.635+33.428 N_f-1.014 N_f^2$ \\
 $3.5$ & $-10.033+0.628 N_f$ & $-339.81+45.844 N_f-1.396 N_f^2$ \\
 $4.0$ & $-12.666+0.788 N_f$ & $-403.372+57.097 N_f-1.751 N_f^2$ \\
\hline \hline
 $\omega$ & $\gamma _q^{(\slashed{q} ,1)}$ & $\gamma _q^{(\slashed{q} ,2)}$ \\ \hline
 $0 \dots 4$ & $22.333 -1.333 N_f$ & $1088.54 -146.397 N_f+3.852 N_f^2$ \\ \hline
\end{tabular}
\end{center}
\caption[]{\label{tab:gam-pt} The NLO and NNLO coefficients of anomalous dimensions $\gamma_m^{(\slashed{q})}$,  $\gamma_m^{(\gamma)}$, $\gamma_q^{(\gamma)}$, $\gamma_q^{(\slashed{q})}$ as a function of $\omega$. Note that  $\gamma_q^{(\slashed{q})}$ is independent of $\omega$.}
\vspace{4mm}
\end{table}

\clearpage


  \section{Simulation Details}
\label{lattice}
Our numerical work is based on RBC-UKQCD data, the lattice details can be found in \cite{RBC:2010qam}.
We compute the propagators using Landau gauge-fixed 2+1, Domain-Wall (Shamir~\cite{Shamir:1993zy})/Iwasaki lattices.
The values of the parameters can be found in~\cite{Garron:2016mva}.\\

In a nutshell, we use two lattice spacings (we refer to them as $24^3$ and $32^3$): 
\begin{eqnarray}
a^{-1} &=& 1.785(5) \ \text{GeV} \ \  \ (24^3)\,, \\
a^{-1} &=& 2.383(9) \ \text{GeV} \ \ \ (32^3\,,
\end{eqnarray}
for each lattice spacing we have three different sea quark masses, $am = 0.005, \ 0.010, \ 0.020$ for the $24^3 \times 64 \times 16$ lattice and $am = 0.004, \ 0.006, \ 0.008$ for the $32^3 \times 64 \times 16 $ lattice.\\

\noindent We take the chiral limit on each lattice spacing using the values
\begin{eqnarray}
am_{res} &=& 0.003152(43) \ \ \ \ (24^3), \\
am_{res} &=& 0.0006664(76) \ \ \ (32^3).
\end{eqnarray}

\noindent Our values for $Z_V$ are \cite{Aoki:2010pe}
\begin{align}
Z_V = 0.71651(46) \ \ \ \ (24^3), \\
Z_V = 0.74475(12) \ \ \ \ (32^2).
\end{align} 

  \section{Extensive study for $\sigma_m$} 
\label{app:extensive}
In this appendix we provide our numerical results for running of the quark mass.
These results are given for $\sigma^{(X)}_{m}(\mu,\mu_0,\omega, \omega_0)$
as defined in Eq.~(\ref{eq:sigma_def}), such that
\be
\label{eq:sigma_2}
Z^{(X)}_m(\mu,\omega) = \sigma^{(X)}_{m}(\mu,\mu_0,\omega, \omega_0) Z^{(X)}_m(\mu_0,\omega_0) \;.
\ee
\subsection{Study of the $\omega$-dependence for fixed energy scales }

In order to study the $\omega$-dependence, we first fix $\mu$ and $\mu_0$ to some reasonable values,
where we expect a rather good control over both the perturbative errors and the lattice artefacts.
We choose $(\mu_0,\mu) =  (2.5 \; \text{GeV}, 1.5 \; \text{GeV})$.
We then vary $\omega$, but first we only consider the `diagonal' case, ie $\omega_0=\omega$.
We show our results in Tables~\ref{tab:g_ssf_2p5to1p5} and ~\ref{tab:q_ssf_2p5to1p5}.
The middle column shows the running itself obtained from the lattice,
while the other three columns on the right show the ratio of the non-perberturbative
running over the perturbative prediction at Leading Order (LO), Next-to-Leading Order (NLO)
and Next-to-Next-to-Leading Order (NNLO). The errors quoted there combine an estimate of
the discretisaiton errors  and the statistical one.
  \begin{table}[th]
    \bc
    \begin{tabular}{ |c| c | c  c  c| } \hline
      $(\mu_0,\mu) =  (2.5,1.5)$ & $ \sigma_m^{(\gamma_\mu)}$  &  $\sigma_m /LO$   &  $\sigma_m /NLO$   &  $\sigma_m /NNLO$ \\
      $\omega=\omega_0\downarrow$ 
      &                  &                      &                           &                   \\
      \hline
      $0.5$ & $1.229(  27)$  &   $1.091(  24)$ &  $1.035(  23)$ &  $1.006(  22)$ \\ 
      $1.0$ & $1.174(  18)$  &   $1.042(  16)$ &  $0.999(  16)$ &  $0.984(  15)$ \\ 
      $1.5$ & $1.184(  17)$  &   $1.051(  15)$ &  $1.016(  15)$ &  $1.009(  15)$ \\ 
      $2.0$ & $1.162(  16)$  &   $1.032(  14)$ &  $1.005(  14)$ &  $1.004(  14)$ \\ 
      $2.5$ & $1.143(  14)$  &   $1.014(  12)$ &  $0.994(  12)$ &  $0.998(  12)$ \\ 
      $3.0$ & $1.124(  12)$  &   $0.998(  11)$ &  $0.982(  11)$ &  $0.990(  11)$ \\ 
      $3.5$ & $1.110(  11)$  &   $0.985(   9)$ &  $0.975(   9)$ &  $0.986(   9)$ \\ 
      $4.0$ & $1.080(   5)$  &   $0.959(   5)$ &  $0.952(   5)$ &  $0.967(   5)$ \\
      \hline   
    \end{tabular}
    \ec
    \caption{Running for the quark mass, $\gamma_\mu$ scheme for
      $(\mu_0,\mu) =  (2.5 \text{GeV}, 1.5 \text{GeV})$.}
      \label{tab:g_ssf_2p5to1p5}
  \end{table}

  \begin{table}[th]
    \bc
    \begin{tabular}{ |c| c | c  c  c| } \hline
      $(\mu_0,\mu) = (2.5,1.5)$ & $ \sigma_m^{(\slashed{q})}$  &  $\sigma_m /LO$   &  $\sigma_m /NLO$   &  $\sigma_m /NNLO$ \\
      $\omega=\omega_0\downarrow$ 
      &                  &                      &                           &                   \\
      \hline
      $0.5$ & $1.194(  23)$  &   $1.059(  21)$ &  $1.012(  20)$ &  $0.992(  20)$ \\ 
      $1.0$ & $1.138(  18)$  &   $1.010(  16)$ &  $0.980(  16)$ &  $0.995(  16)$ \\ 
      $1.5$ & $1.138(  13)$  &   $1.010(  11)$ &  $0.991(  11)$ &  $0.985(  11)$ \\ 
      $2.0$ & $1.117(  13)$  &   $0.992(  12)$ &  $0.981(  12)$ &  $1.000(  12)$ \\ 
      $2.5$ & $1.096(  13)$  &   $0.973(  11)$ &  $0.970(  11)$ &  $0.971(  11)$ \\ 
      $3.0$ & $1.080(  13)$  &   $0.959(  12)$ &  $0.962(  12)$ &  $0.984(  12)$ \\ 
      $3.5$ & $1.068(  11)$  &   $0.948(   9)$ &  $0.956(  10)$ &  $0.963(  10)$ \\ 
      $4.0$ & $1.033(  12)$  &   $0.917(  10)$ &  $0.929(  10)$ &  $0.952(  11)$ \\ 
      \hline   
    \end{tabular}
    \ec
     \caption{Same as Table~\ref{tab:g_ssf_2p5to1p5}. }
      \label{tab:q_ssf_2p5to1p5}
  \end{table}

We observe the non-perturbative running agrees extremely well with NNLO predictions
for all values of $\omega \le 3$. Looking at the perturbative 
convergence and stability, our data seem to favour the region $\omega \sim 2$.

\subsection{Study of the $\omega$-dependence for larger energy ranges }

Here we fix $\mu=2$ GeV and let $\mu_0$ vary over the full range.
Again we only consider the case $\omega_0=\omega$.
The rows $\mu_0=2$ give trivially one, but we leave the results in order
to guide the eyes. We give our results in Table~\ref{tab:g_ssf_mu2GeV}.
\begin{table}
\bc
\begin{tabular}{ |c| c | c  c  c| } \hline
  & $ \sigma_m$  &  $\sigma_m /LO$   &  $\sigma_m /NLO$   &  $\sigma_m /NNLO$ \\
  $\mu_0 \downarrow$ (GeV)    &                  &                      &                           &                   \\
\hline
  $\bm{\omega=0.5}$ &  &  &  &  \\ 
$1.0$ & $0.826(  83)$  &   $1.015( 102)$ &  $1.145( 115)$ &  $1.246( 126)$ \\ 
$1.5$ & $0.884(  17)$  &   $0.949(  19)$ &  $0.982(  19)$ &  $1.001(  20)$ \\ 
$2.0$ & $1$  &   $1$ &  $1$ &  $1$ \\ 
$2.5$ & $1.084(  10)$  &   $1.034(   9)$ &  $1.014(   9)$ &  $1.005(   9)$ \\ 
$3.0$ & $1.153(  17)$  &   $1.062(  15)$ &  $1.029(  15)$ &  $1.014(  15)$ \\ 
$3.5$ & $1.206(  24)$  &   $1.081(  22)$ &  $1.038(  21)$ &  $1.019(  21)$ \\ 
$4.0$ & $1.253(  35)$  &   $1.099(  31)$ &  $1.048(  30)$ &  $1.027(  29)$ \\ 
\hline 
 $\bm{\omega=1.0}$ &  &  &  &  \\ 
$1.0$ & $0.740(  42)$  &   $0.910(  51)$ &  $0.999(  56)$ &  $1.045(  59)$ \\ 
$1.5$ & $0.916(  13)$  &   $0.984(  14)$ &  $1.010(  15)$ &  $1.021(  15)$ \\ 
$2.0$ & $1$  &   $1$ &  $1$ &  $1$ \\ 
$2.5$ & $1.074(   7)$  &   $1.025(   7)$ &  $1.009(   7)$ &  $1.004(   6)$ \\ 
$3.0$ & $1.133(  13)$  &   $1.044(  12)$ &  $1.018(  12)$ &  $1.010(  12)$ \\ 
$3.5$ & $1.179(  21)$  &   $1.057(  19)$ &  $1.024(  18)$ &  $1.014(  18)$ \\ 
$4.0$ & $1.220(  31)$  &   $1.070(  27)$ &  $1.030(  26)$ &  $1.018(  26)$ \\ 
\hline 
 $\bm{\omega=1.5}$ &  &  &  &  \\ 
$1.0$ & $0.764(  35)$  &   $0.939(  43)$ &  $1.010(  46)$ &  $1.032(  47)$ \\ 
$1.5$ & $0.901(  10)$  &   $0.968(  11)$ &  $0.989(  11)$ &  $0.994(  11)$ \\ 
$2.0$ & $1$  &   $1$ &  $1$ &  $1$ \\ 
$2.5$ & $1.067(   6)$  &   $1.017(   6)$ &  $1.005(   6)$ &  $1.003(   6)$ \\ 
$3.0$ & $1.118(  13)$  &   $1.030(  12)$ &  $1.010(  11)$ &  $1.006(  11)$ \\ 
$3.5$ & $1.159(  20)$  &   $1.039(  18)$ &  $1.013(  18)$ &  $1.008(  18)$ \\ 
$4.0$ & $1.196(  31)$  &   $1.049(  27)$ &  $1.018(  26)$ &  $1.012(  26)$ \\ 
\hline 
 $\bm{\omega=2.0}$ &  &  &  &  \\ 
$1.0$ & $0.825(  34)$  &   $1.014(  42)$ &  $1.074(  45)$ &  $1.078(  45)$ \\ 
$1.5$ & $0.910(  10)$  &   $0.978(  11)$ &  $0.994(  11)$ &  $0.995(  11)$ \\ 
$2.0$ & $1$  &   $1$ &  $1$ &  $1$ \\ 
$2.5$ & $1.059(   6)$  &   $1.010(   5)$ &  $1.001(   5)$ &  $1.000(   5)$ \\ 
$3.0$ & $1.106(  12)$  &   $1.019(  11)$ &  $1.003(  11)$ &  $1.002(  11)$ \\ 
$3.5$ & $1.137(  17)$  &   $1.020(  15)$ &  $0.999(  15)$ &  $0.998(  15)$ \\ 
$4.0$ & $1.160(  21)$  &   $1.018(  18)$ &  $0.994(  18)$ &  $0.993(  18)$ \\
\hline  
 \end{tabular}
\ec
\caption{Running for the quark mass, $X=\gamma_\mu, \mu = 2\; \text{GeV}$.
  We only consider $\omega=\omega_0$ and let $\mu_0$ vary  between 1
  and 4 GeV.}
\label{tab:g_ssf_mu2GeV}
\end{table}

\clearpage
\subsection{Study of the running in the non-degenerate $\omega$ case. }

Here we fix again both energy scales $(\mu_0,\mu) = (2.5 \; \text{ GeV},1.5 \; \text{ GeV})$,
and we allow $\omega\ne \omega_0$. Our results are shown in Tables~\ref{tab:g_ssf_non_deg_om}
\ref{tab:g_ssf_non_deg_om2}, for $X=\gamma_\mu$ and in Tables~\ref{tab:q_ssf_non_deg_om} and
\ref{tab:q_ssf_non_deg_om2} for $X=\qslash$.

  \begin{table}
\bc
\begin{tabular}{ |c| c | c  c  c| } \hline
  $ (\mu_0,\mu) = (2.5,1.5)$ & $ \sigma_m$  &  $\sigma_m /LO$   &  $\sigma_m /NLO$   &  $\sigma_m /NNLO$ \\  
  $\omega\downarrow$ 
  &                  &                      &                           &                   \\
 \hline 
$\bm{\omega_0 = 0.5}$ &  &  &  &  \\ 
 $0.5$ & $1.229(  27)$  &   $1.091(  24)$ &  $1.035(  23)$ &  $1.006(  22)$ \\ 
$1.0$ & $1.308(  29)$  &   $1.161(  26)$ &  $1.067(  23)$ &  $1.023(  23)$ \\ 
$1.5$ & $1.373(  30)$  &   $1.219(  27)$ &  $1.094(  24)$ &  $1.038(  23)$ \\ 
$2.0$ & $1.425(  32)$  &   $1.265(  28)$ &  $1.113(  25)$ &  $1.048(  24)$ \\ 
$2.5$ & $1.473(  34)$  &   $1.307(  30)$ &  $1.131(  26)$ &  $1.057(  24)$ \\ 
$3.0$ & $1.515(  35)$  &   $1.345(  31)$ &  $1.145(  27)$ &  $1.065(  25)$ \\ 
$3.5$ & $1.551(  36)$  &   $1.377(  32)$ &  $1.156(  27)$ &  $1.069(  25)$ \\ 
$4.0$ & $1.568(  33)$  &   $1.391(  29)$ &  $1.154(  24)$ &  $1.062(  22)$ \\ 
 \hline 
$\bm{\omega_0 = 1.0}$ &  &  &  &  \\ 
 $0.5$ & $1.103(  18)$  &   $0.979(  16)$ &  $0.969(  16)$ &  $0.967(  16)$ \\ 
$1.0$ & $1.174(  18)$  &   $1.042(  16)$ &  $0.999(  16)$ &  $0.984(  15)$ \\ 
$1.5$ & $1.232(  19)$  &   $1.094(  17)$ &  $1.024(  16)$ &  $0.998(  15)$ \\ 
$2.0$ & $1.279(  20)$  &   $1.136(  18)$ &  $1.042(  16)$ &  $1.008(  16)$ \\ 
$2.5$ & $1.322(  21)$  &   $1.174(  19)$ &  $1.059(  17)$ &  $1.017(  16)$ \\ 
$3.0$ & $1.360(  22)$  &   $1.208(  20)$ &  $1.073(  18)$ &  $1.024(  17)$ \\ 
$3.5$ & $1.392(  22)$  &   $1.236(  20)$ &  $1.083(  17)$ &  $1.028(  17)$ \\ 
$4.0$ & $1.407(  22)$  &   $1.249(  20)$ &  $1.080(  17)$ &  $1.021(  16)$ \\ 
 \hline 
$\bm{\omega_0 = 1.5}$ &  &  &  &  \\ 
 $0.5$ & $1.059(  17)$  &   $0.940(  15)$ &  $0.961(  15)$ &  $0.977(  16)$ \\ 
$1.0$ & $1.128(  16)$  &   $1.001(  14)$ &  $0.992(  14)$ &  $0.994(  14)$ \\ 
$1.5$ & $1.184(  17)$  &   $1.051(  15)$ &  $1.016(  15)$ &  $1.009(  15)$ \\ 
$2.0$ & $1.229(  18)$  &   $1.091(  16)$ &  $1.035(  15)$ &  $1.019(  15)$ \\ 
$2.5$ & $1.271(  19)$  &   $1.128(  17)$ &  $1.051(  16)$ &  $1.028(  16)$ \\ 
$3.0$ & $1.307(  20)$  &   $1.161(  18)$ &  $1.065(  17)$ &  $1.035(  16)$ \\ 
$3.5$ & $1.338(  20)$  &   $1.188(  18)$ &  $1.075(  16)$ &  $1.040(  16)$ \\ 
$4.0$ & $1.353(  13)$  &   $1.201(  12)$ &  $1.073(  10)$ &  $1.033(  10)$ \\ 
 \hline 
$\bm{\omega_0 = 2.0}$ &  &  &  &  \\ 
 $0.5$ & $1.002(  15)$  &   $0.890(  13)$ &  $0.934(  14)$ &  $0.963(  14)$ \\ 
$1.0$ & $1.067(  14)$  &   $0.947(  13)$ &  $0.964(  13)$ &  $0.980(  13)$ \\ 
$1.5$ & $1.119(  15)$  &   $0.993(  14)$ &  $0.987(  13)$ &  $0.994(  14)$ \\ 
$2.0$ & $1.162(  16)$  &   $1.032(  14)$ &  $1.005(  14)$ &  $1.004(  14)$ \\ 
$2.5$ & $1.201(  17)$  &   $1.066(  15)$ &  $1.021(  14)$ &  $1.013(  14)$ \\ 
$3.0$ & $1.236(  18)$  &   $1.097(  16)$ &  $1.034(  15)$ &  $1.020(  15)$ \\ 
$3.5$ & $1.265(  18)$  &   $1.123(  16)$ &  $1.044(  15)$ &  $1.024(  15)$ \\ 
$4.0$ & $1.279(  12)$  &   $1.136(  11)$ &  $1.043(  10)$ &  $1.018(  10)$ \\ 
\hline
 \end{tabular}
 \ec
\caption{Running for the quark mass, $\gamma_\mu$-projector for
      $(\mu_0,\mu) =  (2.5 \text{GeV}, 1.5 \text{GeV})$.}
 \label{tab:g_ssf_non_deg_om}
  \end{table}

\begin{table}
\bc
\begin{tabular}{ |c| c | c  c  c| } \hline
  $ (\mu_0,\mu) = (2.5,1.5)$ & $ \sigma_m$  &  $\sigma_m /LO$   &  $\sigma_m /NLO$   &  $\sigma_m /NLO$ \\  
  $\omega\downarrow$ 
  &                  &                      &                           &                   \\
  \hline
  $\bm{\omega_0 = 2.5}$ &  &  &  &  \\ 
 $0.5$ & $0.952(  12)$  &   $0.845(  10)$ &  $0.907(  11)$ &  $0.947(  12)$ \\ 
$1.0$ & $1.013(  11)$  &   $0.899(  10)$ &  $0.937(  10)$ &  $0.964(  10)$ \\ 
$1.5$ & $1.064(  12)$  &   $0.944(  11)$ &  $0.960(  11)$ &  $0.978(  11)$ \\ 
$2.0$ & $1.105(  13)$  &   $0.981(  11)$ &  $0.978(  11)$ &  $0.988(  11)$ \\ 
$2.5$ & $1.143(  14)$  &   $1.014(  12)$ &  $0.994(  12)$ &  $0.998(  12)$ \\ 
$3.0$ & $1.176(  15)$  &   $1.044(  13)$ &  $1.007(  13)$ &  $1.005(  13)$ \\ 
$3.5$ & $1.203(  14)$  &   $1.068(  13)$ &  $1.016(  12)$ &  $1.008(  12)$ \\ 
$4.0$ & $1.217(   9)$  &   $1.080(   8)$ &  $1.015(   8)$ &  $1.003(   8)$ \\ 
 \hline 
$\bm{\omega_0 = 3.0}$ &  &  &  &  \\ 
 $0.5$ & $0.911(  11)$  &   $0.809(  10)$ &  $0.887(  10)$ &  $0.935(  11)$ \\ 
$1.0$ & $0.970(  10)$  &   $0.861(   9)$ &  $0.915(   9)$ &  $0.952(  10)$ \\ 
$1.5$ & $1.018(  11)$  &   $0.903(   9)$ &  $0.938(  10)$ &  $0.965(  10)$ \\ 
$2.0$ & $1.057(  11)$  &   $0.938(  10)$ &  $0.955(  10)$ &  $0.975(  10)$ \\ 
$2.5$ & $1.092(  12)$  &   $0.970(  10)$ &  $0.970(  10)$ &  $0.983(  10)$ \\ 
$3.0$ & $1.124(  12)$  &   $0.998(  11)$ &  $0.982(  11)$ &  $0.990(  11)$ \\ 
$3.5$ & $1.150(  12)$  &   $1.021(  11)$ &  $0.992(  11)$ &  $0.995(  11)$ \\ 
$4.0$ & $1.163(   7)$  &   $1.032(   6)$ &  $0.990(   6)$ &  $0.988(   6)$ \\ 
 \hline 
$\bm{\omega_0 = 3.5}$ &  &  &  &  \\ 
 $0.5$ & $0.879(  10)$  &   $0.780(   9)$ &  $0.872(  10)$ &  $0.927(  11)$ \\ 
$1.0$ & $0.935(   9)$  &   $0.830(   8)$ &  $0.899(   9)$ &  $0.943(   9)$ \\ 
$1.5$ & $0.982(   9)$  &   $0.871(   8)$ &  $0.922(   9)$ &  $0.957(   9)$ \\ 
$2.0$ & $1.019(  10)$  &   $0.905(   8)$ &  $0.938(   9)$ &  $0.966(   9)$ \\ 
$2.5$ & $1.054(  10)$  &   $0.935(   9)$ &  $0.953(   9)$ &  $0.975(   9)$ \\ 
$3.0$ & $1.084(  11)$  &   $0.962(  10)$ &  $0.965(  10)$ &  $0.982(  10)$ \\ 
$3.5$ & $1.110(  11)$  &   $0.985(   9)$ &  $0.975(   9)$ &  $0.986(   9)$ \\ 
$4.0$ & $1.122(   5)$  &   $0.996(   4)$ &  $0.973(   4)$ &  $0.980(   4)$ \\ 
 \hline 
$\bm{\omega_0 = 4.0}$ &  &  &  &  \\ 
 $0.5$ & $0.846(  10)$  &   $0.751(   9)$ &  $0.853(  10)$ &  $0.915(  11)$ \\ 
$1.0$ & $0.901(   9)$  &   $0.799(   8)$ &  $0.880(   9)$ &  $0.931(   9)$ \\ 
$1.5$ & $0.945(   9)$  &   $0.839(   8)$ &  $0.902(   9)$ &  $0.944(   9)$ \\ 
$2.0$ & $0.982(  10)$  &   $0.871(   9)$ &  $0.918(   9)$ &  $0.954(   9)$ \\ 
$2.5$ & $1.014(  10)$  &   $0.900(   9)$ &  $0.932(   9)$ &  $0.962(  10)$ \\ 
$3.0$ & $1.044(  11)$  &   $0.927(  10)$ &  $0.945(  10)$ &  $0.969(  10)$ \\ 
$3.5$ & $1.068(  10)$  &   $0.948(   9)$ &  $0.954(   9)$ &  $0.973(  10)$ \\ 
 $4.0$ & $1.080(   5)$  &   $0.959(   5)$ &  $0.952(   5)$ &  $0.967(   5)$ \\
 \hline
 \end{tabular}
\ec
\caption{Running for the quark mass, $\gamma_\mu$-projector for
      $(\mu_0,\mu) =  (2.5 \text{GeV}, 1.5 \text{GeV})$ (cont.).}
\label{tab:g_ssf_non_deg_om2}.
\end{table}

\clearpage
\begin{table}
\bc
\begin{tabular}{ |c| c | c  c  c| } \hline
  $ (\mu_0,\mu) = (2.5,1.5)$ & $ \sigma_m$  &  $\sigma_m /LO$   &  $\sigma_m /NLO$   &  $\sigma_m /NNLO$ \\  
  $\omega\downarrow$ 
  &  &  &  &  \\ 
 \hline 
$\bm{\omega_0 = 0.5}$ &  &  &  &  \\ 
 $0.5$ & $1.194(  23)$  &   $1.059(  21)$ &  $1.012(  20)$ &  $0.992(  20)$ \\ 
$1.0$ & $1.294(  26)$  &   $1.148(  23)$ &  $1.052(  21)$ &  $0.986(  20)$ \\ 
$1.5$ & $1.378(  29)$  &   $1.223(  26)$ &  $1.085(  23)$ &  $1.047(  22)$ \\ 
$2.0$ & $1.447(  33)$  &   $1.284(  30)$ &  $1.109(  26)$ &  $1.032(  24)$ \\ 
$2.5$ & $1.513(  38)$  &   $1.343(  34)$ &  $1.133(  29)$ &  $1.081(  27)$ \\ 
$3.0$ & $1.574(  42)$  &   $1.397(  37)$ &  $1.155(  30)$ &  $1.069(  28)$ \\ 
$3.5$ & $1.631(  41)$  &   $1.448(  36)$ &  $1.175(  29)$ &  $1.112(  28)$ \\ 
$4.0$ & $1.661(  40)$  &   $1.474(  35)$ &  $1.177(  28)$ &  $1.084(  26)$ \\ 
 \hline 
$\bm{\omega_0 = 1.0}$ &  &  &  &  \\ 
 $0.5$ & $1.050(  17)$  &   $0.932(  15)$ &  $0.943(  15)$ &  $1.002(  16)$ \\ 
$1.0$ & $1.138(  18)$  &   $1.010(  16)$ &  $0.980(  16)$ &  $0.995(  16)$ \\ 
$1.5$ & $1.212(  20)$  &   $1.076(  17)$ &  $1.010(  16)$ &  $1.057(  17)$ \\ 
$2.0$ & $1.274(  22)$  &   $1.130(  20)$ &  $1.033(  18)$ &  $1.043(  18)$ \\ 
$2.5$ & $1.332(  26)$  &   $1.182(  23)$ &  $1.056(  21)$ &  $1.092(  22)$ \\ 
$3.0$ & $1.386(  29)$  &   $1.230(  26)$ &  $1.076(  22)$ &  $1.081(  22)$ \\ 
$3.5$ & $1.436(  28)$  &   $1.275(  24)$ &  $1.095(  21)$ &  $1.123(  22)$ \\ 
$4.0$ & $1.461(  33)$  &   $1.297(  29)$ &  $1.096(  24)$ &  $1.095(  24)$ \\ 
 \hline 
$\bm{\omega_0 = 1.5}$ &  &  &  &  \\ 
 $0.5$ & $0.985(  11)$  &   $0.874(  10)$ &  $0.924(  10)$ &  $0.933(  10)$ \\ 
$1.0$ & $1.068(  10)$  &   $0.948(   8)$ &  $0.960(   9)$ &  $0.927(   8)$ \\ 
$1.5$ & $1.138(  13)$  &   $1.010(  11)$ &  $0.991(  11)$ &  $0.985(  11)$ \\ 
$2.0$ & $1.196(  16)$  &   $1.061(  15)$ &  $1.013(  14)$ &  $0.972(  13)$ \\ 
$2.5$ & $1.250(  21)$  &   $1.110(  19)$ &  $1.035(  17)$ &  $1.018(  17)$ \\ 
$3.0$ & $1.301(  24)$  &   $1.155(  21)$ &  $1.055(  19)$ &  $1.007(  19)$ \\ 
$3.5$ & $1.348(  21)$  &   $1.197(  19)$ &  $1.074(  17)$ &  $1.047(  17)$ \\ 
$4.0$ & $1.372(  19)$  &   $1.218(  17)$ &  $1.075(  15)$ &  $1.020(  14)$ \\ 
 \hline 
$\bm{\omega_0 = 2.0}$ &  &  &  &  \\ 
 $0.5$ & $0.921(   9)$  &   $0.818(   8)$ &  $0.896(   9)$ &  $0.961(   9)$ \\ 
$1.0$ & $0.998(   9)$  &   $0.886(   8)$ &  $0.930(   8)$ &  $0.954(   8)$ \\ 
$1.5$ & $1.064(  10)$  &   $0.944(   9)$ &  $0.960(   9)$ &  $1.014(  10)$ \\ 
$2.0$ & $1.117(  13)$  &   $0.992(  12)$ &  $0.981(  12)$ &  $1.000(  12)$ \\ 
$2.5$ & $1.168(  17)$  &   $1.037(  15)$ &  $1.002(  15)$ &  $1.047(  16)$ \\ 
$3.0$ & $1.216(  20)$  &   $1.079(  18)$ &  $1.022(  17)$ &  $1.036(  17)$ \\ 
$3.5$ & $1.260(  18)$  &   $1.118(  16)$ &  $1.040(  15)$ &  $1.077(  15)$ \\ 
$4.0$ & $1.283(  19)$  &   $1.139(  17)$ &  $1.041(  16)$ &  $1.051(  16)$ \\  
 \hline
 \end{tabular}
 \ec
 \caption{Same as Table~\ref{tab:g_ssf_non_deg_om2} for the $\qslash$-projector.
   Again we have $(\mu_0,\mu) =  (2.5 \text{GeV}, 1.5 \text{GeV})$.}
\label{tab:q_ssf_non_deg_om}.
\end{table}

\begin{table}
\bc
\begin{tabular}{ |c| c | c  c  c| } \hline
  $ (\mu_0,\mu) = (2.5,1.5)$ & $ \sigma_m$  &  $\sigma_m /LO$   &  $\sigma_m /NLO$   &  $\sigma_m /NNLO$ \\  
  $\omega\downarrow$ 
  &  &  &  &  \\ 
  \hline
  $\bm{\omega_0 = 2.5}$ &  &  &  &  \\ 
 $0.5$ & $0.862(   7)$  &   $0.765(   6)$ &  $0.864(   7)$ &  $0.889(   7)$ \\ 
$1.0$ & $0.935(   7)$  &   $0.830(   6)$ &  $0.899(   7)$ &  $0.884(   7)$ \\ 
$1.5$ & $0.997(   8)$  &   $0.885(   7)$ &  $0.927(   7)$ &  $0.939(   7)$ \\ 
$2.0$ & $1.048(  10)$  &   $0.930(   9)$ &  $0.949(   9)$ &  $0.928(   9)$ \\ 
$2.5$ & $1.096(  13)$  &   $0.973(  11)$ &  $0.970(  11)$ &  $0.971(  11)$ \\ 
$3.0$ & $1.141(  15)$  &   $1.013(  13)$ &  $0.989(  13)$ &  $0.961(  13)$ \\ 
$3.5$ & $1.181(  12)$  &   $1.048(  11)$ &  $1.005(  11)$ &  $0.998(  10)$ \\ 
$4.0$ & $1.203(  21)$  &   $1.068(  19)$ &  $1.007(  18)$ &  $0.974(  17)$ \\ 
 \hline 
$\bm{\omega_0 = 3.0}$ &  &  &  &  \\ 
 $0.5$ & $0.818(   6)$  &   $0.726(   5)$ &  $0.842(   6)$ &  $0.912(   7)$ \\ 
$1.0$ & $0.887(   7)$  &   $0.787(   6)$ &  $0.875(   7)$ &  $0.906(   7)$ \\ 
$1.5$ & $0.945(   7)$  &   $0.839(   6)$ &  $0.903(   6)$ &  $0.962(   7)$ \\ 
$2.0$ & $0.993(   8)$  &   $0.881(   7)$ &  $0.923(   8)$ &  $0.949(   8)$ \\ 
$2.5$ & $1.038(  11)$  &   $0.921(  10)$ &  $0.943(  10)$ &  $0.994(  11)$ \\ 
$3.0$ & $1.080(  13)$  &   $0.959(  12)$ &  $0.962(  12)$ &  $0.984(  12)$ \\ 
$3.5$ & $1.119(  11)$  &   $0.994(  10)$ &  $0.978(  10)$ &  $1.022(  10)$ \\ 
$4.0$ & $1.139(  20)$  &   $1.011(  18)$ &  $0.979(  17)$ &  $0.996(  17)$ \\ 
 \hline 
$\bm{\omega_0 = 3.5}$ &  &  &  &  \\ 
 $0.5$ & $0.780(   6)$  &   $0.692(   5)$ &  $0.822(   6)$ &  $0.859(   7)$ \\ 
$1.0$ & $0.845(   6)$  &   $0.750(   5)$ &  $0.855(   6)$ &  $0.853(   6)$ \\ 
$1.5$ & $0.901(   6)$  &   $0.800(   5)$ &  $0.882(   6)$ &  $0.906(   6)$ \\ 
$2.0$ & $0.947(   8)$  &   $0.840(   7)$ &  $0.902(   7)$ &  $0.894(   7)$ \\ 
$2.5$ & $0.990(  11)$  &   $0.879(  10)$ &  $0.921(  10)$ &  $0.936(  10)$ \\ 
$3.0$ & $1.030(  13)$  &   $0.914(  12)$ &  $0.939(  12)$ &  $0.926(  12)$ \\ 
$3.5$ & $1.068(  11)$  &   $0.948(   9)$ &  $0.956(  10)$ &  $0.963(  10)$ \\ 
$4.0$ & $1.086(  17)$  &   $0.964(  15)$ &  $0.957(  15)$ &  $0.939(  14)$ \\ 
 \hline 
$\bm{\omega_0 = 4.0}$ &  &  &  &  \\ 
$0.5$ & $0.741(   8)$  &   $0.658(   7)$ &  $0.799(   9)$ &  $0.871(   9)$ \\ 
$1.0$ & $0.804(   8)$  &   $0.713(   7)$ &  $0.830(   8)$ &  $0.865(   8)$ \\ 
$1.5$ & $0.856(   9)$  &   $0.760(   8)$ &  $0.856(   9)$ &  $0.919(  10)$ \\ 
$2.0$ & $0.900(  12)$  &   $0.799(  10)$ &  $0.876(  11)$ &  $0.907(  12)$ \\ 
$2.5$ & $0.941(  15)$  &   $0.835(  13)$ &  $0.894(  14)$ &  $0.949(  15)$ \\ 
$3.0$ & $0.979(  17)$  &   $0.869(  15)$ &  $0.912(  16)$ &  $0.940(  16)$ \\ 
$3.5$ & $1.014(  15)$  &   $0.900(  13)$ &  $0.928(  13)$ &  $0.977(  14)$ \\ 
$4.0$ & $1.033(  12)$  &   $0.917(  10)$ &  $0.929(  10)$ &  $0.952(  11)$ \\ 
 \hline
 \end{tabular}
\ec
 \caption{Running for the $\qslash$-projector (cont).}
\label{tab:q_ssf_non_deg_om2}.
\end{table}

  \clearpage
\section{Continuum extrapolation}
\label{app:CL}

\begin{figure}[t]
  \bc
  \begin{tabular}{cc}
    \includegraphics[width=0.50\textwidth]{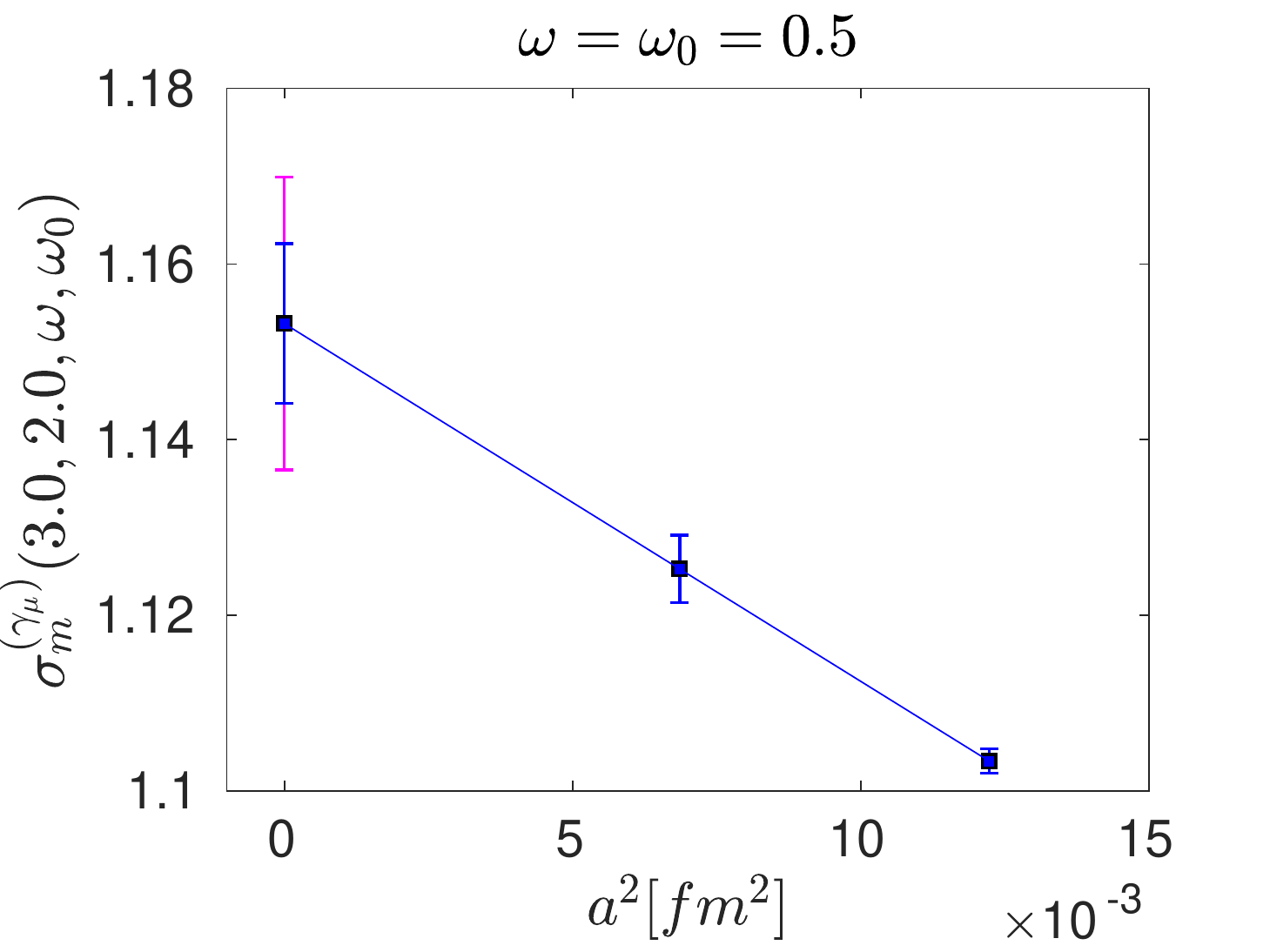} &
    \includegraphics[width=0.50\textwidth]{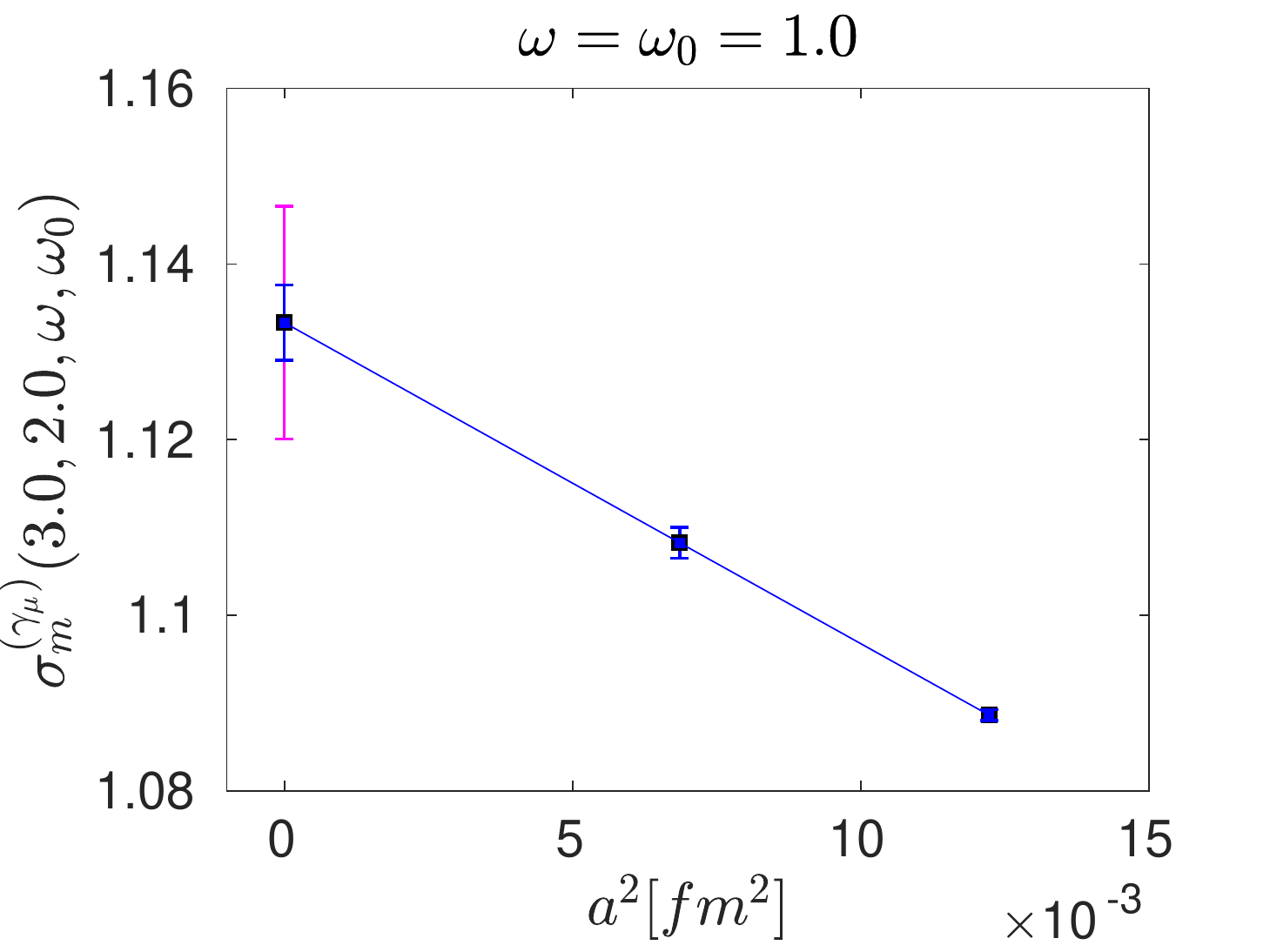} \\
    \includegraphics[width=0.50\textwidth]{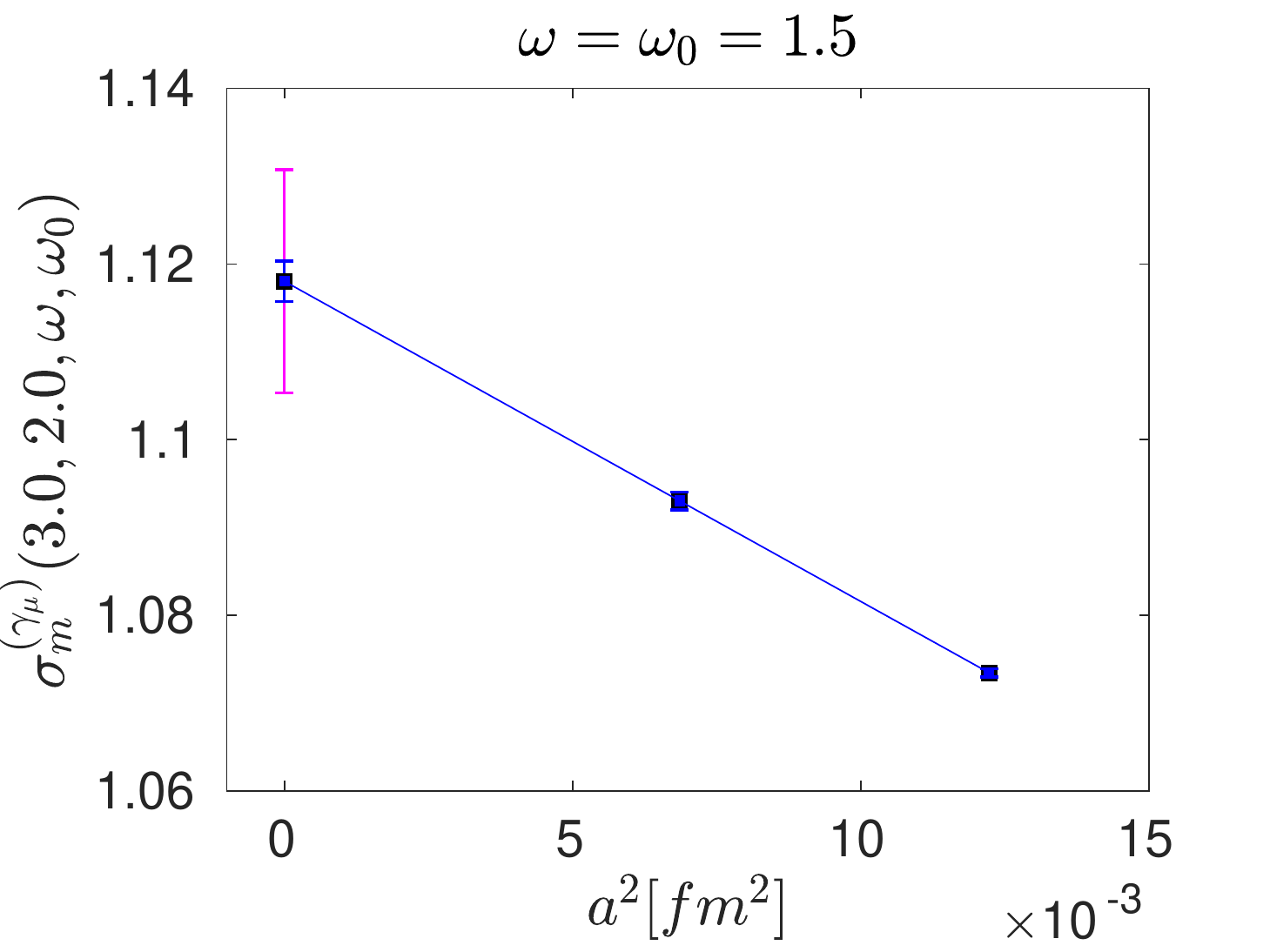} &
    \includegraphics[width=0.50\textwidth]{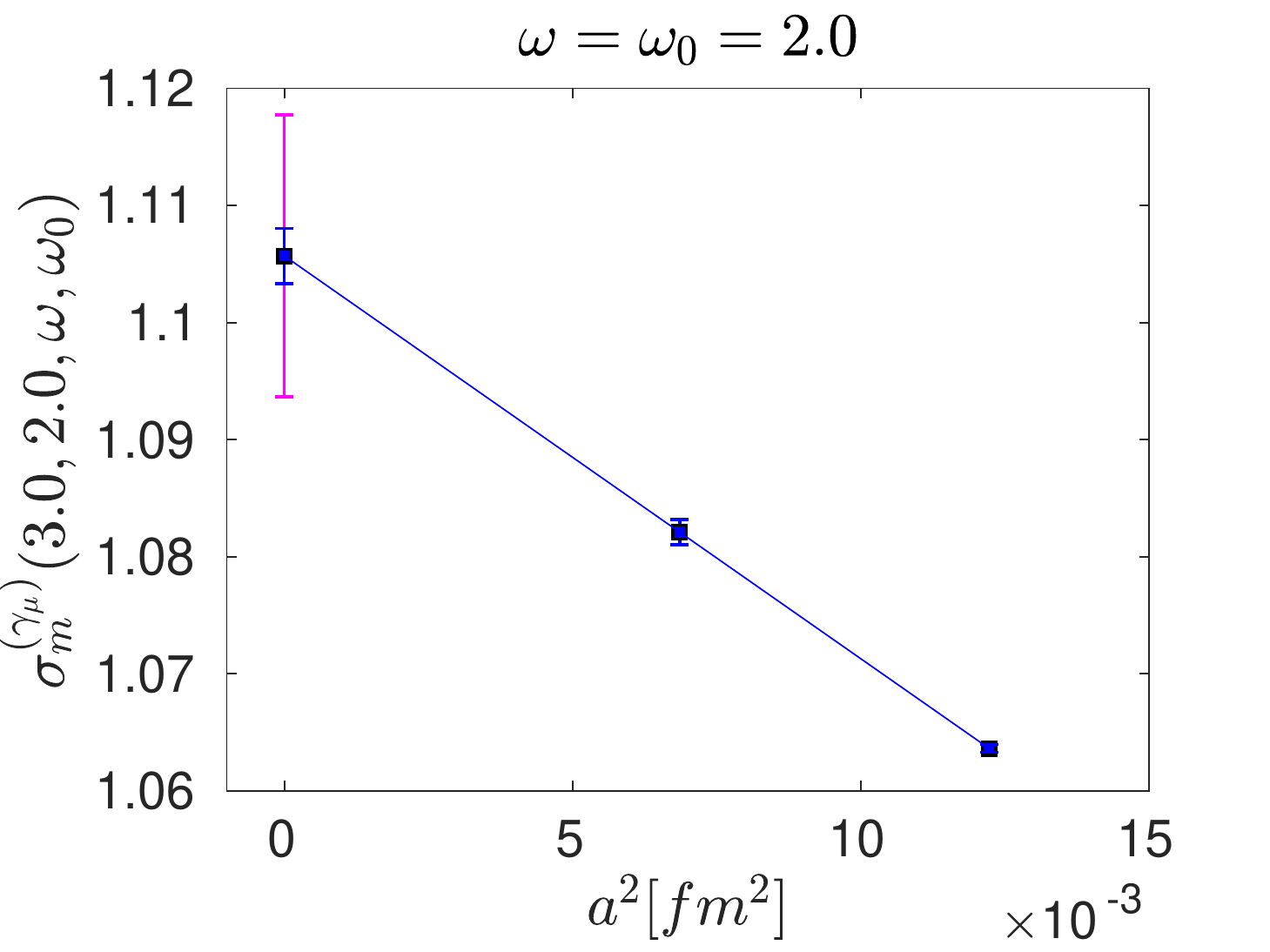} \\
    \end{tabular}
\ec
\caption{Continuum extrapolations for $\Sigma_m$. We choose
$(X,\mu,\mu_0) = (\gamma_\mu, 2\mbox{ GeV}, 3\mbox{ GeV})$
and show our results for the different values of $\omega=\omega_0$ from
0.5 (top left corner) to 2.0 (bottom right corner).
We included an estimation of the error due to the continuum extrapolation (error bar
in magenta, see text).  }
\label{fig:Sigmam_CL_1}
\end{figure}

\begin{figure}[t]
  \bc
  \begin{tabular}{cc}
    \includegraphics[width=0.50\textwidth]{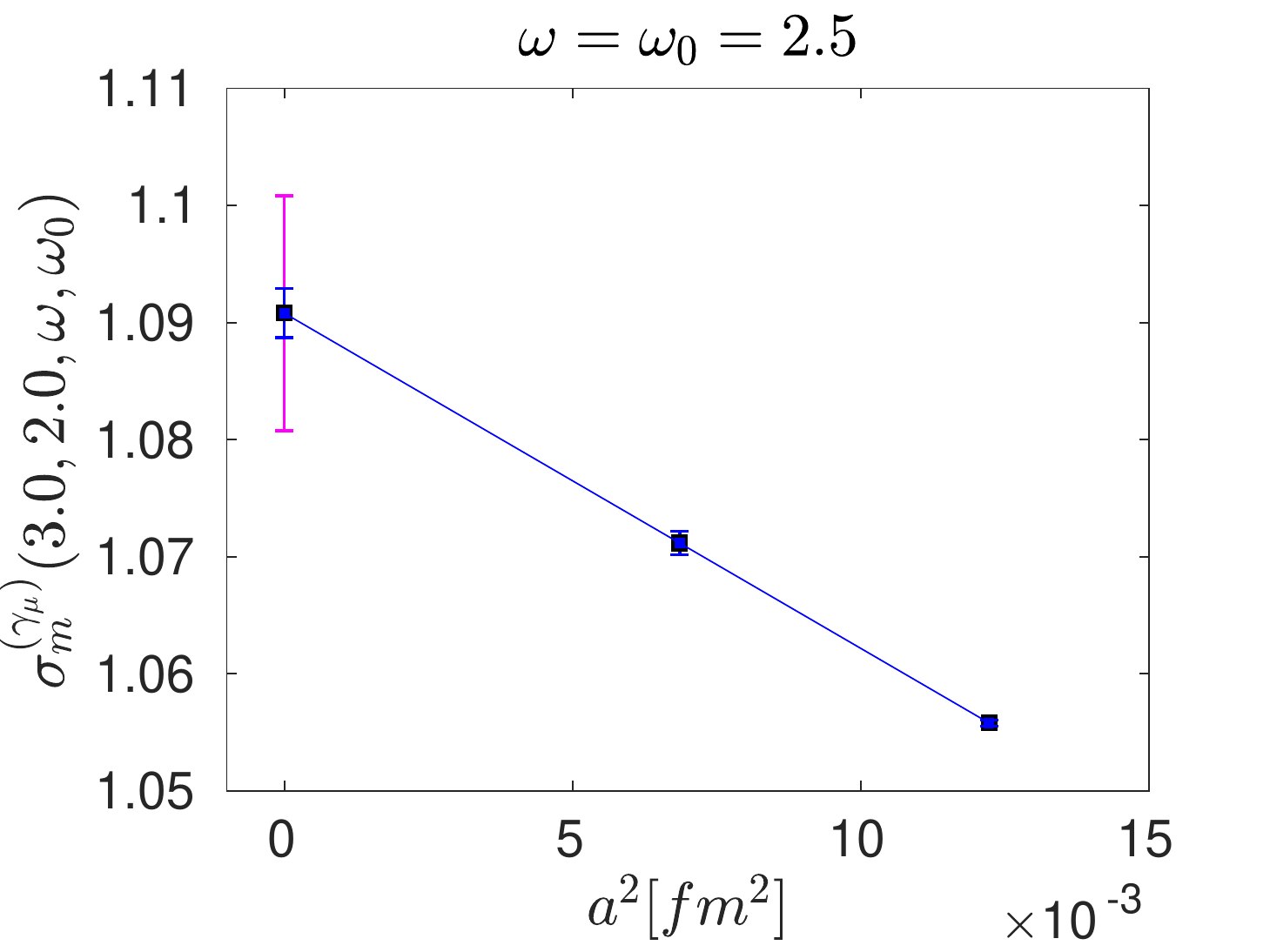} &
    \includegraphics[width=0.50\textwidth]{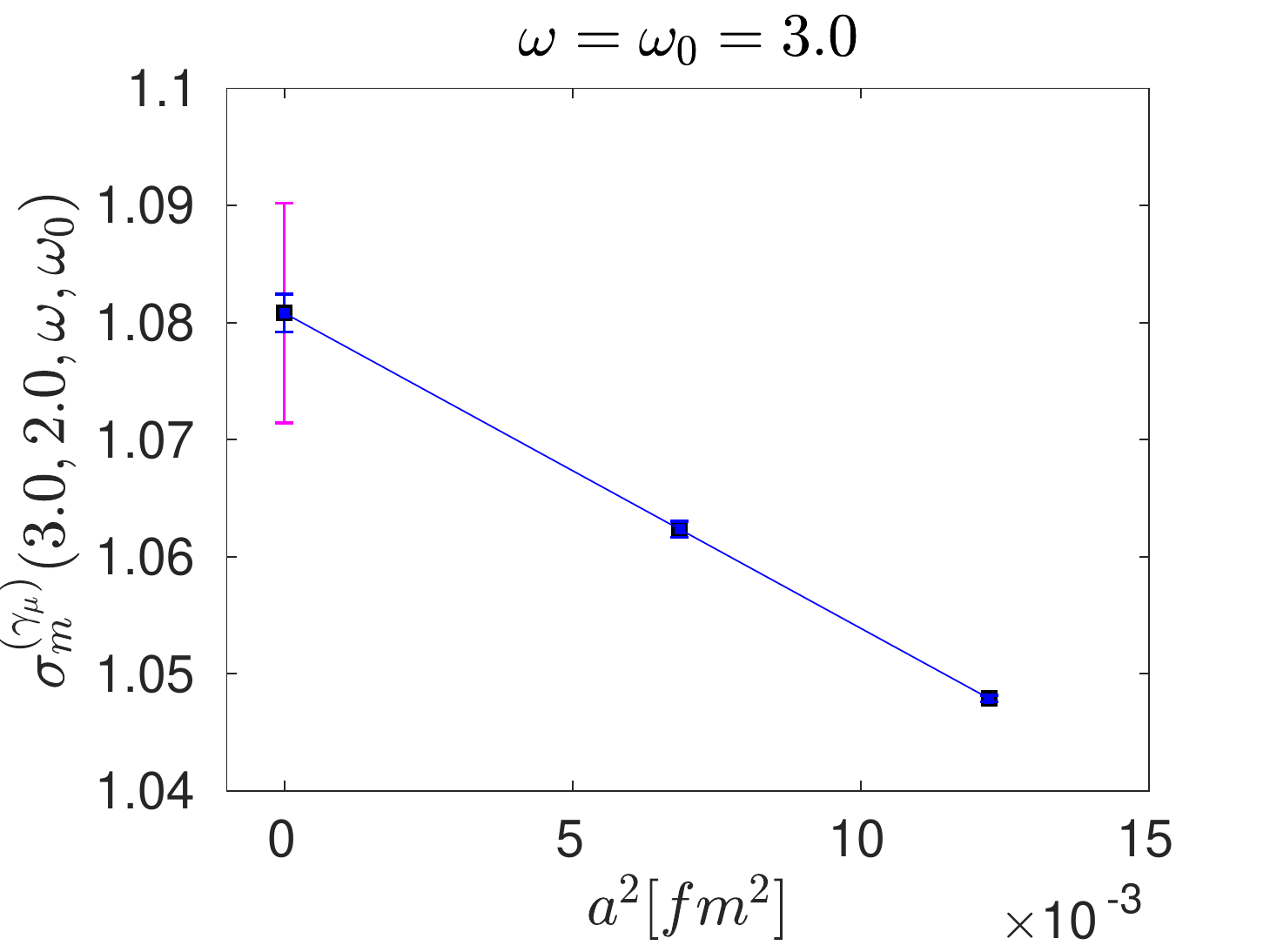} \\
    \includegraphics[width=0.5\textwidth]{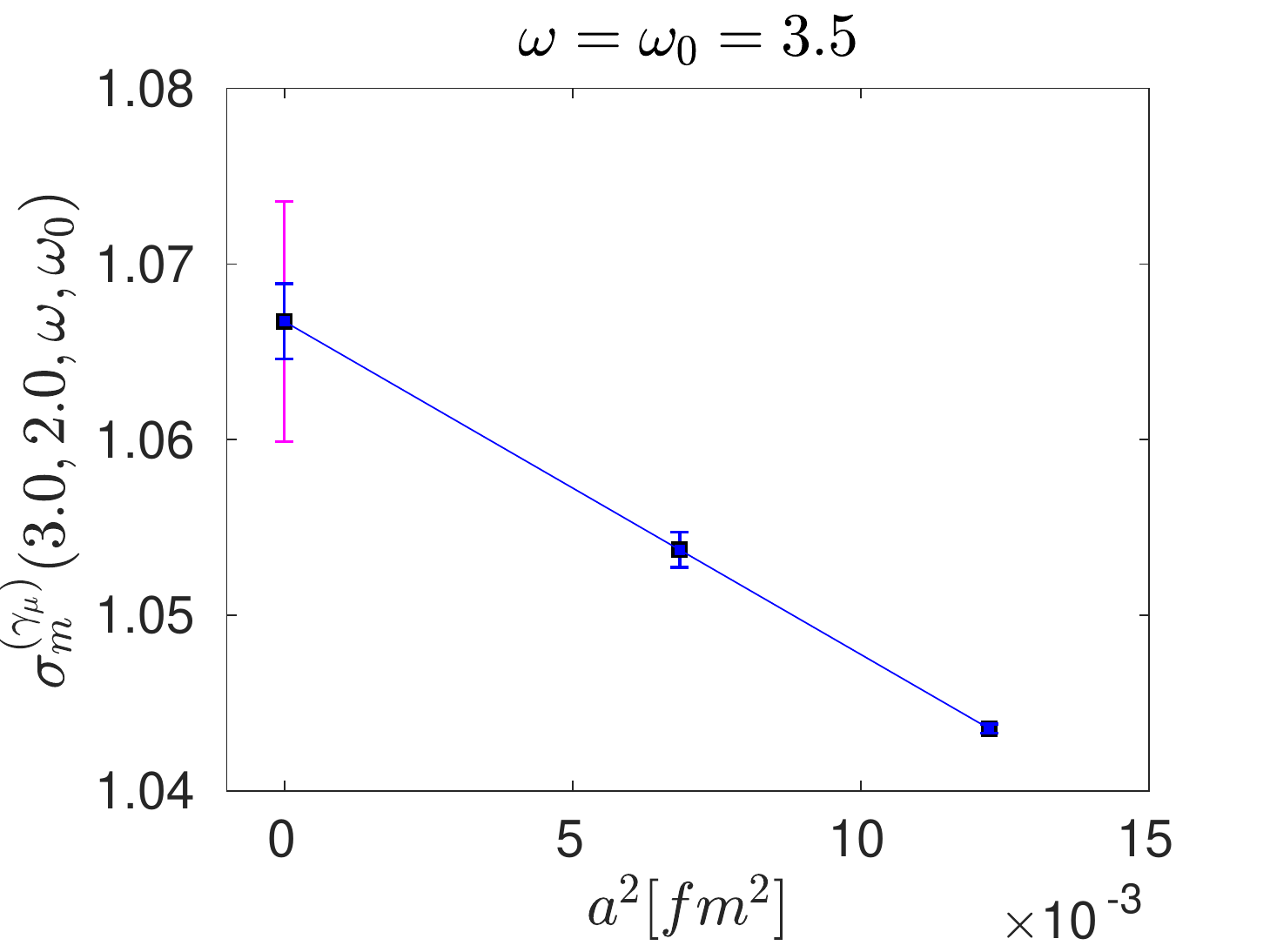} &
    \includegraphics[width=0.5\textwidth]{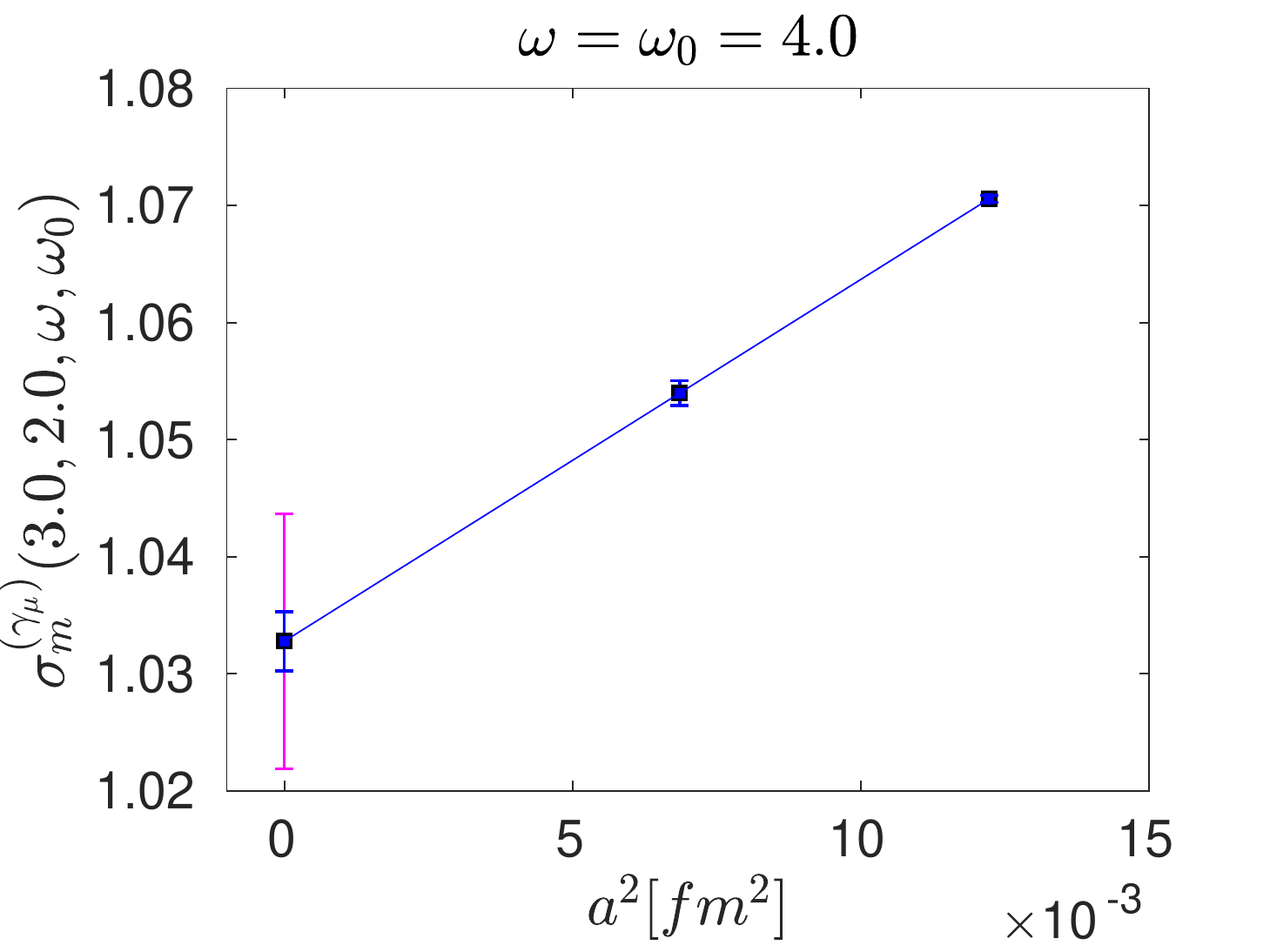} 
    \end{tabular}
\ec
\caption{Same as Fig.~\ref{fig:Sigmam_CL_1} for $\omega=\omega_0 = 2.5, 3.0,\ldots, 4.0$.}
\label{fig:Sigmam_CL_2}
\end{figure}

As an example of continuum extrapolations, we show $\Sigma_m^{(\gamma_\mu)}(\mu,\mu_0, \omega,\omega_0)$
in Figs.~\ref{fig:Sigmam_CL_1} and~\ref{fig:Sigmam_CL_2}. The magenta error bar
is a systematic error, obtained by adding half the difference beween the extrapolated value,
$\sigma_m^{(\gamma_\mu)}(\mu,\mu_0, \omega,\omega_0)$, and the value obtained on the finest lattice.

\clearpage

\end{appendices}

\bibliography{biblio}
\bibliographystyle{utphys}
\end{document}